\documentclass[
longbibliography,
amsmath,amssymb,
aps,
pra,
]{revtex4-2}
\usepackage{graphicx}
\usepackage{dcolumn}
\usepackage{bm}
\usepackage{epsfig}
\usepackage{afterpage}
\usepackage{lineno}

\begin{document}


\title{Phase Diagrams of Kitaev Models for Arbitrary Magnetic-Field Orientations}

\author{F. Y{\i}lmaz}
\email{firat.yilmaz@uni-a.de}
\affiliation{Theoretical Physics III, Center for Electronic Correlations and Magnetism, Institute of Physics, University of Augsburg, Augsburg 86135, Germany}
\author{A. P. Kampf}
\affiliation{Theoretical Physics III, Center for Electronic Correlations and Magnetism, Institute of Physics, University of Augsburg, Augsburg 86135, Germany}
\author{S. K. Yip}
\affiliation{Institute of Physics, Academia Sinica, Taipei 115, Taiwan}
\affiliation{Institute of Atomic and Molecular Sciences, Academia Sinica, Taipei 115, Taiwan}




\date{\today}

\begin{abstract}
The Kitaev model is an exactly solvable quantum spin model within the language of the constrained real fermions. In spite of numerous studies along special magnetic-field orientations, there is a limited amount of knowledge on the complete field-angle characterization, which can provide valuable information on the existence of fractionalized excitations. For this purpose, we first extend previous studies on the field-angle response of the ferromagnetic Kitaev model to its antiferromagnetic version. Yet, the realistic description of the candidate Kitaev materials, within the edge-sharing octahedra paradigm, require additional coupling terms, including a large off-diagonal term $\Gamma$ along with possible anisotropic corrections $\Gamma_p$. It is therefore not sufficient to depend on the topological properties of the bare Kitaev model as the only source for the observed thermal Hall conductivity signals and an understanding of these extended Kitaev models with a complete field response is demanded. Starting from the zero-field phase diagram of realistic K-$\Gamma$-$\Gamma_p$ models, we identify antiferromagnetic zig-zag and (partially) polarized phases as well as two unusual Kitaev(-$\Gamma$) spin-liquid phases. The magnetic field response of these phases for arbitrary field orientations provides a remarkably rich phase diagram. A partially polarized phase is revealed between two ordered phases with a suppressed magnetization, finite fractionalization and finite Chern number. This phase is characterized as an extended Kitaev-$\Gamma$ spin-liquid. To comply our findings with experiments, we reproduce the asymmetry in the extent of the intermediate phases specifically for the two different field directions $\theta = \pm 60^o$.
\begin{description}
\item[PACS numbers]
\end{description}
\end{abstract}

\pacs{Valid PACS appear here}
\maketitle

\section{Introduction}
The Kitaev model is an exactly solvable \cite{kitaev2006anyons} quantum spin model with fractionalized excitations within the family of spin-liquid models. Candidate materials for its $J_{eff} = 1/2$ ferromagnetic (F) realization are the iridates X$_2$IrO$_{3}$ (X: Li \cite{williams2016incommensurate, majumder2018breakdown}, Na \cite{choi2012spin, ye2012direct, singh2012relevance, gretarsson2013magnetic,yamaji2014first,alpichshev2015confinement}) and $\alpha$-RuCl$_3$ \cite{wulferding2020magnon,kubota2015successive, lampen2018anisotropic, plumb2014alpha, baek2017evidence, wang2017magnetic, zheng2017gapless,lampen2018field, banerjee2018excitations,sandilands2015scattering, sandilands2016spin, majumder2015anisotropic,bachus2020thermodynamic,reschke2019terahertz, zhou2022intermediate}. Recently, there is a growing interest in the search of additional materials \cite{liu2018pseudospin,liu2020kitaev,lin2021field} with perhaps antiferromagnetic (AF) Kitaev type couplings such as Na$_3$Co$_2$SbO$_6$-Na$_2$Co$_2$TeO$_6$ \cite{kim2020antiferromagnetic} and the f-electron based rare-earth oxides \cite{jang2019antiferromagnetic}. The latter quantum magnets have significant spin-orbit coupling magnitude and frustration. Detailed analyses of the underlying physics and the material characteristics have been summarized in numerous reviews \cite{motome2020hunting,winter2017models,balents2010spin,savary2016quantum,zhou2017quantum,broholm2020quantum,knolle2019field}. 

The Kitaev model is defined on a honeycomb-lattice with highly anisotropic Ising type exchange couplings at each bond direction.
\begin{eqnarray}\label{KitaevHamiltonian}
H_{K} &=& \sum_{\langle ij \rangle}^{\alpha-bond} K^\alpha S_i^\alpha S_j^\alpha.
\end{eqnarray}
It can host anyonic excitations within the real fermion language. In this approach, each of the Pauli operators is replaced by two composite Majorana fermions $\sigma_j^\alpha = i b^\alpha_j c_j$, $\alpha \in \{ x,y,z\}$ with the local constraint $D_j = b^x_j b^y_j b^z_j c_j = 1$ at site $j$. The basic theoretical work in materializing such an exotic model relies on the Jackeli-Khaliullin mechanism \cite{jackeli2009mott,rau2014generic}. The effective Hamiltonians for transition metals are coupled with additional (extended) edge-sharing \cite{li2017kitaev,motome2020materials} octahedra ligands. In the presence of sufficiently large spin-orbit coupling could eliminate the leading Heisenberg exchange coupling which is typically the dominant term. Thereby the additional Kitaev and other off-diagonal terms become dominant terms. Eventually, most of the recent work has been allocated to the theoretical \cite{janssen2017magnetization,katukuri2014kitaev,laurell2020dynamical, jackeli2009mott,kim2015kitaev} and experimental investigation of the candidate materials. Experimental data suggest that it is highly likely for the candidate materials to have a spin-liquid groundstate (GS) for intermediate magnetic field strengths. This is inferred, in particular, from the continuum of excitations in neutron-scattering experiments \cite{wu2018field}, susceptibility signals for a magnetic instability in the specific-heat data and the magnetic Gr\"uneisen parameter \cite{bachus2020thermodynamic} . A significant amount of work has also focused on alternative mechanisms such as chiral conventional quasiparticles \cite{cookmeyer2018spin,mcclarty2018topological,ye2020phonon,lefranccois2021evidence} or beyond-Kitaev model spin-liquids \cite{gao2019thermal}. In all approaches, the anti/ferromagnetic Kitaev term has to be supplemented by additional interactions for a realistic material specific modeling. Depending on their relative strengths, the direct relation between the thermal Hall coefficient and the Kitaev interaction becomes obscure. 

For an established spin-liquid GS, the thermal Hall conductivity (THC) provides a tool to rule in/out a Kitaev spin-liquid state with a half-quantized neutral edge current as its distinct signature. Such a measurement requires a finite applied magnetic field so that the gapless, topologically trivial Kitaev liquid could go into a gapped, topologically non-trivial phase. Recent measurements on $\alpha$-RuCl$_3$ indicate a large \cite{hentrich2019large} and quantized (if not sample dependent) THC \cite{kasahara2018majorana} within a certain window of in-plane magnetic field strength. This region with a suppressed magnetization is located between staggered ferromagnetic chains (known as the ZZ-$\gamma$ phase) and the field polarized (P) phase. Moreover, the anomalous thermal Hall effect even in the absence of out-of-plane magnetic field components \cite{yokoi2020half} may support the topological origin of a transverse heat current. In this respect, a further characterization of the "intermediate phase" for arbitrary magnetic field and directions could provide valuable information for the search of (non-)Abelian anyons. 

In spite of numerous studies \cite{nasu2018successive,janssen2019heisenberg,ralko2020novel, liang2018intermediate,ido2020correlation,gordon2019theory, catuneanu2018path,liu2018dirac,ronquillo2019signatures} along special field directions such as [001] and [111], there exists limited knowledge on the complete field-angle dependence of $\kappa_{xy}$ in the Kitaev model. For example, in \cite{yokoi2020half}, the half-quantized THC measurements on $\alpha$-RuCl$_3$ has been attributed to the pure Kitaev spin liquid. Yet, it could be a naive approximation if the additional off-diagonal couplings are significant. The studies on the candidate materials indicate the presence of a relatively large spin-orbit coupling $\Gamma$ term. Ref. \cite{takikawa2019impact} has addressed this issue and also the magnetic field effects perturbatively, yet assuming a KSL GS. In this respect, it is crucial to clarify the effect of large $\Gamma \sim \left| K \right| $ \cite{hwang2020identification,yamada2021quantum} as well as arbitrary magnetic field orientations and strengths to understand the validity of existing results based on the putative F Kitaev GS. Naturally, an unbiased method is needed where the ordered phases and the fractional phases are treated on equal footing.
\begin{figure}
\includegraphics[width=0.47\textwidth]{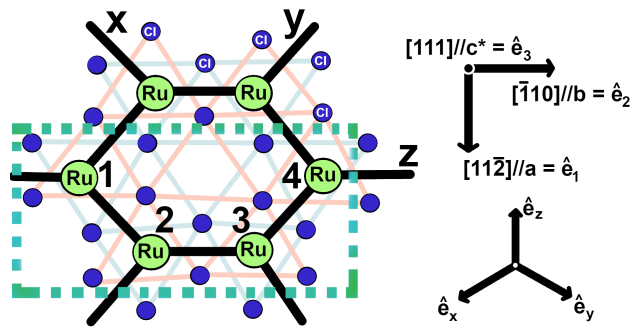}
\caption{(Color online) The crystal structure of $\alpha$-RuCl$_3$. The Ru atoms (green dots) form a honeycomb-lattice (with black bonds) on the plane perpendicular to the crystallographic [111] direction, referred to as the c-axis. The three bonds are labeled as $x,y$ and $z$. The Cl$^-$ ions form edge-sharing octahedra with two sublattices denoted by blue and orange colors. For the Kitaev only models ($\Gamma=0,\Gamma_p = 0$), the unit cell contains two Ru sites connected by the z-bond. For the Kitaev-$\Gamma$ type models ($\Gamma\ne 0,\Gamma_p \ne 0$), a four-site unit cell is necessary to capture the relevant groundstate. It is indicated by a green dotted rectangle with site labels $1-4$.}
\label{pointGroups}
\end{figure}

In this manuscript, we attempt to answer the following two questions: 1. What is the full phase diagram of the antiferromagnetic Kitaev model in the presence of an external magnetic field pointing in arbitrary directions? 2. What are the effects of the additional $\Gamma$-$\Gamma_p$ off-diagonal terms? Regarding the former, we identify a magnetization process with multiple topological distinct phases depending on the field direction and strength. The phase boundaries are characterized by the first Chern number of the fermionic vacuum. Regarding the latter, we examine the phase diagram of the ($K, \Gamma, \Gamma_p$)-model and identified four different zero-field phases: a zig-zag z (ZZ-z), a polarized (P), an extended KSL, and the Kitaev-$\Gamma$ spin-liquid (K$\Gamma$SL) phases. A focus is given to the experimentally relevant ZZ-z phase, in which the ferromagnetic chains are antiferromagnetically aligned perpendicular to the $z$-bond. Our choice of the unit cell in Fig.\ref{pointGroups} allows only ZZ-$z$ phase . This discussion is elaborated further in Sec.\ref{KGGpHsec}. Then, starting from the ZZ-z GS, we examine the magnetic field response as a function of the off-diagonal coupling strength $\Gamma$ . We identify a region with reduced magnetization and decreased energy gap with finite Chern number, named as the partially polarized (PP) phase. This phase is found to host partial fractionalization. 

Throughout this work, we employ a mean-field theory of the Majorana representation with local constraints. The advantage of this approach is that it readily captures the strong correlations inherent within this composite (fractionalized) representation along with the magnetic phases. The detailed derivation of the mean-field equations is provided in the Appendix \ref{MFtheoryDetails}.

The article is divided into three sections. The current introductory section is completed with the additional discussion on the materialization process of the Kitaev model. In section $II$, we identify the thermal Hall coefficients for the pure AF/F Kitaev model for arbitrary magnetic-field orientations and strengths. The critical field strength and the angles for topological phase transitions are determined. In section $III$, the role of the additional off-diagonal terms, $\Gamma,\Gamma_p$ are examined. We complete this section by discussing the implications of our results for the recent thermal Hall experiments on candidate materials.

\subsection{Kitaev materials: $\alpha$-RuCl$_3$}

The experimental side of the search of Kitaev materials has advanced by signatures of massively degenerate, gapless excitations in the iridates XIrO$_3$ and in $\alpha$-RuCl$_3$. The modeling of these specific materials requires additional terms in the spin Hamiltonian. For concernitant materials, the symmetry group is $\mathcal{D}_{3d} = \{\mathcal{I}, 2 \mathcal{C}_3, 3 \mathcal{C}_2, 2 \mathcal{S}_6, 3\mathcal{M} \}$ \cite{jackeli2009mott, rau2014generic, katukuri2014kitaev, janssen2017magnetization}, the symmetry allowed interactions are Kitaev type $K^\alpha S_j^\alpha S_l^\alpha$, Heisenberg type $J \mathbf{S}_j \cdot \mathbf{S}_l$ as well as off-diagonal couplings $\Gamma^\alpha \left( S_j^\beta S_l^\gamma+S_j^\gamma S_l^\beta \right)$, and (when e.g. Ru-Cl environment differs from octahedra) the additional allowed $\Gamma^{\alpha}_p$ terms $\Gamma^{\alpha}_p (S^\beta_j S^\alpha_l+S^\alpha_j S^\beta_l+S^\gamma_j S^\alpha_l+S^\alpha_j S^\gamma_l)$. Therefore, a more general spin coupling term is considered, $\Gamma_{ij}^{\alpha \beta}S_i^\alpha S_j^\beta$ where $\Gamma_{ij}^{\alpha \alpha} \to K,J$ and $\Gamma_{ij}^{\alpha \ne \beta} \to \Gamma, \Gamma_p$. The additional interactions and their role are the subject of current experiments \cite{banerjee2018excitations, lampen2018field, majumder2018breakdown, zheng2017gapless, wang2017magnetic, baek2017evidence, williams2016incommensurate, plumb2014alpha, lampen2018anisotropic,
kubota2015successive}, ab-initio calculations \cite{kim2016revealing}, exact-diagonalization methods \cite{ lee2020magnetic,winter2016challenges,gohlke2018dynamical} and effective low-energy Hubbard Hamiltonians \cite{wang2017theoretical,winter2017models,winter2017breakdown}. 

Here, we focus on $\alpha$-RuCl$_3$ because there are sufficient number of works to compare in the literature to ensure the validity of our approach. The crystal structure of $\alpha-$RuCl$_3$ has a honeycomb structure in Ru-Ru bonds for a cut in the [111] direction (see Fig.\ref{pointGroups}). The Cl atoms are aligned as edge-sharing octahedras. The additional exchange paths through Ru-Cl bonds -in the presence of strong spin-orbit coupling- create a destructive interference for the otherwise leading term $J$, the Heisenberg exchange coupling \cite{jackeli2009mott,rau2014generic}. Therefore, a Kitaev type exchange coupling along with $\Gamma$ and $\Gamma_p$ terms take stage. Experiments in zero magnetic field suggests a long range ZZ-z order for the GS of $\alpha-$RuCl$_3$ \cite{banerjee2018excitations, lampen2018field, majumder2018breakdown, zheng2017gapless, wang2017magnetic, baek2017evidence, williams2016incommensurate, plumb2014alpha, lampen2018anisotropic,
kubota2015successive}. Yet, a finite magnetic field along x-y bonds melts the magnetic order and thereby allows for a transition to a spin-liquid phase \cite{kasahara2018majorana}. 

For the calculations described below, we switch to new orthogonal coordinate system ($\hat{e}_1,\hat{e}_2,\hat{e}_3$) with the following transformation,
\begin{equation} \label{xyzVS123}
\begin{pmatrix}\hat{e}_1\\\hat{e}_2 \\ \hat{e}_3\end{pmatrix}=\frac{1}{\sqrt{6}}\begin{pmatrix}1&1&-2\\-\sqrt{3}&\sqrt{3}&0\\\sqrt{2}&\sqrt{2}&\sqrt{2}\end{pmatrix} \begin{pmatrix}\hat{e}_x\\\hat{e}_y\\\hat{e}_z\end{pmatrix}.
\end{equation}
Here, $\hat{e}_3$ points along the [111] direction, while $\hat{e}_1$, $\hat{e}_2$ are in plane unit vectors perpendicular to x-bond and y-bond, respectively. 

\begin{figure}
\includegraphics[width=0.47\textwidth]{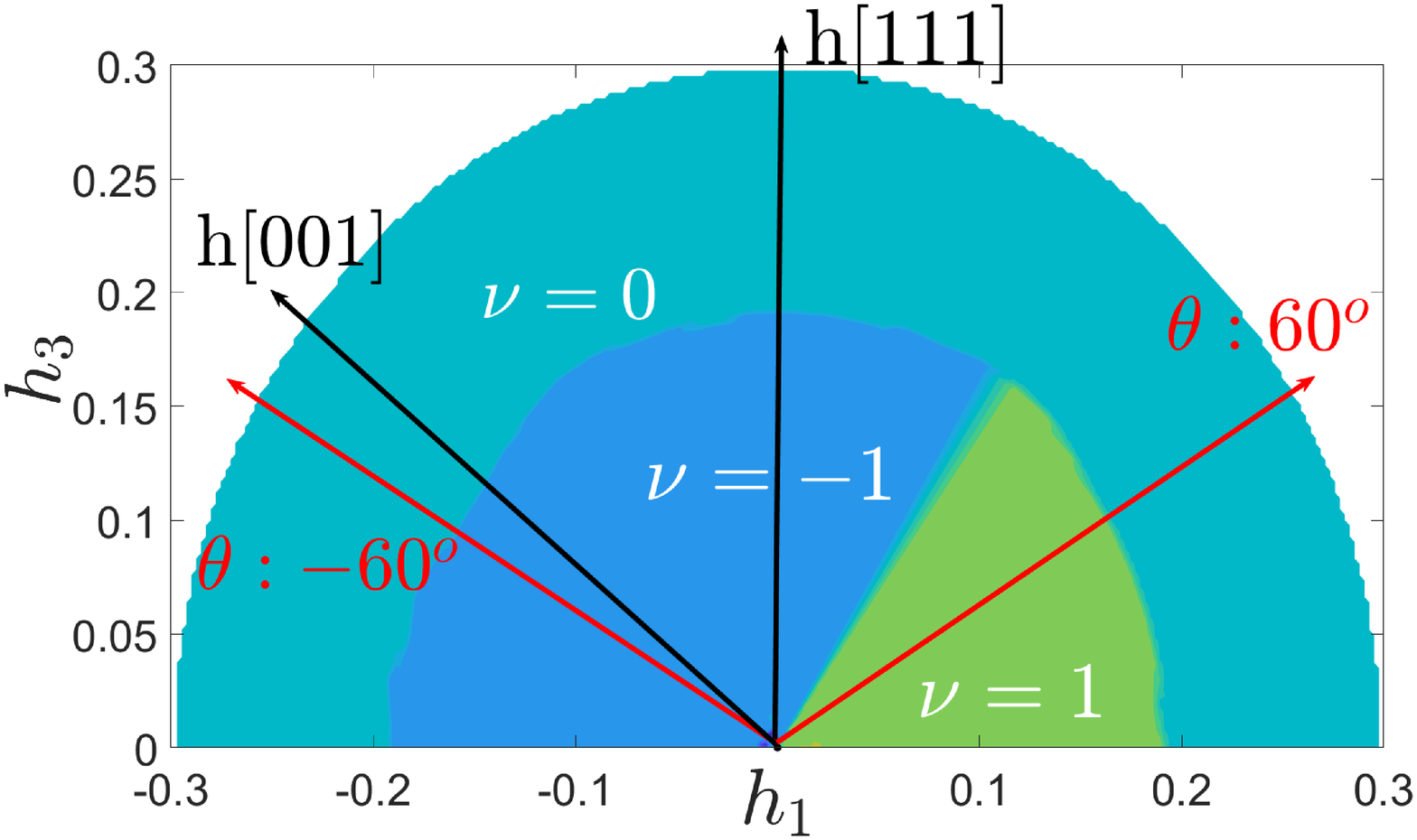}
\caption{(Color online) Ferromagnetic Kitaev model: The Chern number map for arbitrary field direction and strength in the $(h_1-h_3)$ plane, where $h_i = \mathbf{h}\cdot \hat{e}_i$ in units of $\left| K \right|$. Topologically distinct GSs with Chern numbers $\nu = \pm 1$ are identified. The trivial phase $\nu = 0$ corresponds to the high-field polarized phase. The red arrows indicates the experimental magnetic field-angles that is used in Ref.\cite{kasahara2018majorana} with $\theta = \pm 60^o$, where $\theta$ is the clockwise angle w.r.t. the $h_3$ axis. $h[001]$ direction is shown for visualization purposes.}
\label{FKitaev_Chern}
\includegraphics[width=0.47\textwidth]{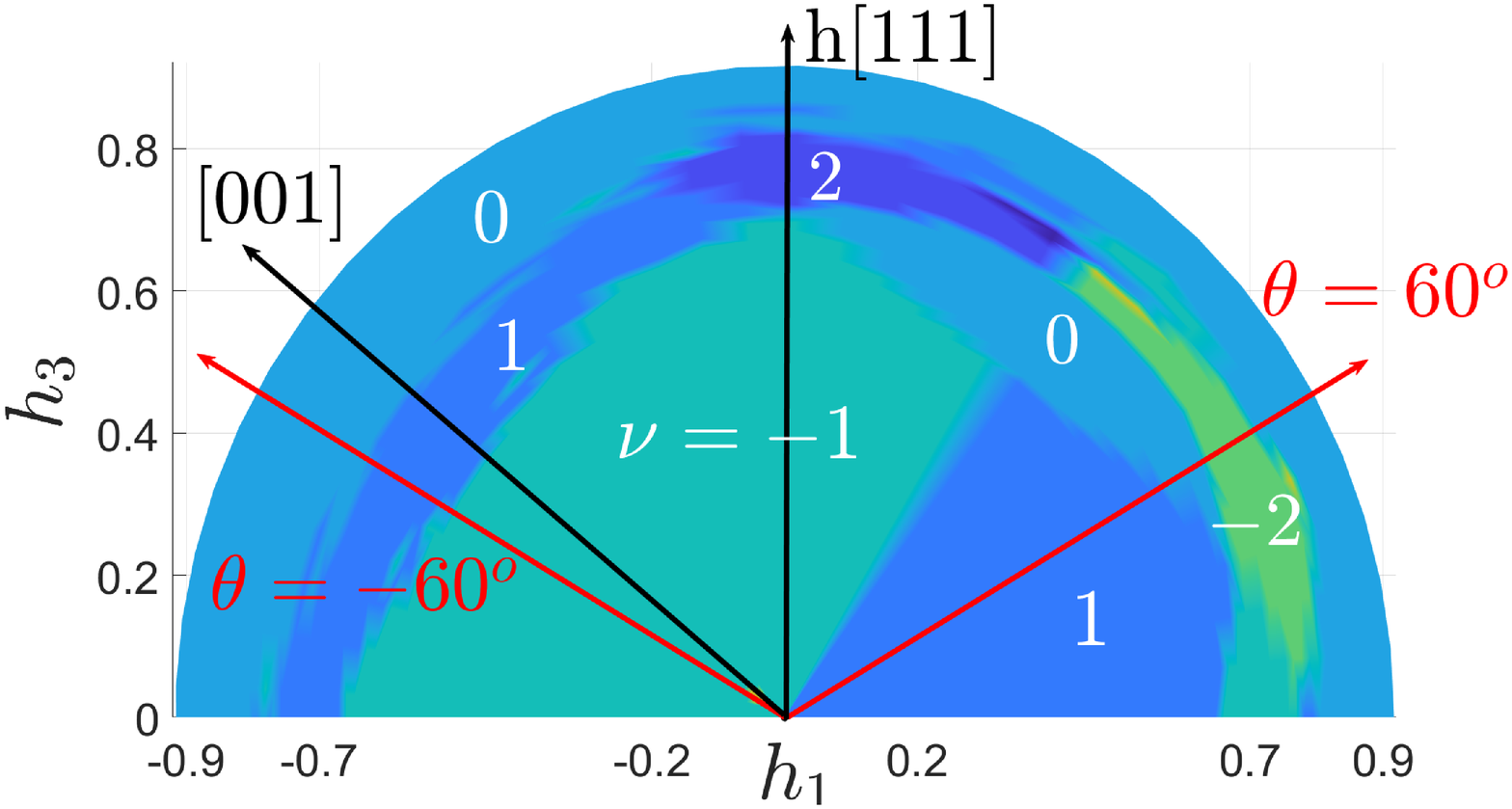}
\caption{(Color online) Antiferromagnetic Kitaev model: The Chern number map for arbitrary field direction and strength in the $(h_1-h_3)$ plane, where $h_i = \mathbf{h}\cdot \hat{e}_i$ in units of $\left| K \right|$. There are in total five topologically distinct GSs possible with Chern numbers $\nu \in \{0,\pm 1, \pm 2 \}$. $\nu = 0$ phase is the field polarized phase. The red arrows indicates the experimental magnetic field-angles that is used in Ref.\cite{kasahara2018majorana} with $\theta = \pm 60^o$, where $\theta$ is the clockwise angle w.r.t. the $h_3$ axis. $h[001]$ direction is shown for visualization purposes.
}
\label{AFKitaev_Chern}
\end{figure}

\section{Kitaev model in an arbitrary magnetic field}
In materials like $\alpha$-RuCl$_3$ and Na$_2$IrO$_3$, the Kitaev exchange coupling along all bond directions are likely to be equal \cite{jackeli2009mott,rau2014generic}. We therefore consider an isotropic Kitaev coupling, $K^\alpha = K$. The isotropic model supports the trivial gapless B-phase \cite{kitaev2006anyons}. It can host two types of fractionalized excitations, the $\mathbb{Z}_2$ vortices and itinerant fermionic excitations, albeit a topologically trivial groundstate. The system can have a topological phase by an external magnetic field along the [111] direction as it opens an energy gap proportional to $\Delta \sim \frac{h^3}{K^2}$ \cite{kitaev2006anyons} where $h$ is the magnetic field strength. The resulting phase is indexed by a finite integer Chern number. Naturally, the concernitant neutral anyonic excitations do not lead to a quantized electric Hall conductivity, they reveal themselves by a quantized response to a temperature gradient \cite{kitaev2006anyons}. A significant connection between the integer valued Chern number and the "half-integer" conductance arises as the momentum space sum is over the half of the Brillouin zone. 

In this section, we consider the Kitaev model in an arbitrarily oriented magnetic field. The Hamiltonian takes the following form ($\mu_B = 1$),
\begin{eqnarray}
H_{K} &=& \sum_{\langle ij \rangle}^{\alpha-bond} K^\alpha S_i^\alpha S_j^\alpha - \sum_{i,\beta} h^\beta S^\beta _i,\\
&=& -\frac{K}{4} \sum_{\langle ij \rangle}^{\alpha-bond} b^\alpha_i c_i b^\alpha_j c_j - \frac{i}{2}\sum_{i,\beta} h^\beta b^\beta _i c_i,
\end{eqnarray}
and $\langle ij \rangle$ indicates the nearest-neighbor sites on the honeycomb-lattice. This model supports topological phase transitions as a function of the magnetic field strengths and orientations (see Appendix \ref{MFtheoryDetails} for the mean-field analysis).

\subsubsection{Ferromagnetic Kitaev model}
We start with the {\bf ferromagnetic Kitaev model} ($K=-1$) which has been extensively studied in the literature before \cite{hickey2019emergence,yokoi2020half}. The Chern number map is shown in Fig.\ref{FKitaev_Chern} for magnetic fields in the a-c plane. The Chern numers are calculated numerically\cite{fukui2005chern}. The radial axe in the selected directions mark the field strengths, $h = \sqrt{h_1^2+h_2^2+h_3^2}$ and $h_i = \mathbf{h}\cdot \hat{e}_i$. The upper half ($h_1-h_3$) plane is sufficient because reversing the field direction simply changes the sign of the Chern number \cite{Comment1}. 

In zero field, the KSL groundstate is gapless. At small field strengths $0 < h\ll 1$, the fermionic vacuum becomes topologically non-trivial \cite{kitaev2006anyons} with $\nu = \pm 1$ \cite{Comment2}. The odd valued Chern numbers indicate the presence of non-Abelian anyons \cite{kitaev2006anyons} whereas even valued Chern numbers imply the presence of Abelian anyons. At higher fields beyond $h^F_c \approx 0.18 \left| K \right|$, the system is topologically trivial with a polarized GS. The sequence of states as a function of field strength reproduces the previous result \cite{hickey2019emergence}. In addition, there is an overall sign reversal (e.g. $\nu = 1 \to -1$) upon crossing the $\theta \approx 35^o$ line in the $h_1-h_3$ plane \cite{kitaev2006anyons,yokoi2020half}. Beware that the angle $\theta$ is defined w.r.t. the $h_3$ axis (clock-wise), e.g. positive angle means positive $h_1$ value. The sign change in $\nu$ can be naturally understood by a sign change of any magnetic field components (in this case $h_z$). It is because the Chern number is proportional to the sign of the energy gap created by the product of the magnetic field components, $\nu \sim \text{sgn}(h_x h_y h_z)$. Using the transpose of Eq.\ref{xyzVS123}, the critical line (in the upper $h_1-h_3$ plane) is identified by the angle where $h_z$ changes sign, this is when $-\sqrt{2} h_1 + h_3 = 0$ or $\theta_c = \tan^{-1} (\frac{h_1}{h_3}) \frac{180}{\pi}= \tan^{-1} (\frac{1}{\sqrt{2}})\frac{180}{\pi}$ is satisfied. The same behavior applies for the antiferromagnetic Kitaev model. 

Our approach captures all the qualitative properties of the field response for the F Kitaev model. On the quantitative side, the critical field strength for the transition to the polarized phase is six times larger ($h^{MF}_c = 0.18 \left| K \right|$) compared to the exact diagonalization results ($h^{ED}_c = 0.03 \left| K \right|$) \cite{yokoi2020half}. 

\subsubsection{Antiferromagnetic Kitaev}
\begin{figure*}
\includegraphics[width=1\textwidth]{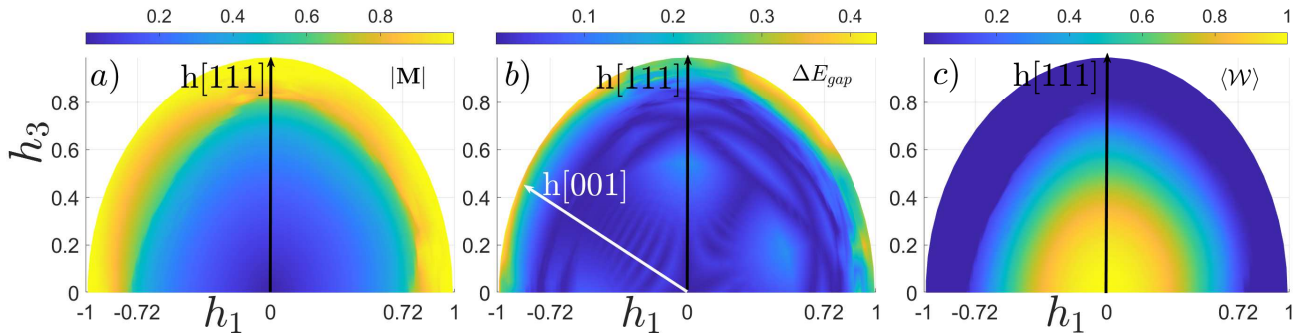}
\caption{(Color online) Characterizing quantities of the {\bf antiferromagnetic Kitaev model}: {\bf a)}: The average magnetic moment, $\left| \mathbf{M} \right| = \left| \mathbf{M}_A + \mathbf{M}_B \right|/2$ within the upper $h_1-h_3$ plane in units of $\left| K \right|$. The subscripts A and B are sublattice indices. The radial direction is the field strength, $h\in \{0,1\}$. {\bf b)}: The energy gap, $\Delta E_{gap} $ within the upper $h_1-h_3$ plane indicates various sequential gap closures and re-openings. {\bf c)}: The MF contracted Wilson loop expectation value, $\langle \hat{\mathcal{W}} \rangle = 2^6\langle S_1^xS_2^yS_3^zS_4^xS_5^yS_6^z \rangle$ on a honeycomb plaquette.}
\label{AFKitaev_ThermodyFncs}
\end{figure*}

The {\bf antiferromagnetic Kitaev model} is examined in two steps within the upper half of the $h_1-h_3$ plane. We first obtain the Chern number map as shown in Fig.\ref{AFKitaev_Chern}. We supplement this result with three additional plots: The average magnetic moment $\left| \mathbf{M} \right| = \left| \mathbf{M}_A +\mathbf{M}_B \right| /2$ (Fig.\ref{AFKitaev_ThermodyFncs}a), the energy gap $\Delta E_{gap} $ (Fig.\ref{AFKitaev_ThermodyFncs}b) and the mean-field Wilson loop expectation value on a honeycomb plaquette, $\langle \hat{\mathcal{W}} \rangle = 2^6 \langle S_1^xS_2^yS_3^zS_4^xS_5^yS_6^z \rangle$ (Fig.\ref{AFKitaev_ThermodyFncs}c). Note that the Wilson loop $\langle \hat{\mathcal{W}} \rangle$ allows to quantify the existence of an emergent gauge field, which is the direct signature of fractionalization or the emergence of local and itinerant Majorana fermions. $\langle \hat{\mathcal{W}} \rangle$ is calculated by a MF contraction $\langle i b^\alpha_i b_j^\alpha \rangle \langle -i c_i c_j \rangle + ...$ and includes terms such as the magnetic moment pairs, i.e. $m^\alpha_i \sim \langle b^\alpha_i c_i \rangle$. We omit all $\sim \langle b^\alpha_i c_i \rangle$ pairs to capture only the gauge-flux effect. 

Considering the overall quantitative response of the AF Kitaev model in Fig.\ref{AFKitaev_ThermodyFncs}, the KSL limit at $h\to 0$ crosses through an intermediate region with increasing field strengths and ultimately reaches a polarized phase. As shown in Fig.\ref{AFKitaev_ThermodyFncs}a, $\left| \mathbf{M} \right|$ gradually increases and saturates at high-fields. Starting with the proximate KSL limit, the magnetic moments gradually increase and saturate to the fully P phase along all directions. In parallel, the Wilson flux $\langle \hat{\mathcal{W}} \rangle$ (see Fig.\ref{AFKitaev_ThermodyFncs}c) decays with increasing field strength indifferent to the field orientation.

We first consider {\bf the evolution along the $h[111]$ ($\hat{e}_3$) direction}, where $h_1 = h_2 = 0$. The role of the field direction is clearly visible in the topological index of the GS, the Chern number, $\nu$ in Fig.\ref{AFKitaev_Chern}. The fermionic vacuum becomes topologically non-trivial ($\nu = -1$) and open an energy gap proportional to $h_x h_y h_z /\mid K\mid^3$. At larger field strengths (see Fig.\ref{AFKitaev_ThermodyFncs}b), there are two band touchings at $h \approx 0.7 \left| K \right| $ and $0.8 \left| K \right| $. Hence, four topologically distinct phases are encountered as a function of the field strength with the Chern numbers: $\nu = 0 \to -1 \to 2 \to 0$. A noteworthy change in the Chern number is $\nu = -1 \to 2$ by 3 units \cite{Zhang2022}, albeit the change is negative. Such a scenario is possible when three Dirac cones \cite{yilmaz2017hofstadter} are involved in this transition. The even valued Chern number is supposed to have Abelian anyons.

The intermediate phase with $\nu=2$ is extended to the $h_1\ne 0$ region as well as the corresponding finite energy gap in Fig.\ref{AFKitaev_ThermodyFncs}b. The topological GS with a Chern number $\nu = 2$ has also been identified in Refs.\cite{ralko2020novel,jiang2020tuning}. It has been asserted that this phase supports multiple Abelian anyonic species \cite{jiang2020tuning}. Yet, this claim has not been confirmed previously. In the high-field limit, the system goes back to a topologically trivial partially polarized phase with $\nu = 0$.
\begin{figure*}
\includegraphics[width=1\textwidth]{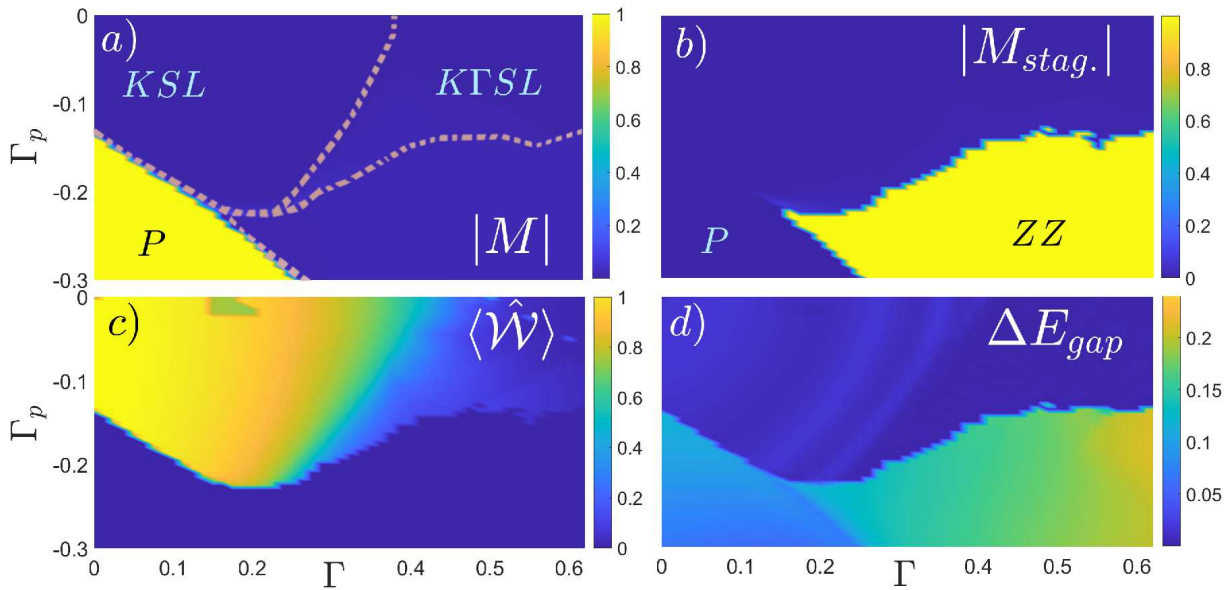}
\caption{(Color online) The GS phase diagram of the K-$\Gamma$-$ \Gamma_p$ model for fixed ferromagnetic Kitaev term, $K= -1$. {\bf a)} Average total magnetization per four-site unit cell, $\left| \mathbf{M} \right| =\left| \sum_{i=1}^4 \mathbf{m}_i \right|/4$. Two distinct regions are readily identified with either vanishing or finite magnetization. The Kitaev spin liquid (KSL), the polarized phase (P), the zig-zag z phase (ZZ-z) and the Kitaev-$\Gamma$ spin liquid phase (K$\Gamma$SL). {\bf b)} The staggered magnetization, $\left| \mathbf{M}_{stag.} \right| =\left| \mathbf{m}_1 + \mathbf{m}_2 -\mathbf{m}_3 -\mathbf{m}_4 \right|/4$ reveals the ZZ-z phase. {\bf c)} The Wilson loop expectation value $\langle \hat{\mathcal{W}} \rangle$ in the mean field factorization as a measure for the total strength of the gauge field. A finite Wilson loop signals the fractionalization of spins as Majorana gauge fields and itinerant Majorana fermions, and thereby distinguishes between KSL and K$\Gamma$SL phases. {\bf d)} The energy gap which distinguishes between the ZZ-z phase and the P-phase. The latter two have vanishing Chern numbers.}
\label{KGGp}
\end{figure*}
The conjectured "multi-species anyon" extends mainly to positive $\theta$ angles as seen in Fig.\ref{AFKitaev_Chern}. Along the negative $\theta$ direction, the $\nu = 2$ region is replaced by a phase with $\nu = 1$. These phases and others are also identified by $\Delta E_{gap} $ map in Fig.\ref{AFKitaev_ThermodyFncs}b. 

{\bf Regarding the $h[001]$ field direction}, it has already been studied in the two references \cite{nasu2018successive,hickey2019emergence}. It was suggested that the groundstate is a gapless spin-liquid until it reaches the field-polarized phase. The results of Nasu et. al \cite{nasu2018successive} rely on a mean-field decoupling between their 'chains', while Hickey and Trebst \cite{hickey2019emergence} employed a numerical diagonalization of a small system. Similarly, we observe a small energy gap (see Fig.\ref{AFKitaev_ThermodyFncs}b) until $h \approx 0.7 \left| K \right|$ with the Chern number $\nu = 1$. Interestingly, $\nu$ reverses sign between $0.7 < h / \left| K \right| < 0.8$ $\nu =-1$. Beyond $h \approx 0.8 \left| K \right|$, the system goes into a trivial phase with a vanishing Chern number. In this respect, our MF approach {\bf provides a different} field evolution along the $h[001]$ direction. The analytical treatment and the thermodynamics of $h[001]$ direction will be provided elsewhere \cite{h001Thermodynamics}.

An interesting transition occurs upon sweeping the magnetic field between the $h[001]$ and $h[111]$ directions (see Fig.\ref{AFKitaev_Chern}). At high fields, the two directions are adiabatically connected with $\nu = 0$. Yet, at lower fields, the two regions are topologically distinct. This transition was {\bf unnoted before}. We observe that the states in the vicinity of the $h[001]$ direction are topologically equivalent to the $h[111]$ direction with an energy gap and the Chern number $\nu = -1$. In this respect, the only distinct regions w.r.t. the $h[111]$ direction along this special $h[001]$ direction. At higher fields, the system goes into a topological transition irrespective of the field direction, yet the Chern numbers are field orientation dependent. 

Before closing this section, we comment on the field directions $\theta = \pm 60^o$ and $\theta = \pm 90^o$ typically explored in experiments \cite{kasahara2018majorana,yokoi2020half}. The experimentally accessible fields up to $50$ Tesla allows $\alpha$-RuCl$_3$ to cross all the phase boundaries for the materials of interest. We therefore indicate the $\theta = \pm 60^o$ directions with red arrows in Fig.s \ref{FKitaev_Chern} and \ref{AFKitaev_Chern}, whereas the $\theta = \pm 90^o$ indicate the $h_1$ axe. It is clear that the field orientation dependence of $\kappa_{xy}$ could be used to trace a AF type Kitaev term in real materials keeping in mind the material specific additional couplings. The change in $\kappa_{xy}$ is expected to be the integer multiplies of $1/2$ in units of $\frac{\pi }{6} k_B^2 T$. Moreover, the topological transitions are accompanied by the sign change in $\kappa_{xy}$ before the system reaches the high-field polarized phase.

\section{Kitaev - Gamma (K-$\Gamma$-$\Gamma_p$) model}
\subsection{Kitaev - Gamma (K-$\Gamma$-$\Gamma_p$) Model in Zero Field}
In order to bridge the gap between experiments and theory, it is indispensable to include the additional terms readily present in candidate materials. The relevant terms are the off-diagonal exchange terms $\Gamma$, $\Gamma_p$ as well as the Heisenberg coupling $J$. In the following, we consider only $\Gamma$ and $\Gamma_p$ terms \cite{gordon2019theory,saha2019hidden,catuneanu2018path,wang2019one,liu2018dirac,rousochatzakis2017classical,samarakoon2018classical,luo2019gapless}, as they are sufficient to capture the ZZ-$\gamma$ phase. The Hamiltonian for the additional terms is,
\begin{eqnarray}
H_{\Gamma,\Gamma_p} &=& \sum_{\langle ij \rangle,\beta \gamma}^{\alpha-bond} \left| \epsilon^{\alpha\beta\gamma} \right| \Big[ \Gamma S_i^\beta S_j^\gamma + \Gamma_p \left( S_i^\alpha S_j^\beta + S_i^\alpha S_j^\gamma \right) \Big].
\end{eqnarray}

\begin{figure*}
\includegraphics[width=1\textwidth]{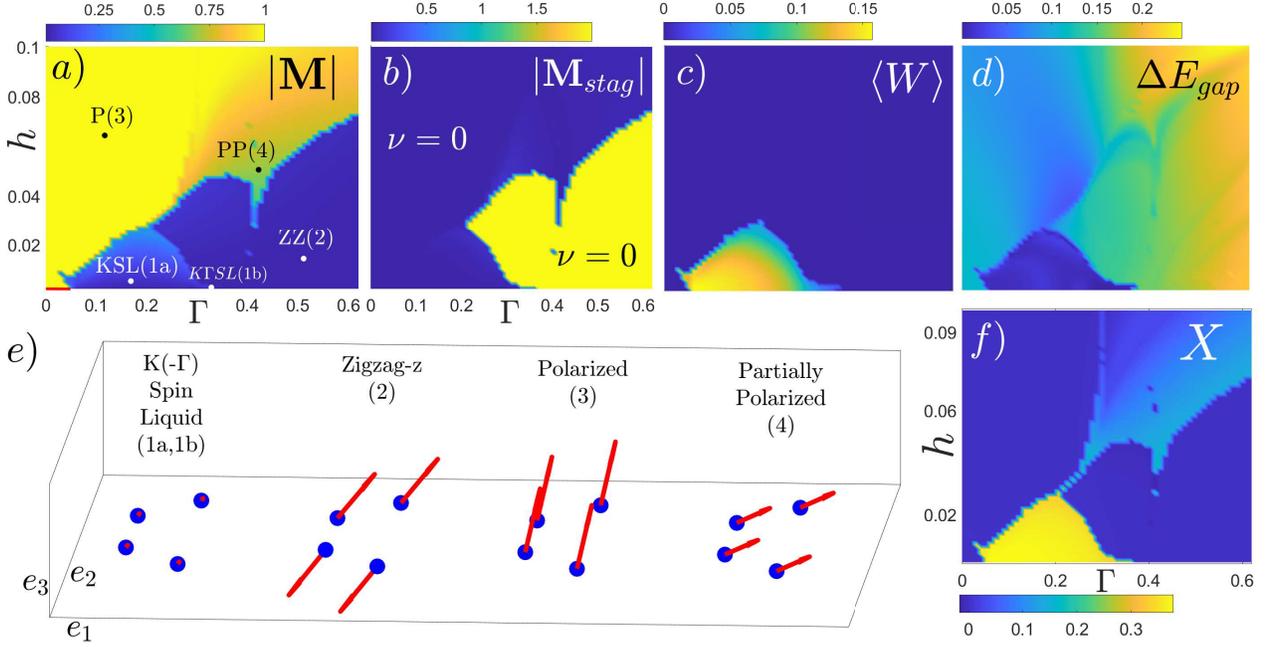}
\caption{(Color online) K-$\Gamma$-$\Gamma_p$ model in a tilted magnetic field for $\theta = 60^o$ w.r.t. c-axis for $K = -1$ and $\Gamma_p = -0.17 \left| K \right|$ in $\Gamma$-$h$ space. {\bf a)} The total magnetization per four-site unit cell, $\left| \mathbf{M} \right| = \left| \sum_{i=1}^4 \mathbf{m}_i \right|/4$. {\bf b)}: The staggered magnetization per four-site unit cell, $\left| \mathbf{M}_{stag} \right| = \left( \left|\mathbf{m}_1 + \mathbf{m}_2 -\mathbf{m}_3-\mathbf{m}_4 \right|/4 \right)$. The staggered definition is used to distinguish the ZZ-z phase. Using plots a and b, one can identify four different phases: KSL, ZZ-z, PP (intermediate) and P-phases. Note that, the ZZ-z phase and the P phase regions have trivial topology, $\nu = 0$.{\bf c)}: The Wilson flux expectation per plaquette. The finite region denoted with yellow-orange colors indicates the KSL states, whereas the remaining regions is either highly suppressed with the absence of fractionalization or negative values indicating another polarized phase. {\bf d)}: The energy gap for the groundstate. It reveals a detailed information on how to distinguish each phase. The energy gap vanishes for the KSL while partially suppressed for the intermediate PP phase. {\bf e)}: The magnetic moment vectors for the four-site unit cell. The points with their numbers are indicated on plot a. {\bf f)}: The mean field parameter for the correlation of the x-type Majorana fermions on the x-bond, $X = \langle i b_j^x b_l^x \rangle_x$. It could be treated as the order parameter for the spin fractionalization.}
\label{KGGpUnderH}
\end{figure*}
The typical strengths of $\Gamma$ and $\Gamma_p$ are understood in various materials yet, there is no agreement \cite{winter2017models} on precise values. For $\alpha$-RuCl$_3$ \cite{wang2017theoretical}, the magnitudes are estimated as $\Gamma \sim \left| K \right|$ of AF-type and a small $\Gamma_p \ll \left| K \right|$ of F-type. For convenience, we partially relax the restriction on $\Gamma,\Gamma_p$ to characterize the GS in a larger parameter space. 

Inspired by $\alpha$-RuCl$_3$, we investigate the GS phase diagram for the $K$-$\Gamma$-$\Gamma_p$ model with a four-site unit cell (see the green dotted rectangle in Fig.\ref{pointGroups}) to capture the relevant phases including the ZZ-z phase. We fix the ferromagnetic Kitaev term ($K = -1$) as the energy unit and examine the phase diagram in $\Gamma$-$\Gamma_p$ space. Fig.\ref{KGGp} a-d show the average magnetic moment ($\left| \mathbf{M} \right| = \left| \sum_{i=1}^4 \mathbf{m}_i \right|/4$ where the index $i$ extends over the four sites unit cell), the staggered magnetic moment ($\left| \mathbf{M}_{stag} \right| =\left|\mathbf{m}_1 + \mathbf{m}_2 -\mathbf{m}_3-\mathbf{m}_4 \right|/4$), the Wilson flux and the energy gap ($\Delta E_{gap}$), respectively. The parameters $\Gamma,\Gamma_p$ cover the range $\Gamma \in [0,0.62] $ and $\Gamma_p \in [-0.3,0] $ in units of $\left| K \right|$. We identify four different phases which are characterized below: KSL, K$\Gamma$SL, P, and the ZZ-z phases. The P \cite{lee2020magnetic} and the ZZ-z phase are distinguished by the directions of the magnetic moments (see Fig.\ref{ZZzpMxyz} in Appendix \ref{KGGpM_is}) as well as by the Wilson flux (Fig.\ref{KGGp}c) and the energy gap (Fig.\ref{KGGp}d). KSL and K$\Gamma$SL phases have suppressed magnetic moments and differ only the average Wilson loop \cite{Comment3}. Yet, the curve separating KSL phase and K$\Gamma$SL phase should not be considered as a sharp boundary. We also identify the characteristic directions of the magnetic moments for each phase in Fig.\ref{KGGpUnderH}d.

Focusing on Fig.\ref{KGGp}a, there is an extended KSL region for relatively small $\left| \Gamma_p \right|$ values. KSL phase is expected to have similar properties as the KSL phase because the Wilson loop of the same magnitude. For larger $\Gamma$ values, there is a crossover to a quantitatively different region named as K$\Gamma$SL \cite{gordon2019theory,wang2019one}. It is also a gapless phase with a suppressed Wilson loop $\mathcal{W}$. For the region where $\Gamma < 0.3 \left| K \right|$ and $\Gamma_p < -0.15 \left| K \right|$, a polarized phase emerges. If the local moments, $m^\alpha \sim \langle i b_i^\alpha c_i \rangle$, were included into the mean-field decomposition of the Wilson loop, a negative $\mathcal{W}$ would be acquired for the P phase \cite{lee2020magnetic}. This phase can be understood by considering the spins as classical vectors for each sublattice, e.g. $\mathbf{S}_A^\alpha= \sin \theta_A \cos \phi_A \hat{x}+ \sin \theta_A \sin \phi_A \hat{y}+ \cos \theta_A \hat{z}$. Exploiting translational invariance, the energy (per unit-cell) then reads
\begin{eqnarray}\label{PolarizedEnergy}
E &=& 3 \left(2\Gamma_p+\Gamma \right) \cos \theta_A \cos \theta_B + \left( K- (2 \Gamma_p +\Gamma) \right) \left( \cos \theta_A \cos \theta_B \sin \theta_A \sin \theta_B \cos (\phi_A-\phi_B)\right).
\end{eqnarray}
The minimization w.r.t. $\{(\theta_A,\phi_A),(\theta_B,\phi_B) \}$ implies all spins to align along the $[111]$ direction. 

{\bf Relevant to the candidate materials}, there is a wide region with a ZZ-z phase as shown in Fig.\ref{KGGp} b-d. It is a gapped, topologically trivial and ordered phase \cite{chaloupka2013zigzag}. The average magnetization $\left| \mathbf{M} \right|$ vanishes while the staggered magnetization is saturated $\left| \mathbf{M}_{stag} \right| = 1$. For a four-site unit cell, the ZZ-z region is extended to larger $\Gamma$ and $\Gamma_p$ magnitudes. $\left| \mathbf{M}_{stag} \right|$ has thus the same value as one would find by assuming from the outset that the magnetic moments are purely classical instead of quantum mechanical. Though our method should be capable in capturing quantum mechanical spin fluctuations and hence a ZZ-z phase with $\left| \mathbf{M}_{stag} \right| < 1$ is in principle allowed, such a phase is not realized within our calculations.

\subsection{Kitaev-Gamma (K-$\Gamma$-$\Gamma_p$) Model with Magnetic Field}\label{KGGpHsec}
In this section, we focus on the magnetic field response of the K-$\Gamma$-$\Gamma_p$ model (see Fig.\ref{KGGp}) that is studied in the previous section for fixed $\Gamma_p = -0.17 \left| K \right|$ in $\Gamma-h$ space. Here we remind that at zero field and as an increasing function of $\Gamma$, there are four phases encountered which are the P phase, the KSL phase, the K$\Gamma$SL phase and the ZZ-z phase. We first orient the magnetic field along the experimentally studied direction with $\theta = 60^o$ in the ac plane \cite{kasahara2018majorana}. In Fig.\ref{KGGpUnderH} the encountered phases are identified as KSL, K$\Gamma$SL, ZZ-z, Partially Polarized (PP) and P phase. The field direction is indicated in Fig.\ref{KGGpUnderH} with $\mathbf{h}$ in the ac ($e_1-e_3$) plane. All the identified phases in $\Gamma-h$ space, shown in Fig.\ref{KGGpUnderH}a, are also illustrated with their local moment vectors in Fig.\ref{KGGpUnderH}e.

\begin{itemize}
\item The P phase in which the magnetization points along the $[111]$ ($\theta=0^o$) direction for small $\Gamma$ values was made evident already by the classical spin argument in Eq.\ref{PolarizedEnergy}. It is the phase labeled with (3) in Fig.\ref{KGGpUnderH}e. A finite field with $\theta = 60^o$ tilts the spins out of the $[111]$ direction. A large enough field strength always takes the system into the P phase, though the critical strength depends on the value of $\Gamma$. Also, the direction of the spins lies between $\theta=0^o$ and $\theta = 60^o$. 
\item At larger $\Gamma$ values, the KSL is bf stabilized and extended till larger field strengths. Because this region has a finite magnetization at a finite $h$, it can only be identified by a suppressed Wilson flux and the finite energy gap, in Fig.\ref{KGGpUnderH}b-c respectively. Depending on the strength of $\Gamma$, KSL can make a phase transition to either the P phase or the ZZ-z phase. There is small area within ZZ phase region with a suppressed Wilson flux in the vicinity of KSL region where the K$\Gamma$SL phase seems to be the stabilized. It is then replaced by the ZZ-z phase at much smaller field strengths.
\item Focusing on the experimentally relevant region with the ZZ-z phase, this classical phase is stable till a finite magnetic field strength. At this point, it must be underlined that we have limited our investigations to the ZZ phases to be ZZ-z only. It is known that \cite{janssen2017magnetization} different zig-zag phases can be favored under external fields, depending on the field directions. The energetic competition between these states is delicate and we prefer to leave that investigation to the future. 
\item Moreover, we indeed observe an intermediate phase before the polarized phase. Because it has a relatively suppressed magnetic moment (see Fig.\ref{KGGpUnderH}a) -indicating the presence of quantum fluctuations- we call it the partially polarized phase. It is a region of finite energy gap (see Fig.\ref{KGGpUnderH}c). The PP phase is present for a wide range of $\Gamma$ values. Even if it seems to be disconnected from the extended KSL region with a vanishing Wilson flux, the existence of finite fractionalization is observed. In Fig.\ref{KGGpUnderH}f, we show one such indicator as the correlation of the x-type Majorana fermions on the x-bond, $X = \langle i b_j^x b_l^x \rangle_x$ (see Fig.\ref{KGGpHThet1a} for other indicators).
\item The anti-symmetric thermal Hall conductivity tensor is proportional to the total Chern number of the fermionic vacuum $\kappa_{xy}^A = (\kappa_{xy}-\kappa_{yx})/2 \sim \nu$. It is the component that is routinely measured in the THC experiments (E.g. see \cite{kasahara2018majorana,yokoi2020half,czajka2021oscillations}). We now relate $\nu$ to the $K$-$\Gamma$-$\Gamma_p$ model in a magnetic field and discuss its implications. Firstly, at h=0 in Fig.\ref{KGGp}, the ZZ-z phase is a collinear phase which can be mapped back to itself by a time-reversal operator and a translation of spins by one unit along the $z$-bond. Naturally, it indicates a trivial Chern number with a vanishing anti-symmetric thermal Hall coefficient, $\kappa_{xy} - \kappa_{yx} \sim \nu = 0$ (We also numerically verify $\nu = 0$ for the ZZ-z phase and the P phase, see Fig.\ref{KGGpUnderH}b). Within the PP phase, the system has a smaller energy gap compared to the ZZ-z phase and the P phase as seen in Fig.\ref{KGGpUnderH}d. We observe a highly non-trivial Chern number distribution within the PP phase depending on $\Gamma$ and $h$. Our conclusion is that the GS wavefunction highly unlikely to have a unit Chern number originated from the pure Kitaev model.
\item There can be two sources of finite Chern number: The fractionalized modes of the extended K$\Gamma$SL or the spin edge states. We plot the fractionalization order parameter $X = \langle i b^x_j b^x_l \rangle_x$ in Fig.\ref{KGGpUnderH}f and $X$ is indeed finite within the PP phase. Thereby it suggests an {affirmative} scenario for the existence of the fractionalized excitations within the intermediate phase in realistic Kitaev models.
\end{itemize}

\begin{figure*}
\includegraphics[width=0.7\textwidth]{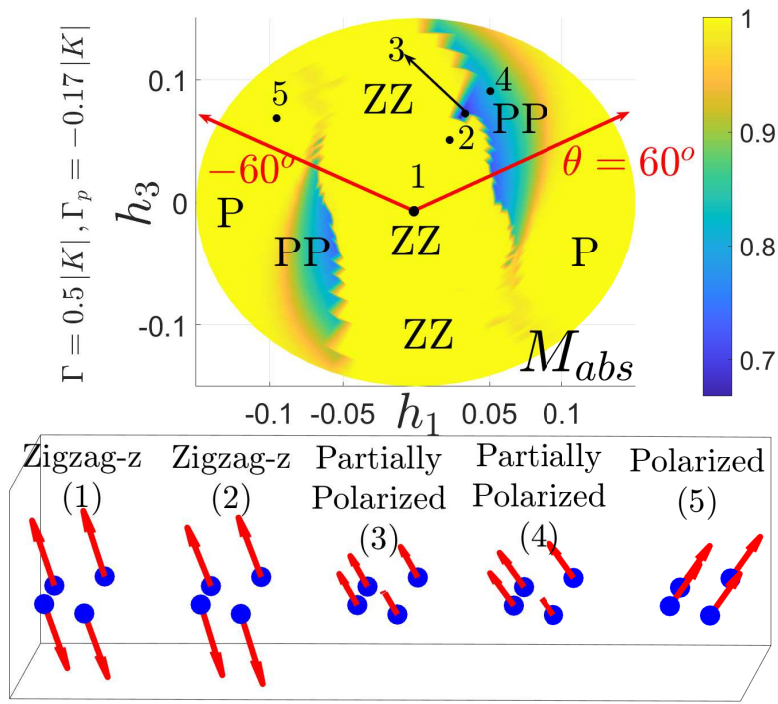}
\caption{(Color online) The density plot the for the absolute magnetization per site ($M_{abs}$) in $h_1-h_3$ space. $M_{abs} =\frac{1}{4 }\sum_{i=1}^4 \left| \mathbf{m}_i \right|$. The model parameters are chosen to host a ZZ-z GS as $K= -1, \Gamma = 0.5 \left| K \right|$ and $\Gamma_p = -0.17 \left| K \right|$. One can identify four different phases: the ZZ-z (1) phase at small field strengths, the canted ZZ-z (2) phase at larger field strength with a wider window along $h_1 = 0$ line, the PP phase (3,4) at intermediate field strengths and finally the P-phase (5) at high-field strengths. The corresponding magnetic moment vectors of each site in a four-site unit cell are labeled from $1$ to $5$.}
\label{ZZzpFieldSweepACMtot1}
\end{figure*}

We now extend our studies to arbitrary field directions. We calculate the absolute magnetization under arbitrary magnetic field (in a-c plane) in Fig.\ref{ZZzpFieldSweepACMtot1}, where $M_{abs} =\frac{1}{4 }\sum_{i=1}^4 \left| \mathbf{m}_i \right|$. We set the couplings as $K=-1,\Gamma=0.5 \left| K \right|, \Gamma_p = -0.17 \left| K \right|$ such that the GS is in ZZ-z phase. The radial direction is the field strength where $\theta$ is the polar angle w.r.t. the c-axis. The ZZ-z phase is stable for small field strengths for all directions. The extended ZZ-z phase can be seen in other studies \cite{janssen2017magnetization,lee2020magnetic} if $\Gamma$ is comparable to the Kitaev term. 
It is also known \cite{janssen2017magnetization} that ZZ-z phase is more stable to a perpendicular magnetic field since the spins can tilt towards the field direction more easily. For $\Gamma <\left| K \right| $, $\mathbf{M}_{stag}$ tend to lie closer to the $x-y$ plane ($\theta = 35^o$), with a small out of plane $z$-component (see Appendix \ref{KGGpM_is} and the figures therein). Similarly, within this context, the perpendicular field mainly points around $\theta = - 55^o$ to $- 60^o$. Consequently, the directions $\theta \in [-60^o, 0]$ stabilizes the ZZ-z phase. In this regard, the THC experiment on $\alpha$-RuCl$_3$ has revealed that the magnetic field responses for $\theta=\pm 60^o$ directions are not symmetrical but the system has a wider intermediate region \cite{yokoi2020half} for $\theta = +60^o$. We indeed verify a wider intermediate region (less stable ZZ-z phase) for $\theta= 60^o$ compared to $- 60^o$. On the contrary, for the field directions closer to the $x-y$ plane ($\theta$ is around $35^o$ or $-150^o$), the ZZ-z phase is highly unstable and paves the way for the PP phase as an intermediate phase. Because the magnetic field is not along $[111]$, the rotational symmetry is already broken and we cannot talk about the nematic phases in literature \cite{lee2020magnetic}.
\section{CONCLUSION}
We sum up this work by referring back to the initially posed questions. Regarding the phase diagram of the AF Kitaev model in a magnetic field: We identified an additional topologically non-trivial groundstate with Abelian anyons (even Chern number $\nu = \pm 2$) in addition to the $\nu = \pm1$ vacuum mainly referred to in the literature as the source of the half-quantized THC. The magnetic field response is sensitive to the magnetic field orientation and strength upon which several topological transitions are encountered for fields in the a-c plane. We emphasize that the gapless phases are hard to trace and demand for special numerical care (e.g. for fields in the $h[001]$ direction).

Regarding the role of the additional off-diagonal $\Gamma$-$\Gamma_p$ terms in the modelling of real materials, we have shown that the pure Kitaev picture changes dramatically for comparable strength of $\Gamma \sim \left| K \right|$ in favor of a K$\Gamma$SL as the gauge-flux expectation value per plaquette vanishes, $\mathcal{W} = 0$. The magnetic field, in this respect, seems to drive $\alpha$-RuCl$_3$ into an intermediate K$\Gamma$SL with coexisting order. Partial fractionalization is still observed in a magnetic field which is clearly different from the Kitaev spin liquid phase, again due to the absence of gauge flux. However, the chosen Chern number calculation method in the intermediate phase does not work efficiently for our purposes. Even though the GS Chern number was understood to be highly different from the pure Kitaev model, we could not draw a final conclusion due to its rapidly changing character in $\Gamma$-h space. In this respect a more precise calculation method is in order. 

Comparing our results with the experimental and theoretical results readily available in the literature, we found a qualitatively satisfying agreement for all limits of $K,\Gamma,\Gamma_p$ and $\mathbf{h}$. Our findings also capture the asymmetry in the thermal Hall response in Ref.\cite{yokoi2020half} for $\theta = \pm 60^o$ magnetic field orientations in a straightforward way. 

This work raises several questions to be addressed in particular about the detailed character of the intermediate phase and its (anyonic) excitations. Unsettled remains the underlying physical mechanism that breaks $\pm K$ symmetry of the pure Kitaev when a magnetic field is present. And it has to be explored how our conceptually simple and comprehensive formalism can be applied to comply with the recently observed oscillations \cite{czajka2021oscillations} in the longitudinal thermal conductivity as a function of inverse magnetic field.
\begin{acknowledgments}
A.P.K. was supported by the Deutsche Forschungsgemeinschaft (DFG, German Research Foundation) through project number 107745057 TRR 80. Part of this work was performed while FSY was in Taiwan, supported by the Ministry of Science and Technology, Taiwan, under grant numbers MOST 107-2112-M-001-035-MY3 and MOST 108-2811-M-001-618.
\end{acknowledgments}

\bibliography{arxivSubmissionbyFirat.bib}

\begin{thebibliography}{91}%
\makeatletter
\providecommand \@ifxundefined [1]{%
 \@ifx{#1\undefined}
}%
\providecommand \@ifnum [1]{%
 \ifnum #1\expandafter \@firstoftwo
 \else \expandafter \@secondoftwo
 \fi
}%
\providecommand \@ifx [1]{%
 \ifx #1\expandafter \@firstoftwo
 \else \expandafter \@secondoftwo
 \fi
}%
\providecommand \natexlab [1]{#1}%
\providecommand \enquote  [1]{``#1''}%
\providecommand \bibnamefont  [1]{#1}%
\providecommand \bibfnamefont [1]{#1}%
\providecommand \citenamefont [1]{#1}%
\providecommand \href@noop [0]{\@secondoftwo}%
\providecommand \href [0]{\begingroup \@sanitize@url \@href}%
\providecommand \@href[1]{\@@startlink{#1}\@@href}%
\providecommand \@@href[1]{\endgroup#1\@@endlink}%
\providecommand \@sanitize@url [0]{\catcode `\\12\catcode `\$12\catcode
  `\&12\catcode `\#12\catcode `\^12\catcode `\_12\catcode `\%12\relax}%
\providecommand \@@startlink[1]{}%
\providecommand \@@endlink[0]{}%
\providecommand \url  [0]{\begingroup\@sanitize@url \@url }%
\providecommand \@url [1]{\endgroup\@href {#1}{\urlprefix }}%
\providecommand \urlprefix  [0]{URL }%
\providecommand \Eprint [0]{\href }%
\providecommand \doibase [0]{https://doi.org/}%
\providecommand \selectlanguage [0]{\@gobble}%
\providecommand \bibinfo  [0]{\@secondoftwo}%
\providecommand \bibfield  [0]{\@secondoftwo}%
\providecommand \translation [1]{[#1]}%
\providecommand \BibitemOpen [0]{}%
\providecommand \bibitemStop [0]{}%
\providecommand \bibitemNoStop [0]{.\EOS\space}%
\providecommand \EOS [0]{\spacefactor3000\relax}%
\providecommand \BibitemShut  [1]{\csname bibitem#1\endcsname}%
\let\auto@bib@innerbib\@empty
\bibitem [{\citenamefont {{K}itaev}(2006)}]{kitaev2006anyons}%
  \BibitemOpen
  \bibfield  {author} {\bibinfo {author} {\bibfnamefont {A.}~\bibnamefont
  {{K}itaev}},\ }\bibfield  {title} {\bibinfo {title} {Anyons in an exactly
  solved model and beyond},\ }\href@noop {} {\bibfield  {journal} {\bibinfo
  {journal} {Annals of Physics}\ }\textbf {\bibinfo {volume} {321}},\ \bibinfo
  {pages} {2} (\bibinfo {year} {2006})}\BibitemShut {NoStop}%
\bibitem [{\citenamefont {Williams}\ \emph {et~al.}(2016)\citenamefont
  {Williams}, \citenamefont {Johnson}, \citenamefont {Freund}, \citenamefont
  {Choi}, \citenamefont {Jesche}, \citenamefont {Kimchi}, \citenamefont
  {Manni}, \citenamefont {Bombardi}, \citenamefont {Manuel}, \citenamefont
  {Gegenwart} \emph {et~al.}}]{williams2016incommensurate}%
  \BibitemOpen
  \bibfield  {author} {\bibinfo {author} {\bibfnamefont {S.}~\bibnamefont
  {Williams}}, \bibinfo {author} {\bibfnamefont {R.}~\bibnamefont {Johnson}},
  \bibinfo {author} {\bibfnamefont {F.}~\bibnamefont {Freund}}, \bibinfo
  {author} {\bibfnamefont {S.}~\bibnamefont {Choi}}, \bibinfo {author}
  {\bibfnamefont {A.}~\bibnamefont {Jesche}}, \bibinfo {author} {\bibfnamefont
  {I.}~\bibnamefont {Kimchi}}, \bibinfo {author} {\bibfnamefont
  {S.}~\bibnamefont {Manni}}, \bibinfo {author} {\bibfnamefont
  {A.}~\bibnamefont {Bombardi}}, \bibinfo {author} {\bibfnamefont
  {P.}~\bibnamefont {Manuel}}, \bibinfo {author} {\bibfnamefont
  {P.}~\bibnamefont {Gegenwart}}, \emph {et~al.},\ }\bibfield  {title}
  {\bibinfo {title} {Incommensurate counterrotating magnetic order stabilized
  by {K}itaev interactions in the layered honeycomb $\alpha$-{L}i$_2${I}r{O}$_
  3$},\ }\href@noop {} {\bibfield  {journal} {\bibinfo  {journal} {Physical
  {R}eview {B}}\ }\textbf {\bibinfo {volume} {93}},\ \bibinfo {pages} {195158}
  (\bibinfo {year} {2016})}\BibitemShut {NoStop}%
\bibitem [{\citenamefont {Majumder}\ \emph {et~al.}(2018)\citenamefont
  {Majumder}, \citenamefont {Manna}, \citenamefont {Simutis}, \citenamefont
  {Orain}, \citenamefont {Dey}, \citenamefont {Freund}, \citenamefont {Jesche},
  \citenamefont {Khasanov}, \citenamefont {Biswas}, \citenamefont {Bykova}
  \emph {et~al.}}]{majumder2018breakdown}%
  \BibitemOpen
  \bibfield  {author} {\bibinfo {author} {\bibfnamefont {M.}~\bibnamefont
  {Majumder}}, \bibinfo {author} {\bibfnamefont {R.}~\bibnamefont {Manna}},
  \bibinfo {author} {\bibfnamefont {G.}~\bibnamefont {Simutis}}, \bibinfo
  {author} {\bibfnamefont {J.}~\bibnamefont {Orain}}, \bibinfo {author}
  {\bibfnamefont {T.}~\bibnamefont {Dey}}, \bibinfo {author} {\bibfnamefont
  {F.}~\bibnamefont {Freund}}, \bibinfo {author} {\bibfnamefont
  {A.}~\bibnamefont {Jesche}}, \bibinfo {author} {\bibfnamefont
  {R.}~\bibnamefont {Khasanov}}, \bibinfo {author} {\bibfnamefont
  {P.}~\bibnamefont {Biswas}}, \bibinfo {author} {\bibfnamefont
  {E.}~\bibnamefont {Bykova}}, \emph {et~al.},\ }\bibfield  {title} {\bibinfo
  {title} {Breakdown of magnetic order in the pressurized {K}itaev iridate
  $\beta$-{L}i$_2${I}r{O}$_ 3$},\ }\href@noop {} {\bibfield  {journal}
  {\bibinfo  {journal} {Physical {R}eview {L}etters}\ }\textbf {\bibinfo
  {volume} {120}},\ \bibinfo {pages} {237202} (\bibinfo {year}
  {2018})}\BibitemShut {NoStop}%
\bibitem [{\citenamefont {Choi}\ \emph {et~al.}(2012)\citenamefont {Choi},
  \citenamefont {Coldea}, \citenamefont {Kolmogorov}, \citenamefont
  {Lancaster}, \citenamefont {Mazin}, \citenamefont {Blundell}, \citenamefont
  {Radaelli}, \citenamefont {Singh}, \citenamefont {Gegenwart}, \citenamefont
  {Choi} \emph {et~al.}}]{choi2012spin}%
  \BibitemOpen
  \bibfield  {author} {\bibinfo {author} {\bibfnamefont {S.}~\bibnamefont
  {Choi}}, \bibinfo {author} {\bibfnamefont {R.}~\bibnamefont {Coldea}},
  \bibinfo {author} {\bibfnamefont {A.}~\bibnamefont {Kolmogorov}}, \bibinfo
  {author} {\bibfnamefont {T.}~\bibnamefont {Lancaster}}, \bibinfo {author}
  {\bibfnamefont {I.}~\bibnamefont {Mazin}}, \bibinfo {author} {\bibfnamefont
  {S.}~\bibnamefont {Blundell}}, \bibinfo {author} {\bibfnamefont
  {P.}~\bibnamefont {Radaelli}}, \bibinfo {author} {\bibfnamefont
  {Y.}~\bibnamefont {Singh}}, \bibinfo {author} {\bibfnamefont
  {P.}~\bibnamefont {Gegenwart}}, \bibinfo {author} {\bibfnamefont
  {K.}~\bibnamefont {Choi}}, \emph {et~al.},\ }\bibfield  {title} {\bibinfo
  {title} {Spin waves and revised crystal structure of honeycomb iridate
  {N}a$_2${I}r{O}$_3$},\ }\href@noop {} {\bibfield  {journal} {\bibinfo
  {journal} {Physical {R}eview {L}etters}\ }\textbf {\bibinfo {volume} {108}},\
  \bibinfo {pages} {127204} (\bibinfo {year} {2012})}\BibitemShut {NoStop}%
\bibitem [{\citenamefont {Ye}\ \emph {et~al.}(2012)\citenamefont {Ye},
  \citenamefont {Chi}, \citenamefont {Cao}, \citenamefont {Chakoumakos},
  \citenamefont {Fernandez-Baca}, \citenamefont {Custelcean}, \citenamefont
  {Qi}, \citenamefont {Korneta},\ and\ \citenamefont {Cao}}]{ye2012direct}%
  \BibitemOpen
  \bibfield  {author} {\bibinfo {author} {\bibfnamefont {F.}~\bibnamefont
  {Ye}}, \bibinfo {author} {\bibfnamefont {S.}~\bibnamefont {Chi}}, \bibinfo
  {author} {\bibfnamefont {H.}~\bibnamefont {Cao}}, \bibinfo {author}
  {\bibfnamefont {B.~C.}\ \bibnamefont {Chakoumakos}}, \bibinfo {author}
  {\bibfnamefont {J.~A.}\ \bibnamefont {Fernandez-Baca}}, \bibinfo {author}
  {\bibfnamefont {R.}~\bibnamefont {Custelcean}}, \bibinfo {author}
  {\bibfnamefont {T.}~\bibnamefont {Qi}}, \bibinfo {author} {\bibfnamefont
  {O.}~\bibnamefont {Korneta}},\ and\ \bibinfo {author} {\bibfnamefont
  {G.}~\bibnamefont {Cao}},\ }\bibfield  {title} {\bibinfo {title} {Direct
  evidence of a zigzag spin-chain structure in the honeycomb lattice: {A}
  neutron and {X}-ray diffraction investigation of single-crystal
  {N}a$_2${I}r{O}$_3$},\ }\href@noop {} {\bibfield  {journal} {\bibinfo
  {journal} {Physical {R}eview {B}}\ }\textbf {\bibinfo {volume} {85}},\
  \bibinfo {pages} {180403} (\bibinfo {year} {2012})}\BibitemShut {NoStop}%
\bibitem [{\citenamefont {Singh}\ \emph {et~al.}(2012)\citenamefont {Singh},
  \citenamefont {Manni}, \citenamefont {Reuther}, \citenamefont {Berlijn},
  \citenamefont {Thomale}, \citenamefont {Ku}, \citenamefont {Trebst},\ and\
  \citenamefont {Gegenwart}}]{singh2012relevance}%
  \BibitemOpen
  \bibfield  {author} {\bibinfo {author} {\bibfnamefont {Y.}~\bibnamefont
  {Singh}}, \bibinfo {author} {\bibfnamefont {S.}~\bibnamefont {Manni}},
  \bibinfo {author} {\bibfnamefont {J.}~\bibnamefont {Reuther}}, \bibinfo
  {author} {\bibfnamefont {T.}~\bibnamefont {Berlijn}}, \bibinfo {author}
  {\bibfnamefont {R.}~\bibnamefont {Thomale}}, \bibinfo {author} {\bibfnamefont
  {W.}~\bibnamefont {Ku}}, \bibinfo {author} {\bibfnamefont {S.}~\bibnamefont
  {Trebst}},\ and\ \bibinfo {author} {\bibfnamefont {P.}~\bibnamefont
  {Gegenwart}},\ }\bibfield  {title} {\bibinfo {title} {Relevance of the
  {H}eisenberg-{K}itaev model for the honeycomb lattice iridates
  {A}$_2${I}r{O}$_3$},\ }\href@noop {} {\bibfield  {journal} {\bibinfo
  {journal} {Physical {R}eview {L}etters}\ }\textbf {\bibinfo {volume} {108}},\
  \bibinfo {pages} {127203} (\bibinfo {year} {2012})}\BibitemShut {NoStop}%
\bibitem [{\citenamefont {Gretarsson}\ \emph {et~al.}(2013)\citenamefont
  {Gretarsson}, \citenamefont {Clancy}, \citenamefont {Singh}, \citenamefont
  {Gegenwart}, \citenamefont {Hill}, \citenamefont {Kim}, \citenamefont
  {Upton}, \citenamefont {Said}, \citenamefont {Casa}, \citenamefont {Gog}
  \emph {et~al.}}]{gretarsson2013magnetic}%
  \BibitemOpen
  \bibfield  {author} {\bibinfo {author} {\bibfnamefont {H.}~\bibnamefont
  {Gretarsson}}, \bibinfo {author} {\bibfnamefont {J.}~\bibnamefont {Clancy}},
  \bibinfo {author} {\bibfnamefont {Y.}~\bibnamefont {Singh}}, \bibinfo
  {author} {\bibfnamefont {P.}~\bibnamefont {Gegenwart}}, \bibinfo {author}
  {\bibfnamefont {J.}~\bibnamefont {Hill}}, \bibinfo {author} {\bibfnamefont
  {J.}~\bibnamefont {Kim}}, \bibinfo {author} {\bibfnamefont {M.}~\bibnamefont
  {Upton}}, \bibinfo {author} {\bibfnamefont {A.}~\bibnamefont {Said}},
  \bibinfo {author} {\bibfnamefont {D.}~\bibnamefont {Casa}}, \bibinfo {author}
  {\bibfnamefont {T.}~\bibnamefont {Gog}}, \emph {et~al.},\ }\bibfield  {title}
  {\bibinfo {title} {Magnetic excitation spectrum of {N}a$_2${I}r{O}$_3$ probed
  with resonant inelastic {X}-ray scattering},\ }\href@noop {} {\bibfield
  {journal} {\bibinfo  {journal} {Physical {R}eview {B}}\ }\textbf {\bibinfo
  {volume} {87}},\ \bibinfo {pages} {220407} (\bibinfo {year}
  {2013})}\BibitemShut {NoStop}%
\bibitem [{\citenamefont {Yamaji}\ \emph {et~al.}(2014)\citenamefont {Yamaji},
  \citenamefont {Nomura}, \citenamefont {Kurita}, \citenamefont {Arita},\ and\
  \citenamefont {Imada}}]{yamaji2014first}%
  \BibitemOpen
  \bibfield  {author} {\bibinfo {author} {\bibfnamefont {Y.}~\bibnamefont
  {Yamaji}}, \bibinfo {author} {\bibfnamefont {Y.}~\bibnamefont {Nomura}},
  \bibinfo {author} {\bibfnamefont {M.}~\bibnamefont {Kurita}}, \bibinfo
  {author} {\bibfnamefont {R.}~\bibnamefont {Arita}},\ and\ \bibinfo {author}
  {\bibfnamefont {M.}~\bibnamefont {Imada}},\ }\bibfield  {title} {\bibinfo
  {title} {First-principles study of the honeycomb-lattice iridates
  {N}a$_2${I}r{O}$_3$ in the presence of strong spin-orbit interaction and
  electron correlations},\ }\href@noop {} {\bibfield  {journal} {\bibinfo
  {journal} {Physical {R}eview {L}etters}\ }\textbf {\bibinfo {volume} {113}},\
  \bibinfo {pages} {107201} (\bibinfo {year} {2014})}\BibitemShut {NoStop}%
\bibitem [{\citenamefont {Alpichshev}\ \emph {et~al.}(2015)\citenamefont
  {Alpichshev}, \citenamefont {Mahmood}, \citenamefont {Cao},\ and\
  \citenamefont {Gedik}}]{alpichshev2015confinement}%
  \BibitemOpen
  \bibfield  {author} {\bibinfo {author} {\bibfnamefont {Z.}~\bibnamefont
  {Alpichshev}}, \bibinfo {author} {\bibfnamefont {F.}~\bibnamefont {Mahmood}},
  \bibinfo {author} {\bibfnamefont {G.}~\bibnamefont {Cao}},\ and\ \bibinfo
  {author} {\bibfnamefont {N.}~\bibnamefont {Gedik}},\ }\bibfield  {title}
  {\bibinfo {title} {Confinement-deconfinement transition as an indication of
  spin-liquid-type behavior in {N}a$_2${I}r{O}$_3$},\ }\href@noop {} {\bibfield
   {journal} {\bibinfo  {journal} {Physical {R}eview {L}etters}\ }\textbf
  {\bibinfo {volume} {114}},\ \bibinfo {pages} {017203} (\bibinfo {year}
  {2015})}\BibitemShut {NoStop}%
\bibitem [{\citenamefont {Wulferding}\ \emph {et~al.}(2020)\citenamefont
  {Wulferding}, \citenamefont {Choi}, \citenamefont {Do}, \citenamefont {Lee},
  \citenamefont {Lemmens}, \citenamefont {Faugeras}, \citenamefont {Gallais},\
  and\ \citenamefont {Choi}}]{wulferding2020magnon}%
  \BibitemOpen
  \bibfield  {author} {\bibinfo {author} {\bibfnamefont {D.}~\bibnamefont
  {Wulferding}}, \bibinfo {author} {\bibfnamefont {Y.}~\bibnamefont {Choi}},
  \bibinfo {author} {\bibfnamefont {S.-H.}\ \bibnamefont {Do}}, \bibinfo
  {author} {\bibfnamefont {C.~H.}\ \bibnamefont {Lee}}, \bibinfo {author}
  {\bibfnamefont {P.}~\bibnamefont {Lemmens}}, \bibinfo {author} {\bibfnamefont
  {C.}~\bibnamefont {Faugeras}}, \bibinfo {author} {\bibfnamefont
  {Y.}~\bibnamefont {Gallais}},\ and\ \bibinfo {author} {\bibfnamefont {K.-Y.}\
  \bibnamefont {Choi}},\ }\bibfield  {title} {\bibinfo {title} {Magnon bound
  states versus anyonic {M}ajorana excitations in the {K}itaev honeycomb magnet
  $\alpha$-{R}u{C}l$_3 $},\ }\href@noop {} {\bibfield  {journal} {\bibinfo
  {journal} {Nature {C}ommunications}\ }\textbf {\bibinfo {volume} {11}},\
  \bibinfo {pages} {1} (\bibinfo {year} {2020})}\BibitemShut {NoStop}%
\bibitem [{\citenamefont {Kubota}\ \emph {et~al.}(2015)\citenamefont {Kubota},
  \citenamefont {Tanaka}, \citenamefont {Ono}, \citenamefont {Narumi},\ and\
  \citenamefont {Kindo}}]{kubota2015successive}%
  \BibitemOpen
  \bibfield  {author} {\bibinfo {author} {\bibfnamefont {Y.}~\bibnamefont
  {Kubota}}, \bibinfo {author} {\bibfnamefont {H.}~\bibnamefont {Tanaka}},
  \bibinfo {author} {\bibfnamefont {T.}~\bibnamefont {Ono}}, \bibinfo {author}
  {\bibfnamefont {Y.}~\bibnamefont {Narumi}},\ and\ \bibinfo {author}
  {\bibfnamefont {K.}~\bibnamefont {Kindo}},\ }\bibfield  {title} {\bibinfo
  {title} {Successive magnetic phase transitions in $\alpha$-{R}u{C}l$_3 $:
  {XY}-like frustrated magnet on the honeycomb lattice},\ }\href@noop {}
  {\bibfield  {journal} {\bibinfo  {journal} {Physical {R}eview {B}}\ }\textbf
  {\bibinfo {volume} {91}},\ \bibinfo {pages} {094422} (\bibinfo {year}
  {2015})}\BibitemShut {NoStop}%
\bibitem [{\citenamefont {Lampen-Kelley}\ \emph {et~al.}(2018)\citenamefont
  {Lampen-Kelley}, \citenamefont {Rachel}, \citenamefont {Reuther},
  \citenamefont {Yan}, \citenamefont {Banerjee}, \citenamefont {Bridges},
  \citenamefont {Cao}, \citenamefont {Nagler},\ and\ \citenamefont
  {Mandrus}}]{lampen2018anisotropic}%
  \BibitemOpen
  \bibfield  {author} {\bibinfo {author} {\bibfnamefont {P.}~\bibnamefont
  {Lampen-Kelley}}, \bibinfo {author} {\bibfnamefont {S.}~\bibnamefont
  {Rachel}}, \bibinfo {author} {\bibfnamefont {J.}~\bibnamefont {Reuther}},
  \bibinfo {author} {\bibfnamefont {J.-Q.}\ \bibnamefont {Yan}}, \bibinfo
  {author} {\bibfnamefont {A.}~\bibnamefont {Banerjee}}, \bibinfo {author}
  {\bibfnamefont {C.~A.}\ \bibnamefont {Bridges}}, \bibinfo {author}
  {\bibfnamefont {H.~B.}\ \bibnamefont {Cao}}, \bibinfo {author} {\bibfnamefont
  {S.~E.}\ \bibnamefont {Nagler}},\ and\ \bibinfo {author} {\bibfnamefont
  {D.}~\bibnamefont {Mandrus}},\ }\bibfield  {title} {\bibinfo {title}
  {Anisotropic susceptibilities in the honeycomb {K}itaev system
  $\alpha$-{R}u{C}l$_3 $},\ }\href@noop {} {\bibfield  {journal} {\bibinfo
  {journal} {Physical {R}eview {B}}\ }\textbf {\bibinfo {volume} {98}},\
  \bibinfo {pages} {100403} (\bibinfo {year} {2018})}\BibitemShut {NoStop}%
\bibitem [{\citenamefont {Plumb}\ \emph {et~al.}(2014)\citenamefont {Plumb},
  \citenamefont {Clancy}, \citenamefont {Sandilands}, \citenamefont {Shankar},
  \citenamefont {Hu}, \citenamefont {Burch}, \citenamefont {Kee},\ and\
  \citenamefont {Kim}}]{plumb2014alpha}%
  \BibitemOpen
  \bibfield  {author} {\bibinfo {author} {\bibfnamefont {K.}~\bibnamefont
  {Plumb}}, \bibinfo {author} {\bibfnamefont {J.}~\bibnamefont {Clancy}},
  \bibinfo {author} {\bibfnamefont {L.}~\bibnamefont {Sandilands}}, \bibinfo
  {author} {\bibfnamefont {V.~V.}\ \bibnamefont {Shankar}}, \bibinfo {author}
  {\bibfnamefont {Y.}~\bibnamefont {Hu}}, \bibinfo {author} {\bibfnamefont
  {K.}~\bibnamefont {Burch}}, \bibinfo {author} {\bibfnamefont {H.-Y.}\
  \bibnamefont {Kee}},\ and\ \bibinfo {author} {\bibfnamefont {Y.-J.}\
  \bibnamefont {Kim}},\ }\bibfield  {title} {\bibinfo {title}
  {$\alpha$-{R}u{C}l$_3 $: {A} spin-orbit assisted {M}ott insulator on a
  honeycomb lattice},\ }\href@noop {} {\bibfield  {journal} {\bibinfo
  {journal} {Physical {R}eview {B}}\ }\textbf {\bibinfo {volume} {90}},\
  \bibinfo {pages} {041112} (\bibinfo {year} {2014})}\BibitemShut {NoStop}%
\bibitem [{\citenamefont {Baek}\ \emph {et~al.}(2017)\citenamefont {Baek},
  \citenamefont {Do}, \citenamefont {Choi}, \citenamefont {Kwon}, \citenamefont
  {Wolter}, \citenamefont {Nishimoto}, \citenamefont {Van Den~Brink},\ and\
  \citenamefont {B{\"u}chner}}]{baek2017evidence}%
  \BibitemOpen
  \bibfield  {author} {\bibinfo {author} {\bibfnamefont {S.-H.}\ \bibnamefont
  {Baek}}, \bibinfo {author} {\bibfnamefont {S.-H.}\ \bibnamefont {Do}},
  \bibinfo {author} {\bibfnamefont {K.-Y.}\ \bibnamefont {Choi}}, \bibinfo
  {author} {\bibfnamefont {Y.~S.}\ \bibnamefont {Kwon}}, \bibinfo {author}
  {\bibfnamefont {A.}~\bibnamefont {Wolter}}, \bibinfo {author} {\bibfnamefont
  {S.}~\bibnamefont {Nishimoto}}, \bibinfo {author} {\bibfnamefont
  {J.}~\bibnamefont {Van Den~Brink}},\ and\ \bibinfo {author} {\bibfnamefont
  {B.}~\bibnamefont {B{\"u}chner}},\ }\bibfield  {title} {\bibinfo {title}
  {Evidence for a field-induced quantum spin liquid in $\alpha$-{R}u{C}l$_3
  $},\ }\href@noop {} {\bibfield  {journal} {\bibinfo  {journal} {Physical
  {R}eview {L}etters}\ }\textbf {\bibinfo {volume} {119}},\ \bibinfo {pages}
  {037201} (\bibinfo {year} {2017})}\BibitemShut {NoStop}%
\bibitem [{\citenamefont {Wang}\ \emph
  {et~al.}(2017{\natexlab{a}})\citenamefont {Wang}, \citenamefont {Reschke},
  \citenamefont {H{\"u}vonen}, \citenamefont {Do}, \citenamefont {Choi},
  \citenamefont {Gensch}, \citenamefont {Nagel}, \citenamefont {R{\~o}{\~o}m},\
  and\ \citenamefont {Loidl}}]{wang2017magnetic}%
  \BibitemOpen
  \bibfield  {author} {\bibinfo {author} {\bibfnamefont {Z.}~\bibnamefont
  {Wang}}, \bibinfo {author} {\bibfnamefont {S.}~\bibnamefont {Reschke}},
  \bibinfo {author} {\bibfnamefont {D.}~\bibnamefont {H{\"u}vonen}}, \bibinfo
  {author} {\bibfnamefont {S.-H.}\ \bibnamefont {Do}}, \bibinfo {author}
  {\bibfnamefont {K.-Y.}\ \bibnamefont {Choi}}, \bibinfo {author}
  {\bibfnamefont {M.}~\bibnamefont {Gensch}}, \bibinfo {author} {\bibfnamefont
  {U.}~\bibnamefont {Nagel}}, \bibinfo {author} {\bibfnamefont
  {T.}~\bibnamefont {R{\~o}{\~o}m}},\ and\ \bibinfo {author} {\bibfnamefont
  {A.}~\bibnamefont {Loidl}},\ }\bibfield  {title} {\bibinfo {title} {Magnetic
  excitations and continuum of a possibly field-induced quantum spin liquid in
  $\alpha$-{R}u{C}l$_3 $},\ }\href@noop {} {\bibfield  {journal} {\bibinfo
  {journal} {Physical {R}eview {L}etters}\ }\textbf {\bibinfo {volume} {119}},\
  \bibinfo {pages} {227202} (\bibinfo {year} {2017}{\natexlab{a}})}\BibitemShut
  {NoStop}%
\bibitem [{\citenamefont {Zheng}\ \emph {et~al.}(2017)\citenamefont {Zheng},
  \citenamefont {Ran}, \citenamefont {Li}, \citenamefont {Wang}, \citenamefont
  {Wang}, \citenamefont {Liu}, \citenamefont {Liu}, \citenamefont {Normand},
  \citenamefont {Wen},\ and\ \citenamefont {Yu}}]{zheng2017gapless}%
  \BibitemOpen
  \bibfield  {author} {\bibinfo {author} {\bibfnamefont {J.}~\bibnamefont
  {Zheng}}, \bibinfo {author} {\bibfnamefont {K.}~\bibnamefont {Ran}}, \bibinfo
  {author} {\bibfnamefont {T.}~\bibnamefont {Li}}, \bibinfo {author}
  {\bibfnamefont {J.}~\bibnamefont {Wang}}, \bibinfo {author} {\bibfnamefont
  {P.}~\bibnamefont {Wang}}, \bibinfo {author} {\bibfnamefont {B.}~\bibnamefont
  {Liu}}, \bibinfo {author} {\bibfnamefont {Z.-X.}\ \bibnamefont {Liu}},
  \bibinfo {author} {\bibfnamefont {B.}~\bibnamefont {Normand}}, \bibinfo
  {author} {\bibfnamefont {J.}~\bibnamefont {Wen}},\ and\ \bibinfo {author}
  {\bibfnamefont {W.}~\bibnamefont {Yu}},\ }\bibfield  {title} {\bibinfo
  {title} {Gapless spin excitations in the field-induced quantum spin liquid
  phase of $\alpha$-{R}u{C}l$_3 $},\ }\href@noop {} {\bibfield  {journal}
  {\bibinfo  {journal} {Physical {R}eview {L}etters}\ }\textbf {\bibinfo
  {volume} {119}},\ \bibinfo {pages} {227208} (\bibinfo {year}
  {2017})}\BibitemShut {NoStop}%
\bibitem [{\citenamefont {Balz}\ \emph {et~al.}(2021)\citenamefont {Balz},
  \citenamefont {Janssen}, \citenamefont {Lampen-Kelley}, \citenamefont
  {Banerjee}, \citenamefont {Liu}, \citenamefont {Yan}, \citenamefont
  {Mandrus}, \citenamefont {Vojta},\ and\ \citenamefont
  {Nagler}}]{lampen2018field}%
  \BibitemOpen
  \bibfield  {author} {\bibinfo {author} {\bibfnamefont {C.}~\bibnamefont
  {Balz}}, \bibinfo {author} {\bibfnamefont {L.}~\bibnamefont {Janssen}},
  \bibinfo {author} {\bibfnamefont {P.}~\bibnamefont {Lampen-Kelley}}, \bibinfo
  {author} {\bibfnamefont {A.}~\bibnamefont {Banerjee}}, \bibinfo {author}
  {\bibfnamefont {Y.}~\bibnamefont {Liu}}, \bibinfo {author} {\bibfnamefont
  {J.-Q.}\ \bibnamefont {Yan}}, \bibinfo {author} {\bibfnamefont
  {D.}~\bibnamefont {Mandrus}}, \bibinfo {author} {\bibfnamefont
  {M.}~\bibnamefont {Vojta}},\ and\ \bibinfo {author} {\bibfnamefont {S.~E.}\
  \bibnamefont {Nagler}},\ }\bibfield  {title} {\bibinfo {title} {Field-induced
  intermediate ordered phase and anisotropic interlayer interactions in
  $\alpha$-{R}u{C}l$_3$},\ }\href@noop {} {\bibfield  {journal} {\bibinfo
  {journal} {Physical Review B}\ }\textbf {\bibinfo {volume} {103}},\ \bibinfo
  {pages} {174417} (\bibinfo {year} {2021})}\BibitemShut {NoStop}%
\bibitem [{\citenamefont {Banerjee}\ \emph {et~al.}(2018)\citenamefont
  {Banerjee}, \citenamefont {Lampen-Kelley}, \citenamefont {Knolle},
  \citenamefont {Balz}, \citenamefont {Aczel}, \citenamefont {Winn},
  \citenamefont {Liu}, \citenamefont {Pajerowski}, \citenamefont {Yan},
  \citenamefont {Bridges} \emph {et~al.}}]{banerjee2018excitations}%
  \BibitemOpen
  \bibfield  {author} {\bibinfo {author} {\bibfnamefont {A.}~\bibnamefont
  {Banerjee}}, \bibinfo {author} {\bibfnamefont {P.}~\bibnamefont
  {Lampen-Kelley}}, \bibinfo {author} {\bibfnamefont {J.}~\bibnamefont
  {Knolle}}, \bibinfo {author} {\bibfnamefont {C.}~\bibnamefont {Balz}},
  \bibinfo {author} {\bibfnamefont {A.~A.}\ \bibnamefont {Aczel}}, \bibinfo
  {author} {\bibfnamefont {B.}~\bibnamefont {Winn}}, \bibinfo {author}
  {\bibfnamefont {Y.}~\bibnamefont {Liu}}, \bibinfo {author} {\bibfnamefont
  {D.}~\bibnamefont {Pajerowski}}, \bibinfo {author} {\bibfnamefont
  {J.}~\bibnamefont {Yan}}, \bibinfo {author} {\bibfnamefont {C.~A.}\
  \bibnamefont {Bridges}}, \emph {et~al.},\ }\bibfield  {title} {\bibinfo
  {title} {Excitations in the field-induced quantum spin liquid state of
  $\alpha$-{R}u{C}l$_3 $},\ }\href@noop {} {\bibfield  {journal} {\bibinfo
  {journal} {npj Quantum Materials}\ }\textbf {\bibinfo {volume} {3}},\
  \bibinfo {pages} {1} (\bibinfo {year} {2018})}\BibitemShut {NoStop}%
\bibitem [{\citenamefont {Sandilands}\ \emph {et~al.}(2015)\citenamefont
  {Sandilands}, \citenamefont {Tian}, \citenamefont {Plumb}, \citenamefont
  {Kim},\ and\ \citenamefont {Burch}}]{sandilands2015scattering}%
  \BibitemOpen
  \bibfield  {author} {\bibinfo {author} {\bibfnamefont {L.~J.}\ \bibnamefont
  {Sandilands}}, \bibinfo {author} {\bibfnamefont {Y.}~\bibnamefont {Tian}},
  \bibinfo {author} {\bibfnamefont {K.~W.}\ \bibnamefont {Plumb}}, \bibinfo
  {author} {\bibfnamefont {Y.-J.}\ \bibnamefont {Kim}},\ and\ \bibinfo {author}
  {\bibfnamefont {K.~S.}\ \bibnamefont {Burch}},\ }\bibfield  {title} {\bibinfo
  {title} {Scattering continuum and possible fractionalized excitations in
  $\alpha$-{R}u{C}l$_3 $},\ }\href@noop {} {\bibfield  {journal} {\bibinfo
  {journal} {Physical {R}eview {L}etters}\ }\textbf {\bibinfo {volume} {114}},\
  \bibinfo {pages} {147201} (\bibinfo {year} {2015})}\BibitemShut {NoStop}%
\bibitem [{\citenamefont {Sandilands}\ \emph {et~al.}(2016)\citenamefont
  {Sandilands}, \citenamefont {Tian}, \citenamefont {Reijnders}, \citenamefont
  {Kim}, \citenamefont {Plumb}, \citenamefont {Kim}, \citenamefont {Kee},\ and\
  \citenamefont {Burch}}]{sandilands2016spin}%
  \BibitemOpen
  \bibfield  {author} {\bibinfo {author} {\bibfnamefont {L.~J.}\ \bibnamefont
  {Sandilands}}, \bibinfo {author} {\bibfnamefont {Y.}~\bibnamefont {Tian}},
  \bibinfo {author} {\bibfnamefont {A.~A.}\ \bibnamefont {Reijnders}}, \bibinfo
  {author} {\bibfnamefont {H.-S.}\ \bibnamefont {Kim}}, \bibinfo {author}
  {\bibfnamefont {K.~W.}\ \bibnamefont {Plumb}}, \bibinfo {author}
  {\bibfnamefont {Y.-J.}\ \bibnamefont {Kim}}, \bibinfo {author} {\bibfnamefont
  {H.-Y.}\ \bibnamefont {Kee}},\ and\ \bibinfo {author} {\bibfnamefont {K.~S.}\
  \bibnamefont {Burch}},\ }\bibfield  {title} {\bibinfo {title} {Spin-orbit
  excitations and electronic structure of the putative {K}itaev magnet
  $\alpha$-{R}u{C}l$_3 $},\ }\href@noop {} {\bibfield  {journal} {\bibinfo
  {journal} {Physical {R}eview {B}}\ }\textbf {\bibinfo {volume} {93}},\
  \bibinfo {pages} {075144} (\bibinfo {year} {2016})}\BibitemShut {NoStop}%
\bibitem [{\citenamefont {Majumder}\ \emph {et~al.}(2015)\citenamefont
  {Majumder}, \citenamefont {Schmidt}, \citenamefont {Rosner}, \citenamefont
  {Tsirlin}, \citenamefont {Yasuoka},\ and\ \citenamefont
  {Baenitz}}]{majumder2015anisotropic}%
  \BibitemOpen
  \bibfield  {author} {\bibinfo {author} {\bibfnamefont {M.}~\bibnamefont
  {Majumder}}, \bibinfo {author} {\bibfnamefont {M.}~\bibnamefont {Schmidt}},
  \bibinfo {author} {\bibfnamefont {H.}~\bibnamefont {Rosner}}, \bibinfo
  {author} {\bibfnamefont {A.}~\bibnamefont {Tsirlin}}, \bibinfo {author}
  {\bibfnamefont {H.}~\bibnamefont {Yasuoka}},\ and\ \bibinfo {author}
  {\bibfnamefont {M.}~\bibnamefont {Baenitz}},\ }\bibfield  {title} {\bibinfo
  {title} {Anisotropic {R}u$^{3+}$ 4d$^5$ magnetism in the $\alpha$-{R}u{C}l$_3
  $ honeycomb system: {S}usceptibility, specific heat, and zero-field {NMR}},\
  }\href@noop {} {\bibfield  {journal} {\bibinfo  {journal} {Physical {R}eview
  {B}}\ }\textbf {\bibinfo {volume} {91}},\ \bibinfo {pages} {180401} (\bibinfo
  {year} {2015})}\BibitemShut {NoStop}%
\bibitem [{\citenamefont {Bachus}\ \emph {et~al.}(2020)\citenamefont {Bachus},
  \citenamefont {Kaib}, \citenamefont {Tokiwa}, \citenamefont {Jesche},
  \citenamefont {Tsurkan}, \citenamefont {Loidl}, \citenamefont {Winter},
  \citenamefont {Tsirlin}, \citenamefont {Valent{\'\i}},\ and\ \citenamefont
  {Gegenwart}}]{bachus2020thermodynamic}%
  \BibitemOpen
  \bibfield  {author} {\bibinfo {author} {\bibfnamefont {S.}~\bibnamefont
  {Bachus}}, \bibinfo {author} {\bibfnamefont {D.}~\bibnamefont {Kaib}},
  \bibinfo {author} {\bibfnamefont {Y.}~\bibnamefont {Tokiwa}}, \bibinfo
  {author} {\bibfnamefont {A.}~\bibnamefont {Jesche}}, \bibinfo {author}
  {\bibfnamefont {V.}~\bibnamefont {Tsurkan}}, \bibinfo {author} {\bibfnamefont
  {A.}~\bibnamefont {Loidl}}, \bibinfo {author} {\bibfnamefont
  {S.}~\bibnamefont {Winter}}, \bibinfo {author} {\bibfnamefont {A.~A.}\
  \bibnamefont {Tsirlin}}, \bibinfo {author} {\bibfnamefont {R.}~\bibnamefont
  {Valent{\'\i}}},\ and\ \bibinfo {author} {\bibfnamefont {P.}~\bibnamefont
  {Gegenwart}},\ }\bibfield  {title} {\bibinfo {title} {Thermodynamic
  perspective on field-induced behavior of $\alpha$-{R}u{C}l$_3 $},\
  }\href@noop {} {\bibfield  {journal} {\bibinfo  {journal} {Physical {R}eview
  {L}etters}\ }\textbf {\bibinfo {volume} {125}},\ \bibinfo {pages} {097203}
  (\bibinfo {year} {2020})}\BibitemShut {NoStop}%
\bibitem [{\citenamefont {Reschke}\ \emph {et~al.}(2019)\citenamefont
  {Reschke}, \citenamefont {Tsurkan}, \citenamefont {Do}, \citenamefont {Choi},
  \citenamefont {Lunkenheimer}, \citenamefont {Wang},\ and\ \citenamefont
  {Loidl}}]{reschke2019terahertz}%
  \BibitemOpen
  \bibfield  {author} {\bibinfo {author} {\bibfnamefont {S.}~\bibnamefont
  {Reschke}}, \bibinfo {author} {\bibfnamefont {V.}~\bibnamefont {Tsurkan}},
  \bibinfo {author} {\bibfnamefont {S.-H.}\ \bibnamefont {Do}}, \bibinfo
  {author} {\bibfnamefont {K.-Y.}\ \bibnamefont {Choi}}, \bibinfo {author}
  {\bibfnamefont {P.}~\bibnamefont {Lunkenheimer}}, \bibinfo {author}
  {\bibfnamefont {Z.}~\bibnamefont {Wang}},\ and\ \bibinfo {author}
  {\bibfnamefont {A.}~\bibnamefont {Loidl}},\ }\bibfield  {title} {\bibinfo
  {title} {Terahertz excitations in $\alpha$-{R}u{C}l$_3 $: {M}ajorana fermions
  and rigid-plane shear and compression modes},\ }\href@noop {} {\bibfield
  {journal} {\bibinfo  {journal} {Physical {R}eview {B}}\ }\textbf {\bibinfo
  {volume} {100}},\ \bibinfo {pages} {100403} (\bibinfo {year}
  {2019})}\BibitemShut {NoStop}%
\bibitem [{\citenamefont {Zhou}\ \emph {et~al.}(2022)\citenamefont {Zhou},
  \citenamefont {Li}, \citenamefont {Matsuda}, \citenamefont {Matsuo},
  \citenamefont {Li}, \citenamefont {Kurita}, \citenamefont {Kindo},\ and\
  \citenamefont {Tanaka}}]{zhou2022intermediate}%
  \BibitemOpen
  \bibfield  {author} {\bibinfo {author} {\bibfnamefont {X.-G.}\ \bibnamefont
  {Zhou}}, \bibinfo {author} {\bibfnamefont {H.}~\bibnamefont {Li}}, \bibinfo
  {author} {\bibfnamefont {Y.~H.}\ \bibnamefont {Matsuda}}, \bibinfo {author}
  {\bibfnamefont {A.}~\bibnamefont {Matsuo}}, \bibinfo {author} {\bibfnamefont
  {W.}~\bibnamefont {Li}}, \bibinfo {author} {\bibfnamefont {N.}~\bibnamefont
  {Kurita}}, \bibinfo {author} {\bibfnamefont {K.}~\bibnamefont {Kindo}},\ and\
  \bibinfo {author} {\bibfnamefont {H.}~\bibnamefont {Tanaka}},\ }\bibfield
  {title} {\bibinfo {title} {Intermediate quantum spin liquid phase in the
  {K}itaev material $\alpha$-rucl$_3 $ under high magnetic fields up to 100
  {T}},\ }\href@noop {} {\bibfield  {journal} {\bibinfo  {journal} {arXiv
  preprint arXiv:2201.04597}\ } (\bibinfo {year} {2022})}\BibitemShut {NoStop}%
\bibitem [{\citenamefont {Liu}\ and\ \citenamefont
  {Khaliullin}(2018)}]{liu2018pseudospin}%
  \BibitemOpen
  \bibfield  {author} {\bibinfo {author} {\bibfnamefont {H.}~\bibnamefont
  {Liu}}\ and\ \bibinfo {author} {\bibfnamefont {G.}~\bibnamefont
  {Khaliullin}},\ }\bibfield  {title} {\bibinfo {title} {Pseudospin exchange
  interactions in d$^7$ cobalt compounds: {P}ossible realization of the
  {K}itaev model},\ }\href@noop {} {\bibfield  {journal} {\bibinfo  {journal}
  {Physical Review B}\ }\textbf {\bibinfo {volume} {97}},\ \bibinfo {pages}
  {014407} (\bibinfo {year} {2018})}\BibitemShut {NoStop}%
\bibitem [{\citenamefont {Liu}\ \emph {et~al.}(2020)\citenamefont {Liu},
  \citenamefont {Chaloupka},\ and\ \citenamefont {Khaliullin}}]{liu2020kitaev}%
  \BibitemOpen
  \bibfield  {author} {\bibinfo {author} {\bibfnamefont {H.}~\bibnamefont
  {Liu}}, \bibinfo {author} {\bibfnamefont {J.}~\bibnamefont {Chaloupka}},\
  and\ \bibinfo {author} {\bibfnamefont {G.}~\bibnamefont {Khaliullin}},\
  }\bibfield  {title} {\bibinfo {title} {{K}itaev spin liquid in 3d transition
  metal compounds},\ }\href@noop {} {\bibfield  {journal} {\bibinfo  {journal}
  {Physical {R}eview {L}etters}\ }\textbf {\bibinfo {volume} {125}},\ \bibinfo
  {pages} {047201} (\bibinfo {year} {2020})}\BibitemShut {NoStop}%
\bibitem [{\citenamefont {Lin}\ \emph {et~al.}(2021)\citenamefont {Lin},
  \citenamefont {Jeong}, \citenamefont {Kim}, \citenamefont {Wang},
  \citenamefont {Huang}, \citenamefont {Masuda}, \citenamefont {Asai},
  \citenamefont {Itoh}, \citenamefont {G{\"u}nther}, \citenamefont {Russina}
  \emph {et~al.}}]{lin2021field}%
  \BibitemOpen
  \bibfield  {author} {\bibinfo {author} {\bibfnamefont {G.}~\bibnamefont
  {Lin}}, \bibinfo {author} {\bibfnamefont {J.}~\bibnamefont {Jeong}}, \bibinfo
  {author} {\bibfnamefont {C.}~\bibnamefont {Kim}}, \bibinfo {author}
  {\bibfnamefont {Y.}~\bibnamefont {Wang}}, \bibinfo {author} {\bibfnamefont
  {Q.}~\bibnamefont {Huang}}, \bibinfo {author} {\bibfnamefont
  {T.}~\bibnamefont {Masuda}}, \bibinfo {author} {\bibfnamefont
  {S.}~\bibnamefont {Asai}}, \bibinfo {author} {\bibfnamefont {S.}~\bibnamefont
  {Itoh}}, \bibinfo {author} {\bibfnamefont {G.}~\bibnamefont {G{\"u}nther}},
  \bibinfo {author} {\bibfnamefont {M.}~\bibnamefont {Russina}}, \emph
  {et~al.},\ }\bibfield  {title} {\bibinfo {title} {Field-induced quantum spin
  disordered state in spin-1/2 honeycomb magnet na2co2teo6},\ }\href@noop {}
  {\bibfield  {journal} {\bibinfo  {journal} {Nature communications}\ }\textbf
  {\bibinfo {volume} {12}},\ \bibinfo {pages} {1} (\bibinfo {year}
  {2021})}\BibitemShut {NoStop}%
\bibitem [{\citenamefont {Kim}\ \emph {et~al.}(2020)\citenamefont {Kim},
  \citenamefont {Jeong}, \citenamefont {Lin}, \citenamefont {Park},
  \citenamefont {Masuda}, \citenamefont {Asai}, \citenamefont {Itoh},
  \citenamefont {Kim}, \citenamefont {Zhou}, \citenamefont {Ma} \emph
  {et~al.}}]{kim2020antiferromagnetic}%
  \BibitemOpen
  \bibfield  {author} {\bibinfo {author} {\bibfnamefont {C.}~\bibnamefont
  {Kim}}, \bibinfo {author} {\bibfnamefont {J.}~\bibnamefont {Jeong}}, \bibinfo
  {author} {\bibfnamefont {G.}~\bibnamefont {Lin}}, \bibinfo {author}
  {\bibfnamefont {P.}~\bibnamefont {Park}}, \bibinfo {author} {\bibfnamefont
  {T.}~\bibnamefont {Masuda}}, \bibinfo {author} {\bibfnamefont
  {S.}~\bibnamefont {Asai}}, \bibinfo {author} {\bibfnamefont {S.}~\bibnamefont
  {Itoh}}, \bibinfo {author} {\bibfnamefont {H.-S.}\ \bibnamefont {Kim}},
  \bibinfo {author} {\bibfnamefont {H.}~\bibnamefont {Zhou}}, \bibinfo {author}
  {\bibfnamefont {J.}~\bibnamefont {Ma}}, \emph {et~al.},\ }\bibfield  {title}
  {\bibinfo {title} {Antiferromagnetic {K}itaev interaction in $ j_{eff}= 1/2$
  cobalt honeycomb materials {N}a$_3 ${C}o$_2 ${S}b{O}$_6 $ and {N}a$_2
  ${C}o$_2 ${T}e{O}$_6$},\ }\href@noop {} {\bibfield  {journal} {\bibinfo
  {journal} {arXiv preprint arXiv:2012.06167}\ } (\bibinfo {year}
  {2020})}\BibitemShut {NoStop}%
\bibitem [{\citenamefont {Jang}\ \emph {et~al.}(2019)\citenamefont {Jang},
  \citenamefont {Sano}, \citenamefont {Kato},\ and\ \citenamefont
  {Motome}}]{jang2019antiferromagnetic}%
  \BibitemOpen
  \bibfield  {author} {\bibinfo {author} {\bibfnamefont {S.-H.}\ \bibnamefont
  {Jang}}, \bibinfo {author} {\bibfnamefont {R.}~\bibnamefont {Sano}}, \bibinfo
  {author} {\bibfnamefont {Y.}~\bibnamefont {Kato}},\ and\ \bibinfo {author}
  {\bibfnamefont {Y.}~\bibnamefont {Motome}},\ }\bibfield  {title} {\bibinfo
  {title} {Antiferromagnetic {K}itaev interaction in f-electron based honeycomb
  magnets},\ }\href@noop {} {\bibfield  {journal} {\bibinfo  {journal}
  {Physical {R}eview {B}}\ }\textbf {\bibinfo {volume} {99}},\ \bibinfo {pages}
  {241106} (\bibinfo {year} {2019})}\BibitemShut {NoStop}%
\bibitem [{\citenamefont {Motome}\ and\ \citenamefont
  {Nasu}(2020)}]{motome2020hunting}%
  \BibitemOpen
  \bibfield  {author} {\bibinfo {author} {\bibfnamefont {Y.}~\bibnamefont
  {Motome}}\ and\ \bibinfo {author} {\bibfnamefont {J.}~\bibnamefont {Nasu}},\
  }\bibfield  {title} {\bibinfo {title} {Hunting {M}ajorana fermions in
  {K}itaev magnets},\ }\href@noop {} {\bibfield  {journal} {\bibinfo  {journal}
  {Journal of the Physical Society of Japan}\ }\textbf {\bibinfo {volume}
  {89}},\ \bibinfo {pages} {012002} (\bibinfo {year} {2020})}\BibitemShut
  {NoStop}%
\bibitem [{\citenamefont {Winter}\ \emph
  {et~al.}(2017{\natexlab{a}})\citenamefont {Winter}, \citenamefont {Tsirlin},
  \citenamefont {Daghofer}, \citenamefont {van~den Brink}, \citenamefont
  {Singh}, \citenamefont {Gegenwart},\ and\ \citenamefont
  {Valenti}}]{winter2017models}%
  \BibitemOpen
  \bibfield  {author} {\bibinfo {author} {\bibfnamefont {S.~M.}\ \bibnamefont
  {Winter}}, \bibinfo {author} {\bibfnamefont {A.~A.}\ \bibnamefont {Tsirlin}},
  \bibinfo {author} {\bibfnamefont {M.}~\bibnamefont {Daghofer}}, \bibinfo
  {author} {\bibfnamefont {J.}~\bibnamefont {van~den Brink}}, \bibinfo {author}
  {\bibfnamefont {Y.}~\bibnamefont {Singh}}, \bibinfo {author} {\bibfnamefont
  {P.}~\bibnamefont {Gegenwart}},\ and\ \bibinfo {author} {\bibfnamefont
  {R.}~\bibnamefont {Valenti}},\ }\bibfield  {title} {\bibinfo {title} {Models
  and materials for generalized {K}itaev magnetism},\ }\href@noop {} {\bibfield
   {journal} {\bibinfo  {journal} {Journal of Physics: Condensed Matter}\
  }\textbf {\bibinfo {volume} {29}},\ \bibinfo {pages} {493002} (\bibinfo
  {year} {2017}{\natexlab{a}})}\BibitemShut {NoStop}%
\bibitem [{\citenamefont {Balents}(2010)}]{balents2010spin}%
  \BibitemOpen
  \bibfield  {author} {\bibinfo {author} {\bibfnamefont {L.}~\bibnamefont
  {Balents}},\ }\bibfield  {title} {\bibinfo {title} {Spin liquids in
  frustrated magnets},\ }\href@noop {} {\bibfield  {journal} {\bibinfo
  {journal} {Nature}\ }\textbf {\bibinfo {volume} {464}},\ \bibinfo {pages}
  {199} (\bibinfo {year} {2010})}\BibitemShut {NoStop}%
\bibitem [{\citenamefont {Savary}\ and\ \citenamefont
  {Balents}(2016)}]{savary2016quantum}%
  \BibitemOpen
  \bibfield  {author} {\bibinfo {author} {\bibfnamefont {L.}~\bibnamefont
  {Savary}}\ and\ \bibinfo {author} {\bibfnamefont {L.}~\bibnamefont
  {Balents}},\ }\bibfield  {title} {\bibinfo {title} {Quantum spin liquids: {A}
  review},\ }\href@noop {} {\bibfield  {journal} {\bibinfo  {journal} {Reports
  on Progress in Physics}\ }\textbf {\bibinfo {volume} {80}},\ \bibinfo {pages}
  {016502} (\bibinfo {year} {2016})}\BibitemShut {NoStop}%
\bibitem [{\citenamefont {Zhou}\ \emph {et~al.}(2017)\citenamefont {Zhou},
  \citenamefont {Kanoda},\ and\ \citenamefont {Ng}}]{zhou2017quantum}%
  \BibitemOpen
  \bibfield  {author} {\bibinfo {author} {\bibfnamefont {Y.}~\bibnamefont
  {Zhou}}, \bibinfo {author} {\bibfnamefont {K.}~\bibnamefont {Kanoda}},\ and\
  \bibinfo {author} {\bibfnamefont {T.-K.}\ \bibnamefont {Ng}},\ }\bibfield
  {title} {\bibinfo {title} {Quantum spin liquid states},\ }\href@noop {}
  {\bibfield  {journal} {\bibinfo  {journal} {Reviews of Modern Physics}\
  }\textbf {\bibinfo {volume} {89}},\ \bibinfo {pages} {025003} (\bibinfo
  {year} {2017})}\BibitemShut {NoStop}%
\bibitem [{\citenamefont {Broholm}\ \emph {et~al.}(2020)\citenamefont
  {Broholm}, \citenamefont {Cava}, \citenamefont {Kivelson}, \citenamefont
  {Nocera}, \citenamefont {Norman},\ and\ \citenamefont
  {Senthil}}]{broholm2020quantum}%
  \BibitemOpen
  \bibfield  {author} {\bibinfo {author} {\bibfnamefont {C.}~\bibnamefont
  {Broholm}}, \bibinfo {author} {\bibfnamefont {R.}~\bibnamefont {Cava}},
  \bibinfo {author} {\bibfnamefont {S.}~\bibnamefont {Kivelson}}, \bibinfo
  {author} {\bibfnamefont {D.}~\bibnamefont {Nocera}}, \bibinfo {author}
  {\bibfnamefont {M.}~\bibnamefont {Norman}},\ and\ \bibinfo {author}
  {\bibfnamefont {T.}~\bibnamefont {Senthil}},\ }\bibfield  {title} {\bibinfo
  {title} {Quantum spin liquids},\ }\href@noop {} {\bibfield  {journal}
  {\bibinfo  {journal} {Science}\ }\textbf {\bibinfo {volume} {367}} (\bibinfo
  {year} {2020})}\BibitemShut {NoStop}%
\bibitem [{\citenamefont {Knolle}\ and\ \citenamefont
  {Moessner}(2019)}]{knolle2019field}%
  \BibitemOpen
  \bibfield  {author} {\bibinfo {author} {\bibfnamefont {J.}~\bibnamefont
  {Knolle}}\ and\ \bibinfo {author} {\bibfnamefont {R.}~\bibnamefont
  {Moessner}},\ }\bibfield  {title} {\bibinfo {title} {A field guide to spin
  liquids},\ }\href@noop {} {\bibfield  {journal} {\bibinfo  {journal} {Annual
  {R}eview of {C}ondensed {M}atter {P}hysics}\ }\textbf {\bibinfo {volume}
  {10}},\ \bibinfo {pages} {451} (\bibinfo {year} {2019})}\BibitemShut
  {NoStop}%
\bibitem [{\citenamefont {Jackeli}\ and\ \citenamefont
  {Khaliullin}(2009)}]{jackeli2009mott}%
  \BibitemOpen
  \bibfield  {author} {\bibinfo {author} {\bibfnamefont {G.}~\bibnamefont
  {Jackeli}}\ and\ \bibinfo {author} {\bibfnamefont {G.}~\bibnamefont
  {Khaliullin}},\ }\bibfield  {title} {\bibinfo {title} {Mott insulators in the
  strong spin-orbit coupling limit: {F}rom {H}eisenberg to a quantum compass
  and {K}itaev models},\ }\href@noop {} {\bibfield  {journal} {\bibinfo
  {journal} {Physical {R}eview {L}etters}\ }\textbf {\bibinfo {volume} {102}},\
  \bibinfo {pages} {017205} (\bibinfo {year} {2009})}\BibitemShut {NoStop}%
\bibitem [{\citenamefont {Rau}\ \emph {et~al.}(2014)\citenamefont {Rau},
  \citenamefont {Lee},\ and\ \citenamefont {Kee}}]{rau2014generic}%
  \BibitemOpen
  \bibfield  {author} {\bibinfo {author} {\bibfnamefont {J.~G.}\ \bibnamefont
  {Rau}}, \bibinfo {author} {\bibfnamefont {E.~K.-H.}\ \bibnamefont {Lee}},\
  and\ \bibinfo {author} {\bibfnamefont {H.-Y.}\ \bibnamefont {Kee}},\
  }\bibfield  {title} {\bibinfo {title} {Generic spin model for the honeycomb
  iridates beyond the {K}itaev limit},\ }\href@noop {} {\bibfield  {journal}
  {\bibinfo  {journal} {Physical {R}eview {L}etters}\ }\textbf {\bibinfo
  {volume} {112}},\ \bibinfo {pages} {077204} (\bibinfo {year}
  {2014})}\BibitemShut {NoStop}%
\bibitem [{\citenamefont {Li}\ \emph {et~al.}(2017)\citenamefont {Li},
  \citenamefont {Li}, \citenamefont {Yu}, \citenamefont {Paramekanti},\ and\
  \citenamefont {Chen}}]{li2017kitaev}%
  \BibitemOpen
  \bibfield  {author} {\bibinfo {author} {\bibfnamefont {F.-Y.}\ \bibnamefont
  {Li}}, \bibinfo {author} {\bibfnamefont {Y.-D.}\ \bibnamefont {Li}}, \bibinfo
  {author} {\bibfnamefont {Y.}~\bibnamefont {Yu}}, \bibinfo {author}
  {\bibfnamefont {A.}~\bibnamefont {Paramekanti}},\ and\ \bibinfo {author}
  {\bibfnamefont {G.}~\bibnamefont {Chen}},\ }\bibfield  {title} {\bibinfo
  {title} {{K}itaev materials beyond iridates: {O}rder by quantum disorder and
  {W}eyl magnons in rare-earth double perovskites},\ }\href@noop {} {\bibfield
  {journal} {\bibinfo  {journal} {Physical {R}eview {B}}\ }\textbf {\bibinfo
  {volume} {95}},\ \bibinfo {pages} {085132} (\bibinfo {year}
  {2017})}\BibitemShut {NoStop}%
\bibitem [{\citenamefont {Motome}\ \emph {et~al.}(2020)\citenamefont {Motome},
  \citenamefont {Sano}, \citenamefont {Jang}, \citenamefont {Sugita},\ and\
  \citenamefont {Kato}}]{motome2020materials}%
  \BibitemOpen
  \bibfield  {author} {\bibinfo {author} {\bibfnamefont {Y.}~\bibnamefont
  {Motome}}, \bibinfo {author} {\bibfnamefont {R.}~\bibnamefont {Sano}},
  \bibinfo {author} {\bibfnamefont {S.}~\bibnamefont {Jang}}, \bibinfo {author}
  {\bibfnamefont {Y.}~\bibnamefont {Sugita}},\ and\ \bibinfo {author}
  {\bibfnamefont {Y.}~\bibnamefont {Kato}},\ }\bibfield  {title} {\bibinfo
  {title} {Materials design of {K}itaev spin liquids beyond the
  {J}ackeli-{K}haliullin mechanism},\ }\href@noop {} {\bibfield  {journal}
  {\bibinfo  {journal} {Journal of Physics: Condensed Matter}\ }\textbf
  {\bibinfo {volume} {32}},\ \bibinfo {pages} {404001} (\bibinfo {year}
  {2020})}\BibitemShut {NoStop}%
\bibitem [{\citenamefont {Janssen}\ \emph {et~al.}(2017)\citenamefont
  {Janssen}, \citenamefont {Andrade},\ and\ \citenamefont
  {Vojta}}]{janssen2017magnetization}%
  \BibitemOpen
  \bibfield  {author} {\bibinfo {author} {\bibfnamefont {L.}~\bibnamefont
  {Janssen}}, \bibinfo {author} {\bibfnamefont {E.~C.}\ \bibnamefont
  {Andrade}},\ and\ \bibinfo {author} {\bibfnamefont {M.}~\bibnamefont
  {Vojta}},\ }\bibfield  {title} {\bibinfo {title} {Magnetization processes of
  zigzag states on the honeycomb lattice: {I}dentifying spin models for
  $\alpha$-{R}u{C}l$_3 $ and {N}a$_2${I}r{O}$_3$},\ }\href@noop {} {\bibfield
  {journal} {\bibinfo  {journal} {Physical {R}eview {B}}\ }\textbf {\bibinfo
  {volume} {96}},\ \bibinfo {pages} {064430} (\bibinfo {year}
  {2017})}\BibitemShut {NoStop}%
\bibitem [{\citenamefont {Katukuri}\ \emph {et~al.}(2014)\citenamefont
  {Katukuri}, \citenamefont {Nishimoto}, \citenamefont {Yushankhai},
  \citenamefont {Stoyanova}, \citenamefont {Kandpal}, \citenamefont {Choi},
  \citenamefont {Coldea}, \citenamefont {Rousochatzakis}, \citenamefont
  {Hozoi},\ and\ \citenamefont {Van Den~Brink}}]{katukuri2014kitaev}%
  \BibitemOpen
  \bibfield  {author} {\bibinfo {author} {\bibfnamefont {V.~M.}\ \bibnamefont
  {Katukuri}}, \bibinfo {author} {\bibfnamefont {S.}~\bibnamefont {Nishimoto}},
  \bibinfo {author} {\bibfnamefont {V.}~\bibnamefont {Yushankhai}}, \bibinfo
  {author} {\bibfnamefont {A.}~\bibnamefont {Stoyanova}}, \bibinfo {author}
  {\bibfnamefont {H.}~\bibnamefont {Kandpal}}, \bibinfo {author} {\bibfnamefont
  {S.}~\bibnamefont {Choi}}, \bibinfo {author} {\bibfnamefont {R.}~\bibnamefont
  {Coldea}}, \bibinfo {author} {\bibfnamefont {I.}~\bibnamefont
  {Rousochatzakis}}, \bibinfo {author} {\bibfnamefont {L.}~\bibnamefont
  {Hozoi}},\ and\ \bibinfo {author} {\bibfnamefont {J.}~\bibnamefont {Van
  Den~Brink}},\ }\bibfield  {title} {\bibinfo {title} {{K}itaev interactions
  between $j= 1/2$ moments in honeycomb {N}a$_2${I}r{O}$_3$ are large and
  ferromagnetic insights from ab initio quantum chemistry calculations},\
  }\href@noop {} {\bibfield  {journal} {\bibinfo  {journal} {New Journal of
  Physics}\ }\textbf {\bibinfo {volume} {16}},\ \bibinfo {pages} {013056}
  (\bibinfo {year} {2014})}\BibitemShut {NoStop}%
\bibitem [{\citenamefont {Laurell}\ and\ \citenamefont
  {Okamoto}(2020)}]{laurell2020dynamical}%
  \BibitemOpen
  \bibfield  {author} {\bibinfo {author} {\bibfnamefont {P.}~\bibnamefont
  {Laurell}}\ and\ \bibinfo {author} {\bibfnamefont {S.}~\bibnamefont
  {Okamoto}},\ }\bibfield  {title} {\bibinfo {title} {Dynamical and thermal
  magnetic properties of the {K}itaev spin liquid candidate
  $\alpha$-{R}u{C}l$_3 $},\ }\href@noop {} {\bibfield  {journal} {\bibinfo
  {journal} {npj Quantum Materials}\ }\textbf {\bibinfo {volume} {5}},\
  \bibinfo {pages} {1} (\bibinfo {year} {2020})}\BibitemShut {NoStop}%
\bibitem [{\citenamefont {Kim}\ \emph {et~al.}(2015)\citenamefont {Kim},
  \citenamefont {Catuneanu}, \citenamefont {Kee} \emph
  {et~al.}}]{kim2015kitaev}%
  \BibitemOpen
  \bibfield  {author} {\bibinfo {author} {\bibfnamefont {H.-S.}\ \bibnamefont
  {Kim}}, \bibinfo {author} {\bibfnamefont {A.}~\bibnamefont {Catuneanu}},
  \bibinfo {author} {\bibfnamefont {H.-Y.}\ \bibnamefont {Kee}}, \emph
  {et~al.},\ }\bibfield  {title} {\bibinfo {title} {{K}itaev magnetism in
  honeycomb {R}u{C}l$_3 $ with intermediate spin-orbit coupling},\ }\href@noop
  {} {\bibfield  {journal} {\bibinfo  {journal} {Physical {R}eview {B}}\
  }\textbf {\bibinfo {volume} {91}},\ \bibinfo {pages} {241110} (\bibinfo
  {year} {2015})}\BibitemShut {NoStop}%
\bibitem [{\citenamefont {Wu}\ \emph {et~al.}(2018)\citenamefont {Wu},
  \citenamefont {Little}, \citenamefont {Aldape}, \citenamefont {Rees},
  \citenamefont {Thewalt}, \citenamefont {Lampen-Kelley}, \citenamefont
  {Banerjee}, \citenamefont {Bridges}, \citenamefont {Yan}, \citenamefont
  {Boone} \emph {et~al.}}]{wu2018field}%
  \BibitemOpen
  \bibfield  {author} {\bibinfo {author} {\bibfnamefont {L.}~\bibnamefont
  {Wu}}, \bibinfo {author} {\bibfnamefont {A.}~\bibnamefont {Little}}, \bibinfo
  {author} {\bibfnamefont {E.~E.}\ \bibnamefont {Aldape}}, \bibinfo {author}
  {\bibfnamefont {D.}~\bibnamefont {Rees}}, \bibinfo {author} {\bibfnamefont
  {E.}~\bibnamefont {Thewalt}}, \bibinfo {author} {\bibfnamefont
  {P.}~\bibnamefont {Lampen-Kelley}}, \bibinfo {author} {\bibfnamefont
  {A.}~\bibnamefont {Banerjee}}, \bibinfo {author} {\bibfnamefont {C.~A.}\
  \bibnamefont {Bridges}}, \bibinfo {author} {\bibfnamefont {J.-Q.}\
  \bibnamefont {Yan}}, \bibinfo {author} {\bibfnamefont {D.}~\bibnamefont
  {Boone}}, \emph {et~al.},\ }\bibfield  {title} {\bibinfo {title} {Field
  evolution of magnons in $\alpha$-{R}u{C}l$_3 $ by high-resolution polarized
  terahertz spectroscopy},\ }\href@noop {} {\bibfield  {journal} {\bibinfo
  {journal} {Physical {R}eview {B}}\ }\textbf {\bibinfo {volume} {98}},\
  \bibinfo {pages} {094425} (\bibinfo {year} {2018})}\BibitemShut {NoStop}%
\bibitem [{\citenamefont {Cookmeyer}\ and\ \citenamefont
  {Moore}(2018)}]{cookmeyer2018spin}%
  \BibitemOpen
  \bibfield  {author} {\bibinfo {author} {\bibfnamefont {J.}~\bibnamefont
  {Cookmeyer}}\ and\ \bibinfo {author} {\bibfnamefont {J.~E.}\ \bibnamefont
  {Moore}},\ }\bibfield  {title} {\bibinfo {title} {Spin-wave analysis of the
  low-temperature thermal {H}all effect in the candidate {K}itaev spin liquid
  $\alpha$-{R}u{C}l$_3 $},\ }\href@noop {} {\bibfield  {journal} {\bibinfo
  {journal} {Physical {R}eview {B}}\ }\textbf {\bibinfo {volume} {98}},\
  \bibinfo {pages} {060412} (\bibinfo {year} {2018})}\BibitemShut {NoStop}%
\bibitem [{\citenamefont {McClarty}\ \emph {et~al.}(2018)\citenamefont
  {McClarty}, \citenamefont {Dong}, \citenamefont {Gohlke}, \citenamefont
  {Rau}, \citenamefont {Pollmann}, \citenamefont {Moessner},\ and\
  \citenamefont {Penc}}]{mcclarty2018topological}%
  \BibitemOpen
  \bibfield  {author} {\bibinfo {author} {\bibfnamefont {P.}~\bibnamefont
  {McClarty}}, \bibinfo {author} {\bibfnamefont {X.-Y.}\ \bibnamefont {Dong}},
  \bibinfo {author} {\bibfnamefont {M.}~\bibnamefont {Gohlke}}, \bibinfo
  {author} {\bibfnamefont {J.}~\bibnamefont {Rau}}, \bibinfo {author}
  {\bibfnamefont {F.}~\bibnamefont {Pollmann}}, \bibinfo {author}
  {\bibfnamefont {R.}~\bibnamefont {Moessner}},\ and\ \bibinfo {author}
  {\bibfnamefont {K.}~\bibnamefont {Penc}},\ }\bibfield  {title} {\bibinfo
  {title} {Topological magnons in {K}itaev magnets at high fields},\
  }\href@noop {} {\bibfield  {journal} {\bibinfo  {journal} {Physical {R}eview
  {B}}\ }\textbf {\bibinfo {volume} {98}},\ \bibinfo {pages} {060404} (\bibinfo
  {year} {2018})}\BibitemShut {NoStop}%
\bibitem [{\citenamefont {Ye}\ \emph {et~al.}(2020)\citenamefont {Ye},
  \citenamefont {Fernandes},\ and\ \citenamefont {Perkins}}]{ye2020phonon}%
  \BibitemOpen
  \bibfield  {author} {\bibinfo {author} {\bibfnamefont {M.}~\bibnamefont
  {Ye}}, \bibinfo {author} {\bibfnamefont {R.~M.}\ \bibnamefont {Fernandes}},\
  and\ \bibinfo {author} {\bibfnamefont {N.~B.}\ \bibnamefont {Perkins}},\
  }\bibfield  {title} {\bibinfo {title} {Phonon dynamics in the {K}itaev spin
  liquid},\ }\href@noop {} {\bibfield  {journal} {\bibinfo  {journal} {Physical
  Review Research}\ }\textbf {\bibinfo {volume} {2}},\ \bibinfo {pages}
  {033180} (\bibinfo {year} {2020})}\BibitemShut {NoStop}%
\bibitem [{\citenamefont {Lefran{\c{c}}ois}\ \emph {et~al.}(2021)\citenamefont
  {Lefran{\c{c}}ois}, \citenamefont {Grissonnanche}, \citenamefont {Baglo},
  \citenamefont {Lampen-Kelley}, \citenamefont {Yan}, \citenamefont {Balz},
  \citenamefont {Mandrus}, \citenamefont {Nagler}, \citenamefont {Kim},
  \citenamefont {Kim} \emph {et~al.}}]{lefranccois2021evidence}%
  \BibitemOpen
  \bibfield  {author} {\bibinfo {author} {\bibfnamefont {{\'E}.}~\bibnamefont
  {Lefran{\c{c}}ois}}, \bibinfo {author} {\bibfnamefont {G.}~\bibnamefont
  {Grissonnanche}}, \bibinfo {author} {\bibfnamefont {J.}~\bibnamefont
  {Baglo}}, \bibinfo {author} {\bibfnamefont {P.}~\bibnamefont
  {Lampen-Kelley}}, \bibinfo {author} {\bibfnamefont {J.}~\bibnamefont {Yan}},
  \bibinfo {author} {\bibfnamefont {C.}~\bibnamefont {Balz}}, \bibinfo {author}
  {\bibfnamefont {D.}~\bibnamefont {Mandrus}}, \bibinfo {author} {\bibfnamefont
  {S.}~\bibnamefont {Nagler}}, \bibinfo {author} {\bibfnamefont
  {S.}~\bibnamefont {Kim}}, \bibinfo {author} {\bibfnamefont {Y.-J.}\
  \bibnamefont {Kim}}, \emph {et~al.},\ }\bibfield  {title} {\bibinfo {title}
  {Evidence of a phonon {H}all effect in the {K}itaev spin liquid candidate
  $\alpha$-rucl$_3$},\ }\href@noop {} {\bibfield  {journal} {\bibinfo
  {journal} {arXiv preprint arXiv:2111.05493}\ } (\bibinfo {year}
  {2021})}\BibitemShut {NoStop}%
\bibitem [{\citenamefont {Gao}\ \emph {et~al.}(2019)\citenamefont {Gao},
  \citenamefont {Hickey}, \citenamefont {Xiang}, \citenamefont {Trebst},\ and\
  \citenamefont {Chen}}]{gao2019thermal}%
  \BibitemOpen
  \bibfield  {author} {\bibinfo {author} {\bibfnamefont {Y.~H.}\ \bibnamefont
  {Gao}}, \bibinfo {author} {\bibfnamefont {C.}~\bibnamefont {Hickey}},
  \bibinfo {author} {\bibfnamefont {T.}~\bibnamefont {Xiang}}, \bibinfo
  {author} {\bibfnamefont {S.}~\bibnamefont {Trebst}},\ and\ \bibinfo {author}
  {\bibfnamefont {G.}~\bibnamefont {Chen}},\ }\bibfield  {title} {\bibinfo
  {title} {Thermal {H}all signatures of non-{K}itaev spin liquids in honeycomb
  {K}itaev materials},\ }\href@noop {} {\bibfield  {journal} {\bibinfo
  {journal} {Physical Review Research}\ }\textbf {\bibinfo {volume} {1}},\
  \bibinfo {pages} {013014} (\bibinfo {year} {2019})}\BibitemShut {NoStop}%
\bibitem [{\citenamefont {Hentrich}\ \emph {et~al.}(2019)\citenamefont
  {Hentrich}, \citenamefont {Roslova}, \citenamefont {Isaeva}, \citenamefont
  {Doert}, \citenamefont {Brenig}, \citenamefont {B{\"u}chner},\ and\
  \citenamefont {Hess}}]{hentrich2019large}%
  \BibitemOpen
  \bibfield  {author} {\bibinfo {author} {\bibfnamefont {R.}~\bibnamefont
  {Hentrich}}, \bibinfo {author} {\bibfnamefont {M.}~\bibnamefont {Roslova}},
  \bibinfo {author} {\bibfnamefont {A.}~\bibnamefont {Isaeva}}, \bibinfo
  {author} {\bibfnamefont {T.}~\bibnamefont {Doert}}, \bibinfo {author}
  {\bibfnamefont {W.}~\bibnamefont {Brenig}}, \bibinfo {author} {\bibfnamefont
  {B.}~\bibnamefont {B{\"u}chner}},\ and\ \bibinfo {author} {\bibfnamefont
  {C.}~\bibnamefont {Hess}},\ }\bibfield  {title} {\bibinfo {title} {Large
  thermal {H}all effect in $\alpha$-{R}u{C}l$_3 $: {E}vidence for heat
  transport by {K}itaev-{H}eisenberg paramagnons},\ }\href@noop {} {\bibfield
  {journal} {\bibinfo  {journal} {Physical {R}eview {B}}\ }\textbf {\bibinfo
  {volume} {99}},\ \bibinfo {pages} {085136} (\bibinfo {year}
  {2019})}\BibitemShut {NoStop}%
\bibitem [{\citenamefont {Kasahara}\ \emph {et~al.}(2018)\citenamefont
  {Kasahara}, \citenamefont {Ohnishi}, \citenamefont {Mizukami}, \citenamefont
  {Tanaka}, \citenamefont {Ma}, \citenamefont {Sugii}, \citenamefont {Kurita},
  \citenamefont {Tanaka}, \citenamefont {Nasu}, \citenamefont {Motome} \emph
  {et~al.}}]{kasahara2018majorana}%
  \BibitemOpen
  \bibfield  {author} {\bibinfo {author} {\bibfnamefont {Y.}~\bibnamefont
  {Kasahara}}, \bibinfo {author} {\bibfnamefont {T.}~\bibnamefont {Ohnishi}},
  \bibinfo {author} {\bibfnamefont {Y.}~\bibnamefont {Mizukami}}, \bibinfo
  {author} {\bibfnamefont {O.}~\bibnamefont {Tanaka}}, \bibinfo {author}
  {\bibfnamefont {S.}~\bibnamefont {Ma}}, \bibinfo {author} {\bibfnamefont
  {K.}~\bibnamefont {Sugii}}, \bibinfo {author} {\bibfnamefont
  {N.}~\bibnamefont {Kurita}}, \bibinfo {author} {\bibfnamefont
  {H.}~\bibnamefont {Tanaka}}, \bibinfo {author} {\bibfnamefont
  {J.}~\bibnamefont {Nasu}}, \bibinfo {author} {\bibfnamefont {Y.}~\bibnamefont
  {Motome}}, \emph {et~al.},\ }\bibfield  {title} {\bibinfo {title} {Majorana
  quantization and half-integer thermal quantum {H}all effect in a {K}itaev
  spin liquid},\ }\href@noop {} {\bibfield  {journal} {\bibinfo  {journal}
  {Nature}\ }\textbf {\bibinfo {volume} {559}},\ \bibinfo {pages} {227}
  (\bibinfo {year} {2018})}\BibitemShut {NoStop}%
\bibitem [{\citenamefont {Yokoi}\ \emph {et~al.}(2021)\citenamefont {Yokoi},
  \citenamefont {Ma}, \citenamefont {Kasahara}, \citenamefont {Kasahara},
  \citenamefont {Shibauchi}, \citenamefont {Kurita}, \citenamefont {Tanaka},
  \citenamefont {Nasu}, \citenamefont {Motome}, \citenamefont {Hickey} \emph
  {et~al.}}]{yokoi2020half}%
  \BibitemOpen
  \bibfield  {author} {\bibinfo {author} {\bibfnamefont {T.}~\bibnamefont
  {Yokoi}}, \bibinfo {author} {\bibfnamefont {S.}~\bibnamefont {Ma}}, \bibinfo
  {author} {\bibfnamefont {Y.}~\bibnamefont {Kasahara}}, \bibinfo {author}
  {\bibfnamefont {S.}~\bibnamefont {Kasahara}}, \bibinfo {author}
  {\bibfnamefont {T.}~\bibnamefont {Shibauchi}}, \bibinfo {author}
  {\bibfnamefont {N.}~\bibnamefont {Kurita}}, \bibinfo {author} {\bibfnamefont
  {H.}~\bibnamefont {Tanaka}}, \bibinfo {author} {\bibfnamefont
  {J.}~\bibnamefont {Nasu}}, \bibinfo {author} {\bibfnamefont {Y.}~\bibnamefont
  {Motome}}, \bibinfo {author} {\bibfnamefont {C.}~\bibnamefont {Hickey}},
  \emph {et~al.},\ }\bibfield  {title} {\bibinfo {title} {Half-integer
  quantized anomalous thermal {H}all effect in the {K}itaev material candidate
  $\alpha$-{R}u{C}l$_3$},\ }\href@noop {} {\bibfield  {journal} {\bibinfo
  {journal} {Science}\ }\textbf {\bibinfo {volume} {373}},\ \bibinfo {pages}
  {568} (\bibinfo {year} {2021})}\BibitemShut {NoStop}%
\bibitem [{\citenamefont {Nasu}\ \emph {et~al.}(2018)\citenamefont {Nasu},
  \citenamefont {Kato}, \citenamefont {Kamiya},\ and\ \citenamefont
  {Motome}}]{nasu2018successive}%
  \BibitemOpen
  \bibfield  {author} {\bibinfo {author} {\bibfnamefont {J.}~\bibnamefont
  {Nasu}}, \bibinfo {author} {\bibfnamefont {Y.}~\bibnamefont {Kato}}, \bibinfo
  {author} {\bibfnamefont {Y.}~\bibnamefont {Kamiya}},\ and\ \bibinfo {author}
  {\bibfnamefont {Y.}~\bibnamefont {Motome}},\ }\bibfield  {title} {\bibinfo
  {title} {Successive {M}ajorana topological transitions driven by a magnetic
  field in the {K}itaev model},\ }\href@noop {} {\bibfield  {journal} {\bibinfo
   {journal} {Physical {R}eview {B}}\ }\textbf {\bibinfo {volume} {98}},\
  \bibinfo {pages} {060416} (\bibinfo {year} {2018})}\BibitemShut {NoStop}%
\bibitem [{\citenamefont {Janssen}\ and\ \citenamefont
  {Vojta}(2019)}]{janssen2019heisenberg}%
  \BibitemOpen
  \bibfield  {author} {\bibinfo {author} {\bibfnamefont {L.}~\bibnamefont
  {Janssen}}\ and\ \bibinfo {author} {\bibfnamefont {M.}~\bibnamefont
  {Vojta}},\ }\bibfield  {title} {\bibinfo {title} {{H}eisenberg-{K}itaev
  physics in magnetic fields},\ }\href@noop {} {\bibfield  {journal} {\bibinfo
  {journal} {Journal of Physics: Condensed Matter}\ }\textbf {\bibinfo {volume}
  {31}},\ \bibinfo {pages} {423002} (\bibinfo {year} {2019})}\BibitemShut
  {NoStop}%
\bibitem [{\citenamefont {Ralko}\ and\ \citenamefont
  {Merino}(2020)}]{ralko2020novel}%
  \BibitemOpen
  \bibfield  {author} {\bibinfo {author} {\bibfnamefont {A.}~\bibnamefont
  {Ralko}}\ and\ \bibinfo {author} {\bibfnamefont {J.}~\bibnamefont {Merino}},\
  }\bibfield  {title} {\bibinfo {title} {Novel chiral quantum spin liquids in
  {K}itaev magnets},\ }\href@noop {} {\bibfield  {journal} {\bibinfo  {journal}
  {Physical {R}eview {L}etters}\ }\textbf {\bibinfo {volume} {124}},\ \bibinfo
  {pages} {217203} (\bibinfo {year} {2020})}\BibitemShut {NoStop}%
\bibitem [{\citenamefont {Liang}\ \emph {et~al.}(2018)\citenamefont {Liang},
  \citenamefont {Jiang}, \citenamefont {Chen}, \citenamefont {Li},\ and\
  \citenamefont {Wang}}]{liang2018intermediate}%
  \BibitemOpen
  \bibfield  {author} {\bibinfo {author} {\bibfnamefont {S.}~\bibnamefont
  {Liang}}, \bibinfo {author} {\bibfnamefont {M.-H.}\ \bibnamefont {Jiang}},
  \bibinfo {author} {\bibfnamefont {W.}~\bibnamefont {Chen}}, \bibinfo {author}
  {\bibfnamefont {J.-X.}\ \bibnamefont {Li}},\ and\ \bibinfo {author}
  {\bibfnamefont {Q.-H.}\ \bibnamefont {Wang}},\ }\bibfield  {title} {\bibinfo
  {title} {Intermediate gapless phase and topological phase transition of the
  {K}itaev model in a uniform magnetic field},\ }\href@noop {} {\bibfield
  {journal} {\bibinfo  {journal} {Physical {R}eview {B}}\ }\textbf {\bibinfo
  {volume} {98}},\ \bibinfo {pages} {054433} (\bibinfo {year}
  {2018})}\BibitemShut {NoStop}%
\bibitem [{\citenamefont {Ido}\ and\ \citenamefont
  {Misawa}(2020)}]{ido2020correlation}%
  \BibitemOpen
  \bibfield  {author} {\bibinfo {author} {\bibfnamefont {K.}~\bibnamefont
  {Ido}}\ and\ \bibinfo {author} {\bibfnamefont {T.}~\bibnamefont {Misawa}},\
  }\bibfield  {title} {\bibinfo {title} {Correlation effects on the
  magnetization process of the {K}itaev model},\ }\href@noop {} {\bibfield
  {journal} {\bibinfo  {journal} {Physical {R}eview {B}}\ }\textbf {\bibinfo
  {volume} {101}},\ \bibinfo {pages} {045121} (\bibinfo {year}
  {2020})}\BibitemShut {NoStop}%
\bibitem [{\citenamefont {Gordon}\ \emph {et~al.}(2019)\citenamefont {Gordon},
  \citenamefont {Catuneanu}, \citenamefont {S{\o}rensen},\ and\ \citenamefont
  {Kee}}]{gordon2019theory}%
  \BibitemOpen
  \bibfield  {author} {\bibinfo {author} {\bibfnamefont {J.~S.}\ \bibnamefont
  {Gordon}}, \bibinfo {author} {\bibfnamefont {A.}~\bibnamefont {Catuneanu}},
  \bibinfo {author} {\bibfnamefont {E.~S.}\ \bibnamefont {S{\o}rensen}},\ and\
  \bibinfo {author} {\bibfnamefont {H.-Y.}\ \bibnamefont {Kee}},\ }\bibfield
  {title} {\bibinfo {title} {Theory of the field-revealed {K}itaev spin
  liquid},\ }\href@noop {} {\bibfield  {journal} {\bibinfo  {journal} {Nature
  {C}ommunications}\ }\textbf {\bibinfo {volume} {10}},\ \bibinfo {pages} {1}
  (\bibinfo {year} {2019})}\BibitemShut {NoStop}%
\bibitem [{\citenamefont {Catuneanu}\ \emph {et~al.}(2018)\citenamefont
  {Catuneanu}, \citenamefont {Yamaji}, \citenamefont {Wachtel}, \citenamefont
  {Kim},\ and\ \citenamefont {Kee}}]{catuneanu2018path}%
  \BibitemOpen
  \bibfield  {author} {\bibinfo {author} {\bibfnamefont {A.}~\bibnamefont
  {Catuneanu}}, \bibinfo {author} {\bibfnamefont {Y.}~\bibnamefont {Yamaji}},
  \bibinfo {author} {\bibfnamefont {G.}~\bibnamefont {Wachtel}}, \bibinfo
  {author} {\bibfnamefont {Y.~B.}\ \bibnamefont {Kim}},\ and\ \bibinfo {author}
  {\bibfnamefont {H.-Y.}\ \bibnamefont {Kee}},\ }\bibfield  {title} {\bibinfo
  {title} {Path to stable quantum spin liquids in spin-orbit coupled correlated
  materials},\ }\href@noop {} {\bibfield  {journal} {\bibinfo  {journal} {npj
  Quantum Materials}\ }\textbf {\bibinfo {volume} {3}},\ \bibinfo {pages} {1}
  (\bibinfo {year} {2018})}\BibitemShut {NoStop}%
\bibitem [{\citenamefont {Liu}\ and\ \citenamefont
  {Normand}(2018)}]{liu2018dirac}%
  \BibitemOpen
  \bibfield  {author} {\bibinfo {author} {\bibfnamefont {Z.-X.}\ \bibnamefont
  {Liu}}\ and\ \bibinfo {author} {\bibfnamefont {B.}~\bibnamefont {Normand}},\
  }\bibfield  {title} {\bibinfo {title} {Dirac and chiral quantum spin liquids
  on the honeycomb lattice in a magnetic field},\ }\href@noop {} {\bibfield
  {journal} {\bibinfo  {journal} {Physical {R}eview {L}etters}\ }\textbf
  {\bibinfo {volume} {120}},\ \bibinfo {pages} {187201} (\bibinfo {year}
  {2018})}\BibitemShut {NoStop}%
\bibitem [{\citenamefont {Ronquillo}\ \emph {et~al.}(2019)\citenamefont
  {Ronquillo}, \citenamefont {Vengal},\ and\ \citenamefont
  {Trivedi}}]{ronquillo2019signatures}%
  \BibitemOpen
  \bibfield  {author} {\bibinfo {author} {\bibfnamefont {D.~C.}\ \bibnamefont
  {Ronquillo}}, \bibinfo {author} {\bibfnamefont {A.}~\bibnamefont {Vengal}},\
  and\ \bibinfo {author} {\bibfnamefont {N.}~\bibnamefont {Trivedi}},\
  }\bibfield  {title} {\bibinfo {title} {Signatures of magnetic-field-driven
  quantum phase transitions in the entanglement entropy and spin dynamics of
  the {K}itaev honeycomb model},\ }\href@noop {} {\bibfield  {journal}
  {\bibinfo  {journal} {Physical {R}eview {B}}\ }\textbf {\bibinfo {volume}
  {99}},\ \bibinfo {pages} {140413} (\bibinfo {year} {2019})}\BibitemShut
  {NoStop}%
\bibitem [{\citenamefont {Takikawa}\ and\ \citenamefont
  {Fujimoto}(2019)}]{takikawa2019impact}%
  \BibitemOpen
  \bibfield  {author} {\bibinfo {author} {\bibfnamefont {D.}~\bibnamefont
  {Takikawa}}\ and\ \bibinfo {author} {\bibfnamefont {S.}~\bibnamefont
  {Fujimoto}},\ }\bibfield  {title} {\bibinfo {title} {Impact of off-diagonal
  exchange interactions on the {K}itaev spin-liquid state of
  $\alpha$-{R}u{C}l$_3 $},\ }\href@noop {} {\bibfield  {journal} {\bibinfo
  {journal} {Physical {R}eview {B}}\ }\textbf {\bibinfo {volume} {99}},\
  \bibinfo {pages} {224409} (\bibinfo {year} {2019})}\BibitemShut {NoStop}%
\bibitem [{\citenamefont {Hwang}\ \emph {et~al.}(2022)\citenamefont {Hwang},
  \citenamefont {Go}, \citenamefont {Seong}, \citenamefont {Shibauchi},\ and\
  \citenamefont {Moon}}]{hwang2020identification}%
  \BibitemOpen
  \bibfield  {author} {\bibinfo {author} {\bibfnamefont {K.}~\bibnamefont
  {Hwang}}, \bibinfo {author} {\bibfnamefont {A.}~\bibnamefont {Go}}, \bibinfo
  {author} {\bibfnamefont {J.~H.}\ \bibnamefont {Seong}}, \bibinfo {author}
  {\bibfnamefont {T.}~\bibnamefont {Shibauchi}},\ and\ \bibinfo {author}
  {\bibfnamefont {E.-G.}\ \bibnamefont {Moon}},\ }\bibfield  {title} {\bibinfo
  {title} {Identification of a {K}itaev quantum spin liquid by magnetic field
  angle dependence},\ }\href@noop {} {\bibfield  {journal} {\bibinfo  {journal}
  {Nature Communications}\ }\textbf {\bibinfo {volume} {13}},\ \bibinfo {pages}
  {1} (\bibinfo {year} {2022})}\BibitemShut {NoStop}%
\bibitem [{\citenamefont {Yamada}\ and\ \citenamefont
  {Fujimoto}(2021)}]{yamada2021quantum}%
  \BibitemOpen
  \bibfield  {author} {\bibinfo {author} {\bibfnamefont {M.~G.}\ \bibnamefont
  {Yamada}}\ and\ \bibinfo {author} {\bibfnamefont {S.}~\bibnamefont
  {Fujimoto}},\ }\bibfield  {title} {\bibinfo {title} {Quantum liquid crystals
  in the finite-field k-$\gamma$ model for $\alpha$-rucl$_3$},\ }\href@noop {}
  {\bibfield  {journal} {\bibinfo  {journal} {arXiv preprint arXiv:2107.03045}\
  } (\bibinfo {year} {2021})}\BibitemShut {NoStop}%
\bibitem [{\citenamefont {Kim}\ \emph {et~al.}(2016)\citenamefont {Kim},
  \citenamefont {Kim},\ and\ \citenamefont {Kee}}]{kim2016revealing}%
  \BibitemOpen
  \bibfield  {author} {\bibinfo {author} {\bibfnamefont {H.-S.}\ \bibnamefont
  {Kim}}, \bibinfo {author} {\bibfnamefont {Y.~B.}\ \bibnamefont {Kim}},\ and\
  \bibinfo {author} {\bibfnamefont {H.-Y.}\ \bibnamefont {Kee}},\ }\bibfield
  {title} {\bibinfo {title} {Revealing frustrated local moment model for
  pressurized hyperhoneycomb iridate: {P}aving the way toward a quantum spin
  liquid},\ }\href@noop {} {\bibfield  {journal} {\bibinfo  {journal} {Physical
  {R}eview {B}}\ }\textbf {\bibinfo {volume} {94}},\ \bibinfo {pages} {245127}
  (\bibinfo {year} {2016})}\BibitemShut {NoStop}%
\bibitem [{\citenamefont {Lee}\ \emph {et~al.}(2020)\citenamefont {Lee},
  \citenamefont {Kaneko}, \citenamefont {Chern}, \citenamefont {Okubo},
  \citenamefont {Yamaji}, \citenamefont {Kawashima},\ and\ \citenamefont
  {Kim}}]{lee2020magnetic}%
  \BibitemOpen
  \bibfield  {author} {\bibinfo {author} {\bibfnamefont {H.-Y.}\ \bibnamefont
  {Lee}}, \bibinfo {author} {\bibfnamefont {R.}~\bibnamefont {Kaneko}},
  \bibinfo {author} {\bibfnamefont {L.~E.}\ \bibnamefont {Chern}}, \bibinfo
  {author} {\bibfnamefont {T.}~\bibnamefont {Okubo}}, \bibinfo {author}
  {\bibfnamefont {Y.}~\bibnamefont {Yamaji}}, \bibinfo {author} {\bibfnamefont
  {N.}~\bibnamefont {Kawashima}},\ and\ \bibinfo {author} {\bibfnamefont
  {Y.~B.}\ \bibnamefont {Kim}},\ }\bibfield  {title} {\bibinfo {title}
  {Magnetic field induced quantum phases in a tensor network study of {K}itaev
  magnets},\ }\href@noop {} {\bibfield  {journal} {\bibinfo  {journal} {Nature
  {C}ommunications}\ }\textbf {\bibinfo {volume} {11}},\ \bibinfo {pages} {1}
  (\bibinfo {year} {2020})}\BibitemShut {NoStop}%
\bibitem [{\citenamefont {Winter}\ \emph {et~al.}(2016)\citenamefont {Winter},
  \citenamefont {Li}, \citenamefont {Jeschke},\ and\ \citenamefont
  {Valent{\'\i}}}]{winter2016challenges}%
  \BibitemOpen
  \bibfield  {author} {\bibinfo {author} {\bibfnamefont {S.~M.}\ \bibnamefont
  {Winter}}, \bibinfo {author} {\bibfnamefont {Y.}~\bibnamefont {Li}}, \bibinfo
  {author} {\bibfnamefont {H.~O.}\ \bibnamefont {Jeschke}},\ and\ \bibinfo
  {author} {\bibfnamefont {R.}~\bibnamefont {Valent{\'\i}}},\ }\bibfield
  {title} {\bibinfo {title} {Challenges in design of {K}itaev materials:
  {M}agnetic interactions from competing energy scales},\ }\href@noop {}
  {\bibfield  {journal} {\bibinfo  {journal} {Physical {R}eview {B}}\ }\textbf
  {\bibinfo {volume} {93}},\ \bibinfo {pages} {214431} (\bibinfo {year}
  {2016})}\BibitemShut {NoStop}%
\bibitem [{\citenamefont {Gohlke}\ \emph {et~al.}(2018)\citenamefont {Gohlke},
  \citenamefont {Moessner},\ and\ \citenamefont
  {Pollmann}}]{gohlke2018dynamical}%
  \BibitemOpen
  \bibfield  {author} {\bibinfo {author} {\bibfnamefont {M.}~\bibnamefont
  {Gohlke}}, \bibinfo {author} {\bibfnamefont {R.}~\bibnamefont {Moessner}},\
  and\ \bibinfo {author} {\bibfnamefont {F.}~\bibnamefont {Pollmann}},\
  }\bibfield  {title} {\bibinfo {title} {Dynamical and topological properties
  of the {K}itaev model in a [111] magnetic field},\ }\href@noop {} {\bibfield
  {journal} {\bibinfo  {journal} {Physical {R}eview {B}}\ }\textbf {\bibinfo
  {volume} {98}},\ \bibinfo {pages} {014418} (\bibinfo {year}
  {2018})}\BibitemShut {NoStop}%
\bibitem [{\citenamefont {Wang}\ \emph
  {et~al.}(2017{\natexlab{b}})\citenamefont {Wang}, \citenamefont {Dong},
  \citenamefont {Yu},\ and\ \citenamefont {Li}}]{wang2017theoretical}%
  \BibitemOpen
  \bibfield  {author} {\bibinfo {author} {\bibfnamefont {W.}~\bibnamefont
  {Wang}}, \bibinfo {author} {\bibfnamefont {Z.-Y.}\ \bibnamefont {Dong}},
  \bibinfo {author} {\bibfnamefont {S.-L.}\ \bibnamefont {Yu}},\ and\ \bibinfo
  {author} {\bibfnamefont {J.-X.}\ \bibnamefont {Li}},\ }\bibfield  {title}
  {\bibinfo {title} {Theoretical investigation of magnetic dynamics in
  $\alpha$-{R}u{C}l$_3$},\ }\href@noop {} {\bibfield  {journal} {\bibinfo
  {journal} {Physical {R}eview {B}}\ }\textbf {\bibinfo {volume} {96}},\
  \bibinfo {pages} {115103} (\bibinfo {year} {2017}{\natexlab{b}})}\BibitemShut
  {NoStop}%
\bibitem [{\citenamefont {Winter}\ \emph
  {et~al.}(2017{\natexlab{b}})\citenamefont {Winter}, \citenamefont {Riedl},
  \citenamefont {Maksimov}, \citenamefont {Chernyshev}, \citenamefont
  {Honecker},\ and\ \citenamefont {Valent{\'\i}}}]{winter2017breakdown}%
  \BibitemOpen
  \bibfield  {author} {\bibinfo {author} {\bibfnamefont {S.~M.}\ \bibnamefont
  {Winter}}, \bibinfo {author} {\bibfnamefont {K.}~\bibnamefont {Riedl}},
  \bibinfo {author} {\bibfnamefont {P.~A.}\ \bibnamefont {Maksimov}}, \bibinfo
  {author} {\bibfnamefont {A.~L.}\ \bibnamefont {Chernyshev}}, \bibinfo
  {author} {\bibfnamefont {A.}~\bibnamefont {Honecker}},\ and\ \bibinfo
  {author} {\bibfnamefont {R.}~\bibnamefont {Valent{\'\i}}},\ }\bibfield
  {title} {\bibinfo {title} {Breakdown of magnons in a strongly spin-orbital
  coupled magnet},\ }\href@noop {} {\bibfield  {journal} {\bibinfo  {journal}
  {Nature {C}ommunications}\ }\textbf {\bibinfo {volume} {8}},\ \bibinfo
  {pages} {1} (\bibinfo {year} {2017}{\natexlab{b}})}\BibitemShut {NoStop}%
\bibitem [{\citenamefont {Hickey}\ and\ \citenamefont
  {Trebst}(2019)}]{hickey2019emergence}%
  \BibitemOpen
  \bibfield  {author} {\bibinfo {author} {\bibfnamefont {C.}~\bibnamefont
  {Hickey}}\ and\ \bibinfo {author} {\bibfnamefont {S.}~\bibnamefont
  {Trebst}},\ }\bibfield  {title} {\bibinfo {title} {Emergence of a
  field-driven {U}(1) spin liquid in the {K}itaev honeycomb model},\
  }\href@noop {} {\bibfield  {journal} {\bibinfo  {journal} {Nature
  {C}ommunications}\ }\textbf {\bibinfo {volume} {10}},\ \bibinfo {pages} {1}
  (\bibinfo {year} {2019})}\BibitemShut {NoStop}%
\bibitem [{\citenamefont {Fukui}\ \emph {et~al.}(2005)\citenamefont {Fukui},
  \citenamefont {Hatsugai},\ and\ \citenamefont {Suzuki}}]{fukui2005chern}%
  \BibitemOpen
  \bibfield  {author} {\bibinfo {author} {\bibfnamefont {T.}~\bibnamefont
  {Fukui}}, \bibinfo {author} {\bibfnamefont {Y.}~\bibnamefont {Hatsugai}},\
  and\ \bibinfo {author} {\bibfnamefont {H.}~\bibnamefont {Suzuki}},\
  }\bibfield  {title} {\bibinfo {title} {Chern numbers in discretized
  {B}rillouin zone: {E}fficient method of computing (spin) {H}all
  conductances},\ }\href@noop {} {\bibfield  {journal} {\bibinfo  {journal}
  {Journal of the Physical Society of Japan}\ }\textbf {\bibinfo {volume}
  {74}},\ \bibinfo {pages} {1674} (\bibinfo {year} {2005})}\BibitemShut
  {NoStop}%
\bibitem [{Com({\natexlab{a}})}]{Comment1}%
  \BibitemOpen
  \bibfield  {title} {\bibinfo {title} {For example, the
  $(h_1,h_3)-(-h_1,-h_3)$ points are same except for a time-reversal symmetry
  (e.g. if $\nu$ is finite, then $\nu \to -\nu$)},\ }\href@noop {} {\
  ({\natexlab{a}})}\BibitemShut {NoStop}%
\bibitem [{Com({\natexlab{b}})}]{Comment2}%
  \BibitemOpen
  \bibfield  {title} {\bibinfo {title} {It should be noted that the conventions
  for the {C}hern number in {K}itaev's original paper and the one used here
  differs by a sign factor.},\ }\href@noop {} {\  ({\natexlab{b}})}\BibitemShut
  {NoStop}%
\bibitem [{\citenamefont {Zhang}\ \emph {et~al.}(2022)\citenamefont {Zhang},
  \citenamefont {Halasz},\ and\ \citenamefont {Batista}}]{Zhang2022}%
  \BibitemOpen
  \bibfield  {author} {\bibinfo {author} {\bibfnamefont {S.-S.}\ \bibnamefont
  {Zhang}}, \bibinfo {author} {\bibfnamefont {G.~B.}\ \bibnamefont {Halasz}},\
  and\ \bibinfo {author} {\bibfnamefont {C.}~\bibnamefont {Batista}},\
  }\bibfield  {title} {\bibinfo {title} {Theory of the {K}itaev model in a
  [111] magnetic field},\ }\href@noop {} {\bibfield  {journal} {\bibinfo
  {journal} {Nature {C}ommunications}\ }\textbf {\bibinfo {volume} {13}}
  (\bibinfo {year} {2022})}\BibitemShut {NoStop}%
\bibitem [{\citenamefont {Y{\i}lmaz}\ and\ \citenamefont
  {Oktel}(2017)}]{yilmaz2017hofstadter}%
  \BibitemOpen
  \bibfield  {author} {\bibinfo {author} {\bibfnamefont {F.}~\bibnamefont
  {Y{\i}lmaz}}\ and\ \bibinfo {author} {\bibfnamefont {M.}~\bibnamefont
  {Oktel}},\ }\bibfield  {title} {\bibinfo {title} {Hofstadter butterfly
  evolution in the space of two-dimensional {B}ravais lattices},\ }\href@noop
  {} {\bibfield  {journal} {\bibinfo  {journal} {Physical {R}eview {A}}\
  }\textbf {\bibinfo {volume} {95}},\ \bibinfo {pages} {063628} (\bibinfo
  {year} {2017})}\BibitemShut {NoStop}%
\bibitem [{\citenamefont {Jiang}\ \emph {et~al.}(2020)\citenamefont {Jiang},
  \citenamefont {Liang}, \citenamefont {Chen}, \citenamefont {Qi},
  \citenamefont {Li},\ and\ \citenamefont {Wang}}]{jiang2020tuning}%
  \BibitemOpen
  \bibfield  {author} {\bibinfo {author} {\bibfnamefont {M.-H.}\ \bibnamefont
  {Jiang}}, \bibinfo {author} {\bibfnamefont {S.}~\bibnamefont {Liang}},
  \bibinfo {author} {\bibfnamefont {W.}~\bibnamefont {Chen}}, \bibinfo {author}
  {\bibfnamefont {Y.}~\bibnamefont {Qi}}, \bibinfo {author} {\bibfnamefont
  {J.-X.}\ \bibnamefont {Li}},\ and\ \bibinfo {author} {\bibfnamefont {Q.-H.}\
  \bibnamefont {Wang}},\ }\bibfield  {title} {\bibinfo {title} {Tuning
  topological orders by a conical magnetic field in the {K}itaev model},\
  }\href@noop {} {\bibfield  {journal} {\bibinfo  {journal} {Physical {R}eview
  {L}etters}\ }\textbf {\bibinfo {volume} {125}},\ \bibinfo {pages} {177203}
  (\bibinfo {year} {2020})}\BibitemShut {NoStop}%
\bibitem [{h00()}]{h001Thermodynamics}%
  \BibitemOpen
  \bibfield  {title} {\bibinfo {title} {F. {Y}{\i}lmaz, {A}. {P}. {K}ampf, {S}.
  {K}. {Y}ip, {T}he manuscript is in progress.},\ }\href@noop {} {\
  }\BibitemShut {NoStop}%
\bibitem [{\citenamefont {Saha}\ \emph {et~al.}(2019)\citenamefont {Saha},
  \citenamefont {Fan}, \citenamefont {Zhang},\ and\ \citenamefont
  {Chern}}]{saha2019hidden}%
  \BibitemOpen
  \bibfield  {author} {\bibinfo {author} {\bibfnamefont {P.}~\bibnamefont
  {Saha}}, \bibinfo {author} {\bibfnamefont {Z.}~\bibnamefont {Fan}}, \bibinfo
  {author} {\bibfnamefont {D.}~\bibnamefont {Zhang}},\ and\ \bibinfo {author}
  {\bibfnamefont {G.-W.}\ \bibnamefont {Chern}},\ }\bibfield  {title} {\bibinfo
  {title} {Hidden plaquette order in a classical spin liquid stabilized by
  strong off-diagonal exchange},\ }\href@noop {} {\bibfield  {journal}
  {\bibinfo  {journal} {Physical {R}eview {L}etters}\ }\textbf {\bibinfo
  {volume} {122}},\ \bibinfo {pages} {257204} (\bibinfo {year}
  {2019})}\BibitemShut {NoStop}%
\bibitem [{\citenamefont {Wang}\ \emph {et~al.}(2019)\citenamefont {Wang},
  \citenamefont {Normand},\ and\ \citenamefont {Liu}}]{wang2019one}%
  \BibitemOpen
  \bibfield  {author} {\bibinfo {author} {\bibfnamefont {J.}~\bibnamefont
  {Wang}}, \bibinfo {author} {\bibfnamefont {B.}~\bibnamefont {Normand}},\ and\
  \bibinfo {author} {\bibfnamefont {Z.-X.}\ \bibnamefont {Liu}},\ }\bibfield
  {title} {\bibinfo {title} {One proximate {K}itaev spin liquid in the
  {K}-{J}-{$\Gamma$} model on the honeycomb lattice},\ }\href@noop {}
  {\bibfield  {journal} {\bibinfo  {journal} {Physical {R}eview {L}etters}\
  }\textbf {\bibinfo {volume} {123}},\ \bibinfo {pages} {197201} (\bibinfo
  {year} {2019})}\BibitemShut {NoStop}%
\bibitem [{\citenamefont {Rousochatzakis}\ and\ \citenamefont
  {Perkins}(2017)}]{rousochatzakis2017classical}%
  \BibitemOpen
  \bibfield  {author} {\bibinfo {author} {\bibfnamefont {I.}~\bibnamefont
  {Rousochatzakis}}\ and\ \bibinfo {author} {\bibfnamefont {N.~B.}\
  \bibnamefont {Perkins}},\ }\bibfield  {title} {\bibinfo {title} {Classical
  spin liquid instability driven by off-diagonal exchange in strong spin-orbit
  magnets},\ }\href@noop {} {\bibfield  {journal} {\bibinfo  {journal}
  {Physical {R}eview {L}etters}\ }\textbf {\bibinfo {volume} {118}},\ \bibinfo
  {pages} {147204} (\bibinfo {year} {2017})}\BibitemShut {NoStop}%
\bibitem [{\citenamefont {Samarakoon}\ \emph {et~al.}(2018)\citenamefont
  {Samarakoon}, \citenamefont {Wachtel}, \citenamefont {Yamaji}, \citenamefont
  {Tennant}, \citenamefont {Batista},\ and\ \citenamefont
  {Kim}}]{samarakoon2018classical}%
  \BibitemOpen
  \bibfield  {author} {\bibinfo {author} {\bibfnamefont {A.~M.}\ \bibnamefont
  {Samarakoon}}, \bibinfo {author} {\bibfnamefont {G.}~\bibnamefont {Wachtel}},
  \bibinfo {author} {\bibfnamefont {Y.}~\bibnamefont {Yamaji}}, \bibinfo
  {author} {\bibfnamefont {D.~A.}\ \bibnamefont {Tennant}}, \bibinfo {author}
  {\bibfnamefont {C.~D.}\ \bibnamefont {Batista}},\ and\ \bibinfo {author}
  {\bibfnamefont {Y.~B.}\ \bibnamefont {Kim}},\ }\bibfield  {title} {\bibinfo
  {title} {Classical and quantum spin dynamics of the honeycomb {$\Gamma$}
  model},\ }\href@noop {} {\bibfield  {journal} {\bibinfo  {journal} {Physical
  {R}eview {B}}\ }\textbf {\bibinfo {volume} {98}},\ \bibinfo {pages} {045121}
  (\bibinfo {year} {2018})}\BibitemShut {NoStop}%
\bibitem [{\citenamefont {Luo}\ \emph {et~al.}(2021)\citenamefont {Luo},
  \citenamefont {Zhao}, \citenamefont {Kee},\ and\ \citenamefont
  {Wang}}]{luo2019gapless}%
  \BibitemOpen
  \bibfield  {author} {\bibinfo {author} {\bibfnamefont {Q.}~\bibnamefont
  {Luo}}, \bibinfo {author} {\bibfnamefont {J.}~\bibnamefont {Zhao}}, \bibinfo
  {author} {\bibfnamefont {H.-Y.}\ \bibnamefont {Kee}},\ and\ \bibinfo {author}
  {\bibfnamefont {X.}~\bibnamefont {Wang}},\ }\bibfield  {title} {\bibinfo
  {title} {Gapless quantum spin liquid in a honeycomb {$\Gamma$} magnet},\
  }\href@noop {} {\bibfield  {journal} {\bibinfo  {journal} {npj Quantum
  Materials}\ }\textbf {\bibinfo {volume} {6}},\ \bibinfo {pages} {1} (\bibinfo
  {year} {2021})}\BibitemShut {NoStop}%
\bibitem [{Com({\natexlab{c}})}]{Comment3}%
  \BibitemOpen
  \bibfield  {title} {\bibinfo {title} {There is a region with a numerical
  error in $\langle \hat{W} \rangle$ plot with a green color for small
  {$\Gamma$}$_p$.},\ }\href@noop {} {\  ({\natexlab{c}})}\BibitemShut {NoStop}%
\bibitem [{\citenamefont {Chaloupka}\ \emph {et~al.}(2013)\citenamefont
  {Chaloupka}, \citenamefont {Jackeli},\ and\ \citenamefont
  {Khaliullin}}]{chaloupka2013zigzag}%
  \BibitemOpen
  \bibfield  {author} {\bibinfo {author} {\bibfnamefont {J.}~\bibnamefont
  {Chaloupka}}, \bibinfo {author} {\bibfnamefont {G.}~\bibnamefont {Jackeli}},\
  and\ \bibinfo {author} {\bibfnamefont {G.}~\bibnamefont {Khaliullin}},\
  }\bibfield  {title} {\bibinfo {title} {Zigzag magnetic order in the iridium
  oxide {N}a$_2${I}r{O}$_3$},\ }\href@noop {} {\bibfield  {journal} {\bibinfo
  {journal} {Physical {R}eview {L}etters}\ }\textbf {\bibinfo {volume} {110}},\
  \bibinfo {pages} {097204} (\bibinfo {year} {2013})}\BibitemShut {NoStop}%
\bibitem [{\citenamefont {Czajka}\ \emph {et~al.}(2021)\citenamefont {Czajka},
  \citenamefont {Gao}, \citenamefont {Hirschberger}, \citenamefont
  {Lampen-Kelley}, \citenamefont {Banerjee}, \citenamefont {Yan}, \citenamefont
  {Mandrus}, \citenamefont {Nagler},\ and\ \citenamefont
  {Ong}}]{czajka2021oscillations}%
  \BibitemOpen
  \bibfield  {author} {\bibinfo {author} {\bibfnamefont {P.}~\bibnamefont
  {Czajka}}, \bibinfo {author} {\bibfnamefont {T.}~\bibnamefont {Gao}},
  \bibinfo {author} {\bibfnamefont {M.}~\bibnamefont {Hirschberger}}, \bibinfo
  {author} {\bibfnamefont {P.}~\bibnamefont {Lampen-Kelley}}, \bibinfo {author}
  {\bibfnamefont {A.}~\bibnamefont {Banerjee}}, \bibinfo {author}
  {\bibfnamefont {J.}~\bibnamefont {Yan}}, \bibinfo {author} {\bibfnamefont
  {D.~G.}\ \bibnamefont {Mandrus}}, \bibinfo {author} {\bibfnamefont {S.~E.}\
  \bibnamefont {Nagler}},\ and\ \bibinfo {author} {\bibfnamefont
  {N.}~\bibnamefont {Ong}},\ }\bibfield  {title} {\bibinfo {title}
  {Oscillations of the thermal conductivity in the spin-liquid state of
  $\alpha$-{R}u{C}l$_3$},\ }\href@noop {} {\bibfield  {journal} {\bibinfo
  {journal} {Nature Physics}\ }\textbf {\bibinfo {volume} {17}},\ \bibinfo
  {pages} {915} (\bibinfo {year} {2021})}\BibitemShut {NoStop}%
\bibitem [{\citenamefont {Feng}\ \emph {et~al.}(2007)\citenamefont {Feng},
  \citenamefont {Zhang},\ and\ \citenamefont {Xiang}}]{feng2007topological}%
  \BibitemOpen
  \bibfield  {author} {\bibinfo {author} {\bibfnamefont {X.-Y.}\ \bibnamefont
  {Feng}}, \bibinfo {author} {\bibfnamefont {G.-M.}\ \bibnamefont {Zhang}},\
  and\ \bibinfo {author} {\bibfnamefont {T.}~\bibnamefont {Xiang}},\ }\bibfield
   {title} {\bibinfo {title} {Topological characterization of quantum phase
  transitions in a spin-1/2 model},\ }\href@noop {} {\bibfield  {journal}
  {\bibinfo  {journal} {Physical {R}eview {L}etters}\ }\textbf {\bibinfo
  {volume} {98}},\ \bibinfo {pages} {087204} (\bibinfo {year}
  {2007})}\BibitemShut {NoStop}%
\bibitem [{\citenamefont {Schmidt}\ \emph {et~al.}(2018)\citenamefont
  {Schmidt}, \citenamefont {Scherer},\ and\ \citenamefont
  {Black-Schaffer}}]{schmidt2018topological}%
  \BibitemOpen
  \bibfield  {author} {\bibinfo {author} {\bibfnamefont {J.}~\bibnamefont
  {Schmidt}}, \bibinfo {author} {\bibfnamefont {D.~D.}\ \bibnamefont
  {Scherer}},\ and\ \bibinfo {author} {\bibfnamefont {A.~M.}\ \bibnamefont
  {Black-Schaffer}},\ }\bibfield  {title} {\bibinfo {title} {Topological
  superconductivity in the extended {K}itaev-{H}eisenberg model},\ }\href@noop
  {} {\bibfield  {journal} {\bibinfo  {journal} {Physical {R}eview {B}}\
  }\textbf {\bibinfo {volume} {97}},\ \bibinfo {pages} {014504} (\bibinfo
  {year} {2018})}\BibitemShut {NoStop}%
\bibitem [{\citenamefont {Schaffer}\ \emph {et~al.}(2012)\citenamefont
  {Schaffer}, \citenamefont {Bhattacharjee},\ and\ \citenamefont
  {Kim}}]{schaffer2012quantum}%
  \BibitemOpen
  \bibfield  {author} {\bibinfo {author} {\bibfnamefont {R.}~\bibnamefont
  {Schaffer}}, \bibinfo {author} {\bibfnamefont {S.}~\bibnamefont
  {Bhattacharjee}},\ and\ \bibinfo {author} {\bibfnamefont {Y.~B.}\
  \bibnamefont {Kim}},\ }\bibfield  {title} {\bibinfo {title} {Quantum phase
  transition in {H}eisenberg-{K}itaev model},\ }\href@noop {} {\bibfield
  {journal} {\bibinfo  {journal} {Physical {R}eview {B}}\ }\textbf {\bibinfo
  {volume} {86}},\ \bibinfo {pages} {224417} (\bibinfo {year}
  {2012})}\BibitemShut {NoStop}%
\bibitem [{\citenamefont {Yoshitake}\ \emph {et~al.}(2017)\citenamefont
  {Yoshitake}, \citenamefont {Nasu}, \citenamefont {Kato},\ and\ \citenamefont
  {Motome}}]{yoshitake2017majorana}%
  \BibitemOpen
  \bibfield  {author} {\bibinfo {author} {\bibfnamefont {J.}~\bibnamefont
  {Yoshitake}}, \bibinfo {author} {\bibfnamefont {J.}~\bibnamefont {Nasu}},
  \bibinfo {author} {\bibfnamefont {Y.}~\bibnamefont {Kato}},\ and\ \bibinfo
  {author} {\bibfnamefont {Y.}~\bibnamefont {Motome}},\ }\bibfield  {title}
  {\bibinfo {title} {Majorana dynamical mean-field study of spin dynamics at
  finite temperatures in the honeycomb {K}itaev model},\ }\href@noop {}
  {\bibfield  {journal} {\bibinfo  {journal} {Physical {R}eview {B}}\ }\textbf
  {\bibinfo {volume} {96}},\ \bibinfo {pages} {024438} (\bibinfo {year}
  {2017})}\BibitemShut {NoStop}%
\end{thebibliography}%

\appendix
\section{Kitaev Spin Liqud (KSL): a MFA}\label{MFtheoryDetails}

For the Kitaev model in Eq.\ref{KitaevHamiltonian}, a solution attempt by Jordan-Wigner slave fermions leads to a quartic interaction term on z-bonds \cite{feng2007topological, nasu2018successive}. However, this approach is not very useful for arbitrary field directions where one has to deal with Pauli strings \cite{nasu2018successive}. We therefore stick to Kitaev's original representation \cite{kitaev2006anyons}. According to this approach, each spin is represented by two Majorana fermions -as real fermionic field-. Consider the following mapping (with $\hbar = 1$),
\begin{equation}\label{KitaevDef}
2 S^\alpha_j = \sigma_j^\alpha = i b^\alpha_j c_j,\quad \alpha \in \{ x,y,z\}.
\end{equation}
Each spin component is thus composite object consisting of two Majorana fermions. However, the size of the Hilbert space is thereby doubled and requires to impose the following quartic constraint at each site to ensure the equivalence and the reality of each ket in the enlarged Hilbert space.
\begin{equation}\label{appConst}
D_j = b^x_j b^y_j b^z_j c_j = 1 \quad or \quad D_j \mid \psi \rangle = \mid \psi \rangle.
\end{equation}
In the new representation the Kitaev Hamiltonian is written as,
\begin{equation}
H_K = \frac{1}{4}\sum_{jl}^{\alpha} K^\alpha b_j^\alpha b_l^\alpha c_j c_l,\quad 
\end{equation}
The spectral decomposition of $H_K$ is obtained by admitting the constants of the motion, the bond operators $\hat{u}_{jl}^\alpha \equiv i b_j^\alpha b_l^\alpha$, $[ H_K, \hat{u}_{jl}^\alpha] = 0$. In other words, each bond gauge-field is a constant of motion and the Hilbert space can be segmented into the gauge sectors of each bond. Choosing all gauge-sector eigenvalues as $\hat{u}_{jl}^\alpha = +1$ for $K^\alpha>0$, the gauge fields are 'frozen' ($b_j^\alpha, b_l^\alpha\to b_{q}^\alpha, b_q^\alpha$) and decoupled from itinerant fermions. Then, the Hamiltonian becomes quadratic,
\begin{eqnarray}
H_K &=& -\frac{i}{4}\sum_{jl}^{\alpha} K^\alpha c_j c_l.
\end{eqnarray}
The itinerant fermion dispersion is identical \cite{kitaev2006anyons} to the dispersion in graphene with Dirac cones, 
\begin{equation}
\epsilon(q) = \pm \left|K^x e^{iqr_x}+K^y e^{iqr_y}+ K^z e^{iqr_z}\right|.
\end{equation}
The bond vectors are $\mathbf{r}_x = - 2 (2 \mathbf{a}_1-\mathbf{a}_2)/3 $, $\mathbf{r}_y = - 2 (2 \mathbf{a}_2-\mathbf{a}_1)/3$ and $\mathbf{r}_z= (2 \mathbf{a}_1+\mathbf{a}_2)/3 $. The Bravias vectors of the honeycomb lattice are $\mathbf{a}_1 = \frac{\sqrt{3}a}{2} (1,\sqrt{3})$ and $\mathbf{a}_2 = \frac{\sqrt{3}a}{2} (-1,\sqrt{3})$ with $a$ being the lattice constant. However, this approach is highly sensitive to proper gauge choices in each sector, therefore it is very limited for further extension to other interactions which are readily present in real materials. Moreover, the additional emerging exchange couplings demands a more all-inclusive approach in which fractional and conventional magnetism can be treated on an equal footing. A mean-field decoupling of Majorana fermions \cite{schmidt2018topological,nasu2018successive,janssen2019heisenberg,schaffer2012quantum,ralko2020novel,yoshitake2017majorana,liang2018intermediate,ido2020correlation} in this respect is a suitable method for two reasons: Firstly, a decoupling scheme provides the desired competition between the conventional and the anomalous pairings. Secondly, the main disadvantage of a typical MF method, i.e. the loss of strong correlations, is partially resolved because the fractionalized excitations are already included through replacing the spin operators by composite Majorana operators.

\begin{widetext}
\begin{eqnarray}
-i b_j^\alpha b_l^\alpha i c_j c_l &\approx& \langle i b_j^\alpha c_j\rangle i b^\alpha_l c_l + i b_j^\alpha c_j \langle i b^\alpha_l c_l\rangle-\langle i b_j^\alpha c_j\rangle\langle i b^\alpha_l c_l\rangle\nonumber \\ 
&-&\langle i b_j^\alpha b_l^\alpha\rangle i c_j c_l + i b_j^\alpha b_l^\alpha \langle -i c_j c_l\rangle-\langle b_j^\alpha b_l^\alpha\rangle\langle -i c_j c_l\rangle,\nonumber \\ 
&\equiv& m_A^\alpha i b^\alpha_l c_l + m_B^\alpha i b_j^\alpha c_j -m_A^\alpha m_B^\alpha
- \Phi^{\alpha\alpha}_\alpha i c_j c_l + W^\alpha i b_j^\alpha b_l^\alpha - \Phi^{\alpha\alpha}_\alpha W^\alpha.
\end{eqnarray}
\end{widetext}

Assuming isotropic bond strengths, $K^\alpha = K$, the Hartree-Fock decoupled mean-field Hamiltonian reads
\begin{eqnarray}\label{MFKitaevDecoupling}
H_K = \frac{K}{4}\sum_{jl}^{\alpha} &&m_A^\alpha i b^\alpha_l c_l + m_B^\alpha i b_j^\alpha c_j \\
&-& \Phi^{\alpha\alpha}_\alpha i c_j c_l + W^\alpha i b_j^\alpha b_l^\alpha - m_A^\alpha m_B^\alpha - \Phi^{\alpha\alpha}_\alpha W^\alpha. \nonumber
\end{eqnarray}
The MF parameters are defined as,
\begin{eqnarray}
m_A^\alpha &=& \langle i b^\alpha_j c_j \rangle,\quad m_B^\alpha = \langle i b^\alpha_l c_l \rangle,\\
W^\gamma &=& -\langle i c_j c_l \rangle_\gamma,\quad \Phi^{\alpha\beta}_\gamma = \langle i b^\alpha_j b^\beta_l \rangle_\gamma.
\end{eqnarray}
Note that $m_{A,B}^\alpha$ are components of the local magnetic moments on site A and B whereas $W^\gamma = W$ and $\Phi^{\alpha\beta}_\gamma$ are the indicators of fractionalization defined on the bond $\gamma$. 
For simplicity, we use shorthand notations for $\Phi_\gamma^{\gamma \gamma}$, e.g. $\Phi_x^{xx} = X$ etc. While the aim is to obtain a quadratic Hamiltonian, there is an issue regarding the quartic constraint on Majorana fermions. The strategy is to rewrite the constraint in terms of two-fermion operators. Multiplying both sides with any two of the four fermion operators, one can rewrite the single quartic constraint in Eq.\ref{appConst} as three quadratic constraint relations,
\begin{eqnarray}
b^z c +b^x b^y&=& 0,\\
b^x c +b^y b^z&=& 0,\\
b^y c +b^z b^x&=& 0,
\end{eqnarray}
where the site index is dropped for simplicity. Based on these relations, the constraints can be introduced into the Hamiltonian with Lagrange multipliers:
\begin{eqnarray}
H_{MF} &=& H_K - H_\lambda, \nonumber \\
H_\lambda &=& \frac{i}{2} \sum^\alpha_j \lambda^\alpha_j \left( b^\alpha_j c_j + \frac{\epsilon^{\alpha \beta \gamma}}{2} b^{\beta}_j b_j^\gamma \right).
\end{eqnarray}

The constrained mean-field Hamiltonian must satisfy the saddle point condition,
\begin{equation}
\frac{\partial \langle H_{MF} \rangle}{\partial \lambda^\alpha} = \frac{\partial E_{tot}(\lambda)}{\partial \lambda^\alpha} = 0,
\end{equation}
which is equivalent to the quadratic constraints to hold for expectation values,
\begin{eqnarray}
\langle ib^z c \rangle &=& -\langle ib^x b^y \rangle ,\\
\langle ib^x c \rangle &=& -\langle ib^y b^z \rangle,\\
\langle ib^y c \rangle &=& -\langle ib^z b^x \rangle.
\end{eqnarray}

For the evaluation of $\langle H_{MF} \rangle$ we consider the general Hamiltonian (H) of the form
\begin{equation}\label{HKorg2} 
H = -i \sum_{j\alpha,l \beta} \epsilon_{j\alpha,l \beta} c_{j \alpha} c_{l \beta}
\end{equation}
on a periodic lattice. Here $c_{j\alpha}$ operators, unlike $c_j$, includes all types of Majorana fermions ($b^x,b^y,b^z,c$). $j,l$ labels are unit cells indices for sublattices A and B, respectively. The sub-indices $\alpha,\beta$ represent the Majorana flavors (e.g. $c_{j,\alpha} \in \{b_j^x,b_j^y,b_j^z,c_j \}$) within the corresponding sublattices. $\epsilon_{j\alpha,l \beta}$ is necessarily anti-symmetric under the interchange $j\alpha \leftrightarrow l \beta$. This Hamiltonian can be diagonalized using the transformation
\begin{equation}\label{KitaevDef2}
c_{j\alpha} = \sqrt{2} \sum_{\mathbf{q},\lambda} u_{j\alpha,\mathbf{q},\lambda} \alpha_{\mathbf{q},\lambda}
\end{equation}
where $\sqrt{2}$ is used to ensure the correct factor in the anti-commutation relations. The coefficients $u_{j\alpha,\mathbf{q},\lambda}$ satisfy the eigenvalue equation,
\begin{equation}\label{EnnApp}
E_{\mathbf{q},\lambda} u_{j\alpha,\mathbf{q},\lambda} = -i \epsilon_{j\alpha,l \beta} u_{l \beta,\mathbf{q},\lambda}.
\end{equation}
$\mathbf{q}$ labels the wavevector and $\lambda \in \{-4,-3,-2,-1,1,2,3,4\}$ labels different solutions at each $\mathbf{q}$. The eigenvectors imply a plane-wave form,
\begin{equation}\label{matElemDiagMat}
u_{j\alpha,\mathbf{q},\lambda} = \frac{1}{\sqrt{N}} e^{i \mathbf{q} \cdot \mathbf{r}_{j\alpha}} \tilde{u}_{\alpha\lambda}(\mathbf{q}).
\end{equation}
Here, $N$ is the number of unit cells, $\mathbf{r}_{j\alpha}$ the position vector of the sublattice A site $j$. For convenience we choose $\sum_\alpha \left| \tilde{u}_{\alpha\lambda}(\mathbf{q}) \right|^2 = 1$. Eqs.\ref{HKorg2} and \ref{KitaevDef2} imply that $\alpha_{\mathbf{q},\lambda}$ satisfies the commutation relation
\begin{equation}
[\alpha_{\mathbf{q},\lambda},H] = E_{\mathbf{q},\lambda} \alpha_{\mathbf{q},\lambda}.
\end{equation}
Explicitly, $\alpha_{\mathbf{q},\lambda} = \frac{1}{\sqrt{2}} \sum_{j,\alpha} u_{j\alpha,\mathbf{q},\lambda}^* c_{j\alpha}$ can be derived from Eq.\ref{HKorg2}. Eq.\ref{KitaevDef2} transforms $H$ into
\begin{equation}\label{HKAsEnergy}
H = \frac{1}{2} \sum_{\mathbf{q},\lambda} E_{\mathbf{q},\lambda} \alpha_{\mathbf{q},\lambda}^\dagger \alpha_{\mathbf{q},\lambda}.
\end{equation}
Taking the complex conjugate of Eq.\ref{EnnApp} tells that the eigenvalues come in pairs: If $u_{l\alpha,\mathbf{q},\lambda}$ is a solution with eigenenergy $ E_{\mathbf{q},\lambda}$, then $u_{l\alpha,-\mathbf{q},-\lambda} = u_{l\alpha,\mathbf{q},\lambda}^*$ (belonging to wavevector $-\mathbf{q}$) has eigenenergy $-E_{\mathbf{q},\lambda}$, and the associated operator $\alpha_{-\mathbf{q},-\lambda}$ is equivalent to $\alpha_{\mathbf{q},\lambda}^\dagger$. Rewriting the diagonalized Hamiltonian in Eq.\ref{HKAsEnergy} in terms of positive energy operators only, we obtain
\begin{eqnarray}
H &=& \frac{1}{2} \sum_{\mathbf{q},\lambda}^{ E_{\mathbf{q},\lambda}>0} \left[ E_{\mathbf{q},\lambda} \alpha_{\mathbf{q},\lambda}^\dagger \alpha_{\mathbf{q},\lambda} - E_{\mathbf{q},\lambda} \alpha_{\mathbf{q},\lambda} \alpha_{\mathbf{q},\lambda}^\dagger\right],\\
&=& \sum_{\mathbf{q},\lambda}^{ E_{\mathbf{q},\lambda}>0} E_{\mathbf{q},\lambda} \alpha_{\mathbf{q},\lambda}^\dagger \alpha_{\mathbf{q},\lambda}-\frac{1}{2} \sum_{\mathbf{q},\lambda}^{ E_{\mathbf{q},\lambda}>0} E_{\mathbf{q},\lambda} \label{a26}
\end{eqnarray}
where the sums extend over only the positive energy eigenstates. $\alpha_{\mathbf{q},\lambda}$ operators satisfy the usual anticommutation relations $\{\alpha_{\mathbf{q},\lambda_1},\alpha^\dagger_{\mathbf{q}',\lambda_2}\}=\delta_{\mathbf{q},\mathbf{q}'} \delta_{\lambda_1,\lambda_2}$. The groundstate, therefore, is defined as $\alpha_ {\mathbf{q},1} | GS \rangle = 0$. This representation is useful for finite temperature extensions.

{\bf Considering this work}, $\epsilon_{j\alpha,l\beta}$ corresponds to the matrix elements of the mean field Hamiltonian in Eq.\ref{MFKitaevDecoupling}. The Hamiltonian has a $8 \times 8$ matrix representation $\mathcal{H}_{\mathbf{q}}$ in the momentum space. Our convention is,
\begin{equation}
H_{MF} = \sum_\mathbf{q} \psi_\mathbf{q}^\dagger \mathcal{H}_\mathbf{q} \psi_\mathbf{q},
\end{equation}
where $ \psi_\mathbf{q} = \left( b^x_\mathbf{q}, b^y_\mathbf{q}, b^z_\mathbf{q}, c_\mathbf{q}, \bar{b}^x_\mathbf{q}, \bar{b}^y_\mathbf{q}, \bar{b}^z_\mathbf{q}, \bar{c}_\mathbf{q}\right)$ indicates the Fourier transform of the Majorana operators (or $c_{\mathbf{q} \alpha}$ where each operator being represented by index $\alpha$). $\mathcal{H}_\mathbf{q} $ can be diagonalized as $\Gamma_{\mathbf{q}}^\dagger \mathcal{H}_{\mathbf{q}} \Gamma_{\mathbf{q}} = D_{\mathbf{q}}$ where the diagonalizing matrix $\Gamma_{\mathbf{q}}$ has the form,
\begin{equation}
\Gamma_{\mathbf{q}}= \big( \mid \mathbf{q},-4 \rangle \mid \mathbf{q},-3 \rangle \mid \mathbf{q},-2 \rangle\mid \mathbf{q},-1 \rangle\mid \mathbf{q},1\rangle\mid \mathbf{q},2 \rangle\mid \mathbf{q},3 \rangle\mid \mathbf{q},4 \rangle \big),
\end{equation}
and $\mid \mathbf{q},\lambda \rangle = \alpha_{\mathbf{q},\lambda}^\dagger \mid 0 \rangle$ are $8 \times 1$ vectors where $u_{j\alpha,\mathbf{q}\lambda} = \langle j \alpha \mid \mathbf{q},\lambda \rangle$. Thereby the matrix elements of $\Gamma_{\mathbf{q}}$ are $\tilde{u}_{\alpha\lambda}(\mathbf{q})$ with $\alpha,\lambda$ being the entries for rows and columns, respectively.

The mean field analysis is concluded by the self-consistency relations for the MF parameters. For bond correlations, a general self-consistency reads
\begin{eqnarray}
\langle i c_{j\alpha} c_{l\beta} \rangle_\gamma &=& \frac{2i}{N} \sum_{\mathbf{q},\lambda}^{E_{\mathbf{q},\lambda}>0} e^{i \mathbf{q} \cdot \mathbf{r}_{\gamma}} \tilde{u}_{\alpha\lambda}(\mathbf{q}) \tilde{u}^*_{\beta\lambda}(\mathbf{q}),
\end{eqnarray}
where the site indices $j,l$ determine the bond direction $\mathbf{r}_{\gamma} = \mathbf{r}_{j \alpha} - \mathbf{r}_{l\beta}$. The parameters $\alpha,\beta, \gamma$ for the bond correlations are defined as,
\begin{eqnarray}
X &\to& \alpha = 1, \beta = 5, \gamma = x,\\
Y &\to& \alpha = 2, \beta = 6, \gamma = y,\\
Z &\to& \alpha = 3, \beta = 7, \gamma = z,\\
W^\theta&\to& \alpha = 4, \beta = 8, \gamma = \theta \quad \text{with a "-" sign}.
\end{eqnarray}
On-site correlations (e.g. magnetic moments) are,
\begin{eqnarray}
\langle i c_{j\alpha} c_{j\beta} \rangle &=& \frac{2i}{N} \sum_{\mathbf{q},\lambda}^{E_{\mathbf{q},\lambda}>0} \tilde{u}_{\alpha\lambda}(\mathbf{q}) \tilde{u}^*_{\beta\lambda}(\mathbf{q}).
\end{eqnarray}
The parameters are defined as 
\begin{eqnarray}
m^x_A &\to& \alpha = 1, \beta = 4, \quad m^x_B \to \alpha = 5, \beta = 8\\
m^y_A &\to& \alpha = 2, \beta = 4, \quad m^y_B \to \alpha =6, \beta = 8\\
m^z_A &\to& \alpha = 3, \beta = 4, \quad m^z_B \to \alpha = 7, \beta = 8 \quad \text{etc.}
\end{eqnarray}
The self-consistency cycle is completed by the recognizing that $\tilde{u}_{\alpha\lambda}(\mathbf{q}) $ are functions of the MF parameters.

\begin{figure*}
\includegraphics[width=0.65\textwidth]{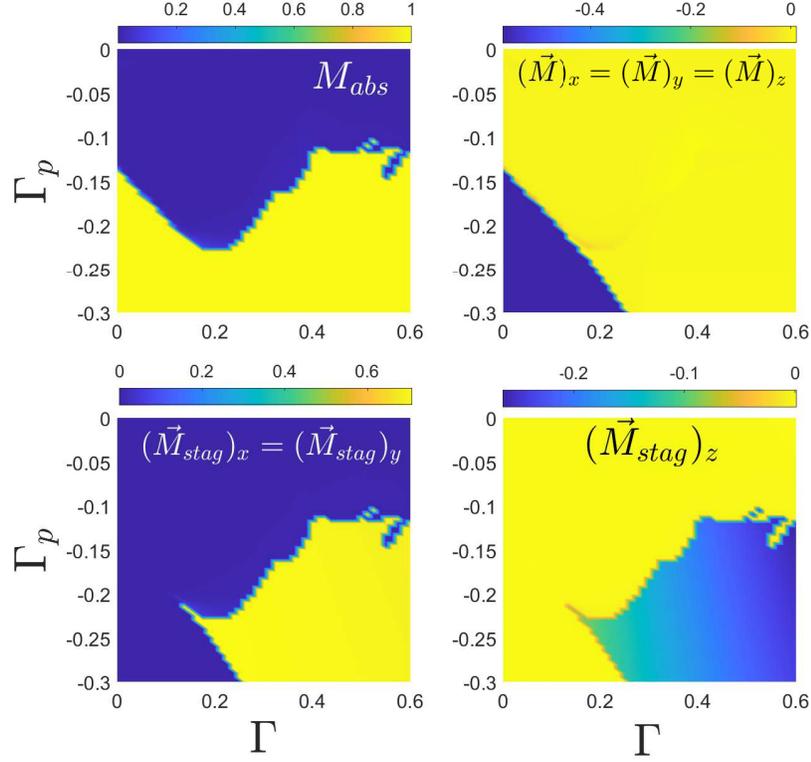}
\caption{(Color online) The absolute ($ M_{abs}$), the components of the average ($ M^\alpha$) and the components of the staggered ($ M^\alpha_{stag}$) magnetization for the four-site unit cell in $\Gamma-\Gamma_p$ space. $ M_{abs} = \sum_{i=1}^4 \left| \mathbf{m}_i \right| /4$, $(\mathbf{M})_\alpha = \sum_{i=1}^4 (\mathbf{m}_i)_\alpha/4$ and $ (\mathbf{M_{stag}})_\alpha= \left( m_1 + m_2 -m_3 -m_4\right)_\alpha /4$. The average magnetization components, shown in the first row, are equal ($M^\alpha = M_0$) and clearly indicates a Ferromagnetic polarized regime. The staggered magnetization components (second row) signals the presence of an AF zig-zag phase, where the $x-y$ components are equal and different from $z$-component, $M_{abs}^x=M_{abs}^y\ne M_{abs}^z$. Moreover, they have an opposite sign.}
\label{ZZzpMxyz}
\end{figure*}

\section{The details of the K-$\Gamma$-$\Gamma_p$ model under magnetic field}\label{KGGpM_is}
We provide the magnetic moment details regarding the zig-zag z-phase. The ZZ z-phase is characterized by 1d ferromagnetic chains connected by antiferromagnetic z-bond, $\mathbf{m}_1 = \mathbf{m}_2 = -\mathbf{m}_3=-\mathbf{m}_4$. In addition, $m_1^{x,y}=m_2^{x,y} \ne m_{1,2}^z$ and the same goes for the sites $3$ and $4$ with an opposite direction. In Fig.\ref{ZZzpMxyz}, it can be understood either by each component for the average- and the staggered magnetization. The z-bonds of ZZ-z phase are aligned antiferromagnetically. It is classically known \cite{janssen2017magnetization} that these spins, for $\Gamma = 0$, lie completely in the $x-y$ plane. It is the plane containing the $\theta = 35^o$ line. As an increasing function of $\Gamma$, they acquire a non-zero $z$-component and finally for $\Gamma/\left| K \right| \to \infty$ the spins point along $[111]$. This behavior is clearly visible in Fig.\ref{ZZzpMxyz}, the bottom plots for $(\mathbf{M}_{stag})^{x,y,z}$. The $x$ and $y$ components of the staggered magnetization have opposite signs with the $z$-component.

\begin{figure*}
\includegraphics[width=0.9\textwidth]{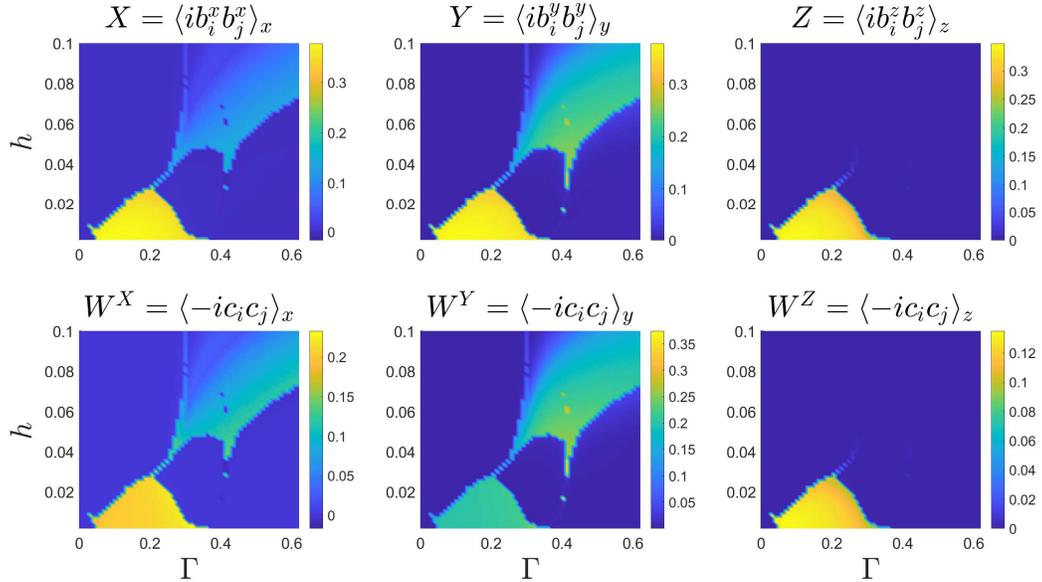}
\caption{(Color online) The fractionalization correlations averaged over the four-site unit-cell. A non-zero value indicates the presence of fractionalization. The partially polarized phase, known as the "intermediate phase" has partial fractionalization.}
\label{KGGpHThet1a}
\end{figure*}

In Fig. \ref{KGGpHThet1a}, the curious PP intermediate phase is investigated further. The interesting question is if this PP phase host any type of fractionalization. The positive answer is supported by the finite bond correlations for $X,Y$ as well as $W^X$ and $W^Y$ throughout the PP regime.

\end{document}